\documentclass[12pt]{iopart}
\eqnobysec
  
\usepackage{amsfonts}
\usepackage{amssymb}
\usepackage{graphicx}
\usepackage{pstricks} 
\usepackage{axodraw}

\newcommand{\be}{\begin{equation}}
\newcommand{\bea}{\begin{eqnarray}}
\newcommand{\ee}{\end{equation}}
\newcommand{\eea}{\end{eqnarray}}
\newcommand{\bpi}{\begin{picture}}
\newcommand{\bce}{\begin{center}}
\newcommand{\epi}{\end{picture}}
\newcommand{\ece}{\end{center}}

\newcommand{\Rxi}{R_\xi}

\newcommand{\sw}{s_{\rm{\scriptscriptstyle{W}}}}
\newcommand{\cw}{c_{\rm{\scriptscriptstyle{W}}}}
\newcommand{\gw}{g_{\rm{\scriptscriptstyle{W}}}}

\newcommand{\tw}{\theta_{\rm{\scriptscriptstyle{W}}}}
\newcommand{\Mw}{M_W}
\newcommand{\Mz}{M_Z}
\newcommand{\Mv}{M_V}

\newcommand{\sumv}{\sum_{V}}
\newcommand{\sumvp}{\sum_{V'}}
\newcommand{\sums}{\sum_{\rm S}}
\newcommand{\sumsp}{\sum_{\rm S'}}

\newcommand{\mom}{(k_1,k_2)}
\newcommand{\momo}{(q)}
\newcommand{\momt}{(q-k,k)}

\newcommand{\sumi}{\sum_{{\rm class}\ (i)}}
\newcommand{\sumii}{\sum_{{\rm class}\ (ii)}}

\newcommand{\ua}{u^{\scriptscriptstyle{A}}}
\newcommand{\uz}{u^{\scriptscriptstyle{Z}}}
\newcommand{\uv}{u^{\scriptscriptstyle{V}}}

\newcommand{\uvp}{u^{\scriptscriptstyle{V'}}}
\newcommand{\cv}{C_{V}}
\newcommand{\cvp}{C_{V'}}
\newcommand{\cs}{C_{\rm S}}
\newcommand{\csp}{C_{\rm S'}}
\newcommand{\uno}{(-1)^{\rm S}}

\newcommand{\rS}{{\rm S}}

\newcommand{\up}{u^+}
\newcommand{\um}{u^-}

\newcommand{\Wm}{{W^-}}
\newcommand{\phip}{{\phi^+}}
\newcommand{\phim}{{\phi^-}}
\newcommand{\sV}{V}
\newcommand{\Vp}{{V'}}
\newcommand{\schi}{{\chi}}
\newcommand{\sH}{H}

\begin{document}

\begin{flushright}
ECT*-04-02
\end{flushright}

\title[Electroweak pinch technique to all orders]
{Electroweak pinch technique to all orders}

\author{Daniele Binosi}

\address{ECT*\\ Villa Tambosi, Strada delle Tabarelle 286,
I-38050 Villazzano (Trento), Italy}

\ead{\mailto{binosi@ect.it}}

\begin{abstract}
The  generalization  of the  pinch  technique  to  all orders  in  the
electroweak sector of the Standard Model within the class of 
the renormalizable 't Hooft gauges, is presented.  
In particular, both the all-order PT gauge-boson-- and scalar--fermions
vertices, as well as the diagonal and mixed gauge-boson and scalar 
self-energies are explicitly constructed.
This is achieved through the generalization to the Standard Model
of the procedure recently applied to the QCD case, which consist of two steps: 
({\it i}) the   
identification of special Green's functions, which serve as a common
kernel to all  self-energy and vertex diagrams, and ({\it ii}) the 
study of the (on-shell) Slavnov-Taylor identities  they satisfy.   
It is then shown  that the  ghost, scalar and scalar--gauge-boson Green's  functions appearing  
in these  identities capture  precisely the  result  of the  pinching  action at 
arbitrary order. It turns  out that
the aforementioned 
Green's functions  play a crucial role, their
net effect being the non-trivial modification of the ghost, scalar and scalar--gauge-boson 
diagrams of the gauge-boson-- or scalar--fermions
vertex we have started from, in such a way as  to dynamically generate the
characteristic ghost and scalar sector of the background field method. 
The pinch technique gauge-boson and scalar self-energies are also explicitly constructed by
resorting to the method of the background-quantum identities.
\end{abstract}

\submitto{\JPG}

\maketitle

\section{Introduction}

The possibility  of defining a consistent perturbative  expansion of a
non-Abelian  gauge theory in  the continuum, is  intimately connected
with the prescription  of a gauge fixing procedure; the latter will in fact
remove the redundant degrees of freedom
originating from gauge  invariance, thus allowing for the  derivation of a
self-consistent set of Feynman rules. 
At this point however, a new type of redundancy appears, for as the
Green's functions  of the theory, which constitutes the building  blocks of
the  perturbative  expansion, will  carry  a  great  deal of  unphysical
information, their dependence  on  the gauge  fixing parameter  
($\xi$ in $R_\xi$ gauges, $\xi_Q$ in the background field gauge, $n_\mu$ in axial gauges, etc.)
being a paradigmatic example. 
As long as one deals strictly with physical quantities
(such as $S$-matrix elements), this gauge fixing parameter dependence 
is not a problem at all, since it is never to be seen; on its turn, the latter 
fact suggests therefore that large cancellations,
driven by powerful field theoretical mechanisms, take place 
among the various Green's functions of the theory.

A tool to unveil such cancellations has been the   Pinch   Technique   (PT)
\cite{Cornwall:1982zr}, a (diagrammatic)
technique by which a given physical amplitude is reorganized into  sub-amplitudes,   
which  have  the  same  kinematic properties as conventional $n$-point functions
(self-energies, vertices and boxes) being in addition endowed with 
important physical properties, such as the independence of  the
gauge-fixing  scheme  and parameters  chosen  to  quantize the  theory,
gauge-invariance, [{\it  i.e.}, the PT Green's functions satisfy  simple 
tree-level-like Ward identities 
(WIs) instead of the usual complicated non linear Slavnov-Taylor identities (STIs)
involving ghost  fields], the display of only  
{\it  physical}  thresholds, and, finally, a good behavior at high energies. 
The aforementioned reorganization has been achieved diagrammatically at the one \cite{Cornwall:1982zr}
and two \cite{Papavassiliou:1999az} loops, 
by recognizing that {\it longitudinal momenta} circulating inside vertex and 
box diagrams can change the topology of the latter by ``pinching out" internal fermion lines, generating in this way propagator-like terms; 
these terms are then reassigned to conventional 
self-energies graphs in order to give rise to the effective PT Green's functions,
which manifestly possess the properties (generally associated to physical observables) 
described above.  

The conceptual and phenomenological  
advantages of being able to work with such special Green's 
functions in a non-Abelian field theoretical context [such as the Standard Model (SM)],
are to be found in those physical circumstances where one has to go beyond the confines of fixed order perturbation theory, to look for a systematic rearrangement/resummation
of the perturbative series.
An exemplification of this situation,  that captures simultaneously the multitude of problems involved, is 
given by the problem of computing  transition
amplitudes in  the vicinity of resonances, where the
tree-level  propagator  of  the  particle mediating  the  interaction,
$\Delta= (s-M^2)^{-1}$, becomes  singular as  the center-of-mass
energy  $\sqrt{s}\sim  M$ \cite{Veltman:th}. The  standard  way  for  regulating  this
physical  kinematic  singularity is  to  use  a  Breit-Wigner type  of
propagator, which essentially amounts to the replacement $(s-M^2)^{-1}
\to  (s-M^2+iM\Gamma)^{-1}$,  where  $\Gamma$  is  the  width  of  the
unstable (resonating) particle. The field-theoretical  mechanism which
enables this  replacement is the  Dyson resummation of  the (one-loop)
self-energy  $\Pi(s)$ of  the unstable  particle, which  leads  to the
substitution $(s-M^2)^{-1} \to [s-M^2+\Pi(s)]^{-1}$ [the running width
of  the particle is  then defined  as $M\Gamma(s)  ={\rm Im}\,\Pi(s)$].
This means then that the Breit-Wigner  
procedure is in fact equivalent
to a {\it reorganization}  of the perturbative series: the Dyson 
summation of the self-energy $\Pi(s)$ amounts to 
removing a particular piece
from  each order  of the  perturbative expansion,  since from  all the
Feynman graphs contributing to a given order $n$ we only pick 
the part
which contains $n$  self-energy bubbles  $\Pi(s)$, 
and  then take  $n \to
\infty$.  
Given  that non-trivial cancellations  involving the various
Green's function is  generally taking place at any  given order of the
conventional perturbative  expansion, the act of removing  one of them
from each order  can distort those cancellations, finally introducing
spurious gauge fixing parameter dependences in the resummed propagator
$\Pi(s)$; this  is indeed what
happens  when constructing  non-Abelian running  widths in general, 
and the SM ones in particular.   
The  application of the PT  ensures   that  all  unphysical
contributions contained inside $\Pi(s)$ have been identified and properly
discarded, {\it before} $\Pi(s)$ undergoes resummation 
\cite{Papavassiliou:1995fq}.
Thus, at one-loop order, the resummation formalism
based  on  the  PT  accomplishes  the
simultaneous reconciliation  of crucial physical  requirements such as
gauge independence, gauge invariance,
renormalization-group  invariance,  and  the optical  and  equivalence
theorems  \cite{Cornwall:1974km}.

As a second example, we consider the proper definition of form factors in non-Abelian theories. This definition  
poses in general many problems, basically related to the gauge 
independence/invariance of the final answer \cite{Fujikawa:1972fe}.
The application of the PT in this context, has allowed for an unambiguous   
definition of such quantities, 
some representative SM examples being the 
magnetic dipole and electric 
quadrupole moments of the $W$ \cite{Papavassiliou:1993ex}, 
the top-quark magnetic moment \cite{Papavassiliou:1994qe}, 
and the neutrino charge radius \cite{Bernabeu:2000hf}. 
Most notably, 
the gauge independent, renormalization group invariant, 
and target independent SM neutrino charge radius constructed through the PT
constitutes a genuine {\it physical} observable,   
since it can be 
extracted (at least in principle) from experiments 
\cite{Bernabeu:2002nw}.

Other interesting applications include 
the gauge-invariant formulation of the process independent part of the 
$\rho$ parameter at one- \cite{Degrassi:1993kn} 
and two-loops \cite{Papavassiliou:1996hj}, 
various finite temperature calculations 
\cite{Nadkarni:1988ti},
the correct definition of non-Abelian effective charges \cite{Papavassiliou:1997fn}, 
a novel approach to the comparison of electroweak data with 
theory \cite{Hagiwara:1994pw},
resonant CP violation \cite{Pilaftsis:1997dr},
the construction of the two-loop PT quark self-energy \cite{Binosi:2001hy}, 
and, more recently, the issue of (supersymmetric) particle mixings
\cite{Yamada:2001px,Espinosa:2002cd}, the determination of the gauge independent
form factors for M\o{}ller scattering and their relation to the running of the weak mixing angle \cite{Ferroglia:2003wa}, and 
the discussion of the PT as a physical renormalization
scheme for GUTs \cite{Binger:2003by}.

One question that has not been answered yet, is the one regarding the possibility of
generalizing the PT (and thus many of the aforementioned results) to all orders 
in the context of spontaneously broken theories in general, and of the electroweak sector of the SM in particular. 
With respect to the QCD case, two are the main difficulties of working in the spontaneous symmetry breaking scenario, which makes the all-order generalization of the PT procedure a much more challeging excercise. The first one is related to the proliferation  of Feynman
diagrams due to the richer particle spectrum of such theories; the second one is related to the complications arising  from the presence of Goldtone's bosons, which implies that the BRST symmetry (and therefore the STIs) will now be realized through them. Nevertheless, by unveiling the intertwining between the PT and the  
BRST symmetry underlying the theory, we will show that the all-order generalization becomes possible
along the same lines put forward in
\cite{Binosi:2002ft} for the QCD case.

To this end, in Section~\ref{Pro}, we introduce our
notations and conventions together with a brief review of the
Batalin-Vilkovisky and Nielsen formalism, which will be later used
in deriving and analyzing the PT Green's functions. Then, 
in Section~\ref{EwPT}, we will review the PT in the case of non-conserved
currents, and proceed to isolate all the possible sources of pinching
momenta, {\it i.e.}, the  tree-level (gauge-bosons) 
propagators and the trilinear vertices of the type ${\cal V}^3$ and 
${\cal S}^2{\cal V}$ (where ${\cal V}$ and ${\cal S}$ stands for
gauge-bosons and scalars respectively, see below).

Each one of these sources is then treated separately in Section~\ref{Canc}. 
We first show
how the $\Rxi$ Feynman gauge ($\Rxi$FG) $\xi=1$  
is reached by establishing a connection between the PT and 
the Nielsen identities formalism; in this way the longitudinal momenta coming from the
propagators are eliminated.
Then we concentrate on the pinching momenta of the trilinear vertices 
by reexamining the PT algorithm in the light of the BRST symmetry, and
arguing that the original one-loop PT rearrangements  are  
but lower-order  manifestations  of a  fundamental
cancellation taking place between  graphs of distinct kinematic nature
when computing the divergence  of a special four-point function that will
also be isolated.

Sections~\ref{Gbff} and~\ref{Sff} are somewhat more technical and present in
great detail the construction of the (all orders) PT
gauge-boson-- and scalar--fermion-fermion vertices both in the charged
as well as in the neutral sector. 
Once  the effective Green's functions have been constructed, 
they will be  compared to the  corresponding Green's functions  obtained in
the Feynman  gauge of  the background field method (BFM)  
\cite{Dewitt:1967ub}. 
The  latter is  a  special gauge-fixing 
procedure, implemented  at the  level of  the generating functional.  In 
particular, it preserves  the symmetry of  the action under ordinary  gauge
transformations  with respect to  the background (classical) gauge field
$\widehat{{\cal V}}_{\mu}$,  while the quantum gauge fields ${\cal V}_{\mu}$ appearing in
the  loops transform  homogeneously under  the gauge  group,  {\it i.e.},
as ordinary matter  fields which happened  
to be assigned to  the adjoint representation 
\cite{Weinberg:kr}.   As a  result  of the  background gauge symmetry, the 
$n$-point  functions  $\langle 0  | T \left[ \widehat{{\cal V}}_{\mu_1}(x_1)
\widehat{{\cal V}}_{\mu_2}(x_2)\dots  
\widehat{{\cal V}}_{\mu_n}(x_n) \right] |0 \rangle$
are gauge-invariant, in the sense that they 
satisfy naive, QED-like WIs.
Notice however that they are {\it not} gauge-independent, 
because they 
depend {\it explicitly} on the quantum gauge-fixing parameter 
$\xi_Q$ used  to define the tree-level  propagators of  the 
quantum gluons. In theories with 
spontaneous symmetry breaking this dependence on $\xi_Q$ gives rise 
to {\it unphysical} thresholds inside these Green's functions 
for $\xi_Q \neq 1$, a fact
which limits their usefulness for resummation purposes 
\cite{Papavassiliou:1995fq}.
Only the case of the background Feynman gauge (BFG)
({\it i.e.}, BFM with $\xi_Q =1$) 
is free from unphysical poles, 
and it has been shown that the results of these Green's functions collapse to those of the PT, 
at one loop 
\cite{Denner:1994nn} (full SM)
and at two loops 
\cite{Binosi:2002bs} (SM with massless fermions). 
As we will see, this 
correspondence between the PT Green's functions and the ones
obtained using the BFG persists  to all
orders in the full SM case, in a complete parallel to the QCD case \cite{Binosi:2002ft}. 
We would like to stress that in deriving such a correspondence, at no point we
will employ an {\it a priori} knowledge of the BFM. 
Instead both its special ghost sectors, as well as the
different vertices involving one background and two quantum fields, will arise {\it
  dynamically} and, at the same time, will be projected out to the special value $\xi_Q=1$.
Once this equality between the Green's functions obtained using either schemes has been established and correctly interpreted (see Section~\ref{Conc}), 
it will provide a valuable book-keeping scheme, since  the BFM Feynman rules in the Feynman gauge can be directly employed in the construction of the effective PT Green's functions.

In Section~\ref{Rec} we construct explicitly the PT two-point functions using the
BQIs together with the results on the vertices previously proved.
Finally, the paper ends with our conclusions and two appendices, where the STIs
and BQIs used in our proof are listed. 

\section{\label{Pro}Prolegomena}

\subsection{The electroweak lagrangian}

In order to define the relevant quantities and set up the
notation used throughout the paper, we begin by writing the classical
(gauge invariant) Standard Model (SM) lagrangian as
\be
{\cal L}^{\rm cl}_{{\rm SM}}={\cal L}_{{\rm YM}}+{\cal L}_{{\rm H}}+{\cal
L}_{{\rm F}}.
\ee
The gauge invariant 
${SU}(2)_{{W}}
\otimes{U}(1)_{{Y}}$ 
Yang-Mills
part ${\cal L}_{\rm YM}$ consists of an isotriplet $W^a_\mu$ (with
$a={1,2,3}$) associated with the weak isospin generators
$T^a_{{W}}$, and 
an isosinglet $W^4_\mu$ with weak hypercharge $Y_{{W}}$
associated to the group factor $U(1)_{{Y}}$; it reads
\bea
\fl
{\cal L}_{\rm YM}&=&-\frac14F^a_{\mu\nu}F^{a\,\mu\nu} \nonumber \\
\fl
&=& -\frac14\left(\partial_\mu W^a_\mu-\partial_\nu W^a_\mu+\gw
f^{abc}W^b_\mu W^c_\nu\right)^2-\frac14\left(\partial_\mu W^4_\nu
-\partial_\nu W^4_\mu\right)^2. 
\eea 
The Higgs-boson part ${\cal L}_{\rm H}$ involves a complex 
${SU}(2)_{{W}}$
scalar doublet field $\Phi$ and its complex (charge) conjugate
$\widetilde\Phi$, given by
\bea
\Phi=\left(
\begin{array}{c} 
\phi^+\\
\frac1{\sqrt2}\left(H+i\chi\right)
\end{array}
\right),
&\hspace{2.0cm}&
\widetilde\Phi\equiv i\tau_2\Phi^*=\left(
\begin{array}{c} 
\frac1{\sqrt2}\left(H-i\chi\right)\\
-\phi^-
\end{array}
\right).
\eea
Here $H$ denotes the physical Higgs field, while $\phi^\pm$ and $\chi$
represents respectively the charged and neutral unphysical (Goldstone's) 
degrees of freedom.
Then ${\cal L}_{\rm H}$ takes the form
\be
{\cal L}_{\rm
H}=\left(D_\mu\Phi\right)^\dagger\left(D^\mu\Phi\right)-V(\Phi),
\ee
with the covariant derivative $D_\mu$ defined as
\be
D_\mu=\partial_\mu-i\gw T^a_{\scriptstyle{\rm W}}W^a_\mu+ig_1\frac{Y_{\scriptscriptstyle{\rm
W}}}2W^4_\mu,
\ee
and the Higgs potential as
\be
V(\Phi)=\frac\lambda4\left(\Phi^\dagger\Phi\right)^2
-\mu^2\left(\Phi^\dagger\Phi\right).
\ee
The SM leptons (we neglect the quark sector in what follows) 
are grouped into left-handed doublets
\be
\Psi^{L}_i=P^{L}\Psi_i=\left(
\begin{array}{c} 
\nu_i^{L}\\
\ell_i^{L}
\end{array}
\right), 
\ee
which transform
under the fundamental representation of ${SU}(2)_{{W}}
\otimes{U}(1)_{{Y}}$, and right-handed singlets
(which comprise only the charged leptons)
\be
\psi^{ R}_i=P^{R}\psi_i=\ell^{R}_i
\ee
transforming with
respect to the Abelian subgroup ${U}(1)_{{Y}}$ only.
In the previous formulas, $i$ is the generation index, and the
projection operators are defined according to $P^{{L,R}}=(1\mp\gamma_5)/2$.
In this way the leptonic part of ${\cal L}_{\rm{ F}}$ reads
\be
{\cal L}_{\rm{ F}}=\sum_i\left(i\overline\Psi^{L}_i\gamma^\mu
D_\mu\Psi^{L}_i+i\overline\psi^{R}_i\gamma^\mu
D_\mu\psi^{R}_i-\overline\Psi^{L}_iG^\ell_i\psi^{R}_i\Phi
+{\rm h.c.}\right),
\ee
with $G^\ell_i$ the Yukawa coupling.

The Higgs field $H$ will give mass to all the Standard Model
fields, by acquiring
a vacuum expectation value 
$v$; in particular the masses of the gauge fields are generated
after absorbing the massless would-be Goldstone bosons $\phi^\pm$ and
$\chi$. The physical massive gauge-bosons $W^\pm,\,Z$ and the
(massless) photon $A$ 
are then obtained by diagonalizing the mass matrix, and reads
\be
W^\pm_\mu=\frac1{\sqrt2}\left(W^1_\mu\mp iW^2_\mu\right),
\hspace{1.0cm}
\left( 
\begin{array}{c}
Z_\mu \\
A_\mu
\end{array}
\right)=
\left(
\begin{array}{cc}
\cw & \sw \\
-\sw & \cw
\end{array}\right)
\left(
\begin{array}{c}
W^3_\mu \\
W^4_\mu
\end{array}
\right),
\ee
where 
\be
\cw=\cos\tw=\frac{\gw}{\sqrt{g_1^2+\gw^2}}=
\frac{M_W}{M_Z},
\hspace{1.5cm}
\sw=\sin\tw=\sqrt{1-\cw^2},
\ee
with $\tw$ the weak mixing angle. 

For quantizing the theory, a gauge fixing term must be added to the
classical Lagrangian ${\cal L}^{\rm cl}_{{\rm SM}}$. 
To avoid tree-level mixing between gauge and
scalar fields, a renormalizable $R_\xi$ gauge of the 't Hooft type is
most commonly chosen; this is specified by one gauge parameter for
each gauge-boson, and defined through the linear gauge fixing
functions
\bea
{\cal F}^\pm
&=&\partial^\mu W^\pm_\mu\mp i\xi_{ W}\Mw\phi^\pm,
\nonumber \\
{\cal F}^{Z}&=&
\partial^\mu Z_\mu- \xi_{Z}\Mz\chi,
\nonumber \\
{\cal F}^{A}&=&
\partial^\mu A_\mu, 
\eea
yielding to the $R_\xi$ gauge fixing Lagrangian
\bea
{\cal L}_{{\rm GF}}&=&
\xi_{\scriptscriptstyle
W}B^+B^-+B^+{\cal F}^-+B^-{\cal F}^+
+\frac12\xi_{Z} \left(B^{
Z}\right)^2+B^{Z}{\cal F}^{Z}
\nonumber \\
&+&\frac12\xi_{A} 
\left(B^{A}\right)^2+
B^{A}{\cal F}^{A}.
\eea

The fields $B^\pm,\ B^{Z}$ and
$B^{A}$ represent auxiliary, non propagating
fields: they are the so called Nakanishy-Lautrup Lagrange multipliers
for the gauge condition, and they can be eliminated through their
equations of motion
\be
B^\pm= -\frac1{\xi_{\scriptscriptstyle W}}{\cal F}^\pm, \qquad
B^{Z}= -\frac1{\xi_{Z}}
{\cal F}^{Z}, \qquad
B^{A}= -\frac1{\xi_{ A}}
{\cal F}^{A},
\ee
which lead to the usual gauge fixing Lagrangian
\be
{\cal L}_{{\rm GF}}=-\frac1{\xi_{W}}{\cal
F}^+{\cal F}^--
\frac1{2\xi_{Z}}
\left({\cal F}^{Z}\right)^2
-\frac1{2\xi_{A}}\left({\cal F}^{
A}\right)^2.
\ee

The Faddeev-Popov ghost sector corresponding to the above gauge fixing
Lagrangian reads then 
\be
{\cal L}_{{\rm FPG}}=-\bar\up s{\cal F}^+-\bar\um s{\cal F}^-
-\bar\uz s{\cal F}^Z-\bar\ua s{\cal F}^A,
\ee
where $s$ is the BRST operator (see below).
The ghost Lagrangian contains kinetic terms for the Faddeev-Popov
fields, which allows to introduce them as dynamical fields of the
theory.

Summarizing, the complete Standard Model Lagrangian in the $R_\xi$
gauges is given by
\be
{\cal L}_{{\rm SM}}={\cal L}^{\rm cl}_{{\rm SM}}+{\cal L}_{{\rm F}}+
{\cal L}_{{\rm GF}}+{\cal L}_{{\rm FPG}}.
\label{SMlag}
\ee
The full set of Feynman rules derived from this Lagrangian
(together with the BFM gauge fixing procedure and the corresponding
Feynman rules) can be found in \cite{Denner:1995xt}, and will be used
throughout the paper.

\subsection{The Batalin-Vilkovisky formalism}

Due to the presence of the gauge fixing and Faddeev-Popov ghost terms,
the SM lagrangian of Eq.(\ref{SMlag}), is no longer gauge invariant;
however it is invariant under the BRST symmetry, whose transformations
for the SM fields read
\bea
s W^{\pm}_\mu&=&\partial_\mu u^{\pm}\mp i\gw
W^{\pm}_\mu\sumv\cv\uv
\pm i\gw \sumv\cv V_\mu u^{\pm},
\nonumber \\
sV_\mu&=&\partial_\mu u^{V}+i\gw\cv
\left(W^+_\mu\um-W^-_\mu\up\right),
\nonumber \\ 
s \phi^\pm&=&\pm\frac{i\gw}2\left(H\pm i\chi+v\right)u^{\pm}
\pm i\gw\phi^\pm\sumv\cv'\uv, \nonumber \\
s\chi&=&\frac{\gw}2\left(\phi^+\um+\phi^-\up\right)-\frac{\gw}{2\cw}\left(
H+v\right)\uz, \nonumber \\
s H&=&\frac{i\gw}2\left(\phi^+\um-\phi^-\up\right)+\frac{\gw}{2\cw}\chi\uz,
\nonumber \\
s\nu^{ L}_i&=&\frac{i\gw}{\sqrt 2}\ell_i^{L}u^++\frac{i\gw}{2\cw}
\nu^{L}_i\uz, \nonumber \\
s\ell^{L}_i&=&\frac{i\gw}{\sqrt 2}\nu_i^{L}u^-
-i\gw\ell^{L}_i\sumv\cv'\uv, \qquad
s\ell^{R}=i\gw\ell^{R}\left(\sw\ua+\frac{\sw^2}{\cw}\uz\right), 
\nonumber \\
s u^{\pm}&=&\pm i\gw u^{\pm}\sumv\cv\uv, \qquad
s\uv=i\gw\cv u^-u^+, \nonumber \\
s\bar u^{\pm} &=& -\frac1{\xi_W}{\cal F}^{\mp}, \qquad s\bar
\uv = -\frac1{\xi_{V}}{\cal F}^{V}, \nonumber \\
s^2\Phi &=& 0,\qquad \Phi = {\rm any\ SM\ field,}
\label{BRST}
\eea 
where $V=A,Z$, and we have defined
\be
C_{V}=\left\{
\begin{array}{ll}
\sw, &\quad {\rm if}\ \ V=A, \\
-\cw, & \quad {\rm if}\ \ V=Z.
\end{array}
\right. \qquad 
C_{V}'=\left\{
\begin{array}{ll}
-\sw, &\quad {\rm if}\ \ V=A, \\
\frac{\cw^2-\sw^2}{2\cw}, & \quad {\rm if}\ \ V=Z.
\end{array}
\right. 
\label{CV}
\ee

To take full advantage of the presence of the BRST symmetry, in the
Batalin-Vilkovisky formalism \cite{Batalin:1977pb} one introduces for each SM field $\Phi$ a
corresponding anti-field $\Phi^*$, and couples them through the
Lagrangian (for details see also 
\cite{Grassi:1999tp,Binosi:2002ez})
\be
{\cal L}_{\rm BRST}=\sum_\Phi\Phi^* s\,\Phi.
\label{LBRST}
\ee

Then the BRST invariance of the SM action, or, that is the same, the
unitarity of the $S$-matrix and the gauge independence of the physical
observables, are encoded into the master equation
\be
{\mathfrak S}(\Gamma)=0,
\label{me0}
\ee
where
\be
{\mathfrak S}(\Gamma)=\int\!d^4x\,\sum_\Phi\frac{\delta^R\Gamma}{\delta\Phi}
\frac{\delta^L\Gamma}{\delta\Phi^*}.
\label{me}
\ee
In Eq.(\ref{me}), the sum runs over all the SM fields, $R$ and $L$
denote the right and left differentiation respectively, and finally
$\Gamma$ represents the effective action [which depends on the antifields
through Eq.(\ref{LBRST})]. 
This equation can be used to derive the complete set of
non-linear STIs to  all orders in the perturbative theory, via the
repeated application of functional differentiation 
(see again \cite{Grassi:1999tp,Binosi:2002ez}).

However, the important point here is that
the STI functional (\ref{me}) can be written down in
the BFM formalism. To this end, one introduces a set of background sources
$\Omega$ associated to each SM field that will be split
into its background ($\widehat\Phi$) and quantum ($\Phi$) parts. Then
the master equation will read \cite{Grassi:1999tp}
\be
{\mathfrak S}'(\Gamma')=0,
\ee
where
\be
{\mathfrak S}'(\Gamma')={\mathfrak S}(\Gamma')+
\int\!d^4x\,\sum_\Phi\Omega\left(\frac{\delta^R\Gamma}{\delta\widehat\Phi}-
\frac{\delta^R\Gamma}{\delta\Phi}\right),
\label{mebfm}
\ee
and $\Gamma'$ denotes the effective action depending on the background sources
$\Omega$ ($\Gamma\equiv\Gamma'|_{\Omega=0}$).

Differentiation of the above STI functional with respect to the
background sources and background or quantum fields, gives then rise
to identities relating 1PI functions involving  background fields with the
ones involving quantum fields: these background-quantum identities
(BQIs) can be then used as a tool to relate the PT answer to the BFM ones
\cite{Binosi:2002ft,Binosi:2002bs,Binosi:2002ez}.

Finally a technical remark. When deriving STIs and BQIs in the
Batalin-Vilkovisky formalism, we will always work with the {\it
  minimal} generating functional $\Gamma$, where all the ``trivial pairs"
have been removed
\cite{Barnich:2000zw}. In the case of a linear gauge
fixing, such as the one at hand, this is equivalent to
working with the ``reduced'' functional $\Gamma$, defined by subtracting
from the complete generating functional $\Gamma^{\scriptscriptstyle{\rm
C}}$ the local term $\int\!d^4x{\cal L}_{\rm GF}$ corresponding to the
gauge fixing part of the Lagrangian. 
One should then keep in mind that the Green's functions
generated by the minimal effective action $\Gamma$, or the complete one
$\Gamma^{\rm C}$, are {\it not} equal \cite{Gambino:1999ai}. 
At tree-level, one has for example that
\bea
\Gamma_{W^\pm_\mu W^\mp_\nu}^{(0)}(q)&=&\Gamma^{{\rm C}\,(0)}_{W^\pm_\mu
W^\mp_\nu}(q)+\frac i{\xi_{\rm{\scriptscriptstyle{W}}}}q_\mu q_\nu \nonumber \\
&=&-i\left[\left(q^2-\Mw^2\right)g_{\mu\nu}-q_\mu q_\nu\right],
\nonumber  \\
\Gamma_{\phi^\pm\phi^\mp}^{(0)}(q)&=&\Gamma_{\phi^\pm\phi^\mp}^{{\rm C}\,(0)}+
i\xi_{\rm{\scriptscriptstyle{W}}}\Mw^2 \nonumber  \\
&=& iq^2.
\label{uns}
\eea  
At higher orders the difference depends only on
the renormalization of the $W$ field and of the gauge parameter. 
It should also be noticed that, since we have eliminated the classical
gauge-fixing fermion from the generating functional $\Gamma$, we allow for
tree-level mixing between the scalar and the gauge-boson sector, with
\be
\Gamma^{(0)}_{W^\pm_\mu\phi^\mp}(q)=\pm iM_Wq_\mu, \qquad 
\Gamma^{(0)}_{Z_\mu\chi}(q)=-M_Zq_\mu.
\ee
 
\subsection{Nielsen Identities}

By enlarging the BRST symmetry, we can construct a tool to control
the dependence of the Green's functions on the gauge parameter $\xi_i$
in a completely algebraic way.

We first of all start observing that, according to the fact that
terms that are total BRST
variation do not contribute between physical states, the sum of the
gauge fixing 
and Faddeev-Popov lagrangians can be rewritten as
\bea
\fl
{\cal L}_{{\rm GF}}+{\cal L}_{{\rm FPG}} &=&
s\Bigg(\frac12\xi_W(\bar u^+B^++\bar u^-B^-)+\bar u^+{\cal F}^++\bar
u^-{\cal F}^- + \frac12\sumv\xi_V\bar u^VB^V\nonumber \\
&+&\sumv\bar u^V{\cal
  F}^V\Bigg).
\label{BRSTinv}
\eea

To gain control over the $\xi_i$ parameter dependence of the Green's
functions we then promote the latter to be (static) fields and introduce
their corresponding BRST sources $\eta_i$ (with $i=W,\, V$) such that
\be
s\xi_i=\eta_i,\qquad\qquad s\eta_i=0.
\label{ext}
\ee
After doing this, Eq.(\ref{BRSTinv}) is no longer valid, and to
preserve the BRST invariance of the SM Lagrangian one has to add to
the the sum ${\cal L}_{{\rm GF}}+{\cal L}_{{\rm FPG}}$ the following
term
\be
{\cal L}_{\rm N}= -\frac1{2\xi_W}\eta_W\left(\bar u^+{\cal F}^++\bar
u^-{\cal F}^-\right)-\sumv\frac1{2\xi_V}\eta_V\left(\bar u^V
{\cal F}^V\right),
\ee
which will control the couplings of the sources $\eta_i$ with the SM
fields, giving the corresponding Feynman rules.
For all practical calculations one can set $\eta_i=0$ thus recovering
both 
the unextended BRST transformations of Eq.(\ref{BRST}),
as well as the master equation of
(\ref{me0}). However when $\eta_i\neq0$, the master equation reads
\be
{\mathfrak S}_\eta(\Gamma)=0,
\ee
where
\be 
{\mathfrak S}_\eta(\Gamma) = {\mathfrak
S}(\Gamma)+\eta_i\partial_{\xi_i}\Gamma.
\ee
Thus, after differentiating this new master equation and setting $\eta_i$
to zero, we get
\be
\left.\partial_{\xi_i}\Gamma\right\vert_{\eta_i=0}=-\left.\left(
\int\!d^4x\,\partial_{\eta_i}\sum_\Phi\frac{\delta^R\Gamma}{\delta\Phi}
\frac{\delta^L\Gamma}{\delta\Phi^*}\right)\right\vert_{\eta_i=0}.
\label{NI}
\ee

Establishing the above functional equation, allows (via the
repeated application of functional differentiation) to control the
gauge parameter dependence of the different Green's functions
appearing in the theory (but, unlike the PT, cannot be used to construct gauge
invariant and gauge 
fixing parameter independent Green's functions).
These relations are known in the literature 
under the name of Nielsen identities (NIs)~\cite{Nielsen:1975fs}.

Notice, finally, that the extension of the BRST symmetry through
Eq.(\ref{ext}) is just a technical trick to gain control over the gauge
parameter dependence of the various Green's functions appearing in the
theory; thus, unlike the STIs generated from Eq.(\ref{me}),
Eq.(\ref{NI}) does not have to be preserved in the renormalization
process, that will in general deform it (see \cite{Gambino:1999ai} and
references therein). We will briefly 
return to this issue in Section \ref{NISec}.

\section{\label{EwPT}Electroweak PT}

A general $S$-matrix element of a $2\to 2$ process
can be written following 
the standard Feynman rules as 
\be
T(s,t,m_i)\ =\ T_1(s,\xi)\ +\ T_2(s,m_i,\xi)\ +\
T_3(s,t,m_i,\xi),
\label{Arx}
\ee
Evidently the Feynman diagrams impose 
a decomposition of  $T(s,t,m_i)$ into three distinct sub-amplitudes 
$T_1$, $T_2$, and $T_3$, with a very characteristic kinematic
structure, {\it i.e.} a very particular dependence on the the 
Mandelstam kinematic
variables and the masses. Thus, $T_1$ 
is the conventional self-energy contribution, which
only depends on
the momentum transfer $s$, $T_2$ corresponds to vertex diagrams
which in general depend also on the masses of the external particles, 
whereas $T_3$ is a box-contribution, having in addition a non-trivial
dependence on the Mandelstam variable $t$. However, all these 
sub-amplitudes, in addition to their dependence of the physical 
kinematic variables, also display a non-trivial dependence on
the unphysical gauge fixing parameter parameter $\xi$. 
Of course 
we know that the BRST symmetry guarantees that 
the total $T(s,t,m_i)$ is independent of 
$\xi$, {\it i.e.} $d T/d \xi =0$; thus, in general, a  
set of delicate gauge-cancellations will take place. 
The PT framework provides a very particular realization of this
cancellations.
Specifically, the transition
amplitude above can be decomposed as
\cite{Cornwall:1982zr} 
\be
T(s,t,m_i)\ =\ \widehat{T}_1(s)\ +\ \widehat{T}_2(s,m_i)\ +\
\widehat{T}_3(s,t,m_i),
\label{TPT}
\ee
{\it i.e.}, in terms of three individually gauge-invariant and gauge fixing
parameter independent 
quantities:
a propagator-like part ($\widehat{T}_1$), a vertex-like piece
($\widehat{T}_2$),
and a part containing box graphs ($\widehat{T}_3$). The key observation
that allow to reach this important result is that vertex and box
graphs contain in general 
pieces, which are kinematically akin to self-energy graphs
of the transition amplitude.
The PT is a systematic way of extracting such pieces and
appending them to the conventional self-energy graphs.
In the same way, effective gauge invariant
vertices may be constructed, if
after subtracting from the conventional vertices the
propagator-like pinch parts we add the vertex-like pieces, if any, 
coming from boxes. 
The remaining purely box-like contributions are then
also gauge invariant.
 
In what follows we will consider for concreteness 
the $S$-matrix element for a $2\to2$ fermion elastic scattering process 
$f'(p'_1)\bar f'(p'_2)\to 
f(p_1)\bar f(p_2)$ we set $q=p'_2-p'_1 = p_2-p_1$, with $s=q^2$   
the square of the momentum transfer. One could equally well
study the annihilation channel, in which case $s$ would be
the center-of-mass energy. If not stated explicitly we will always
assume that the initial and final fermions are the same ({\it i.e.},
no mixing will be considered).

In order to identify the pieces which are
to be reassigned, in the original PT algorithm
all one had to do is to resort to the fundamental
WIs of the theory, triggered when the longitudinal
momenta $k_{\mu}$ appearing inside Feynman diagrams eventually
reach the elementary gauge-boson--fermions vertex involving
one on-shell fermion carrying momentum $p_1$
and one off-shell quark, carrying momentum $p_1+k$. In particular,
in a theory with conserved currents (QCD or the SM with massless
matter content) the WI triggered will be of the form  
\be
k_\mu\gamma^\mu=S^{-1}(p_1+k)+S^{-1}(p_1).
\label{ccWI}
\ee 
Depending  on the  order and  the topology of  the diagram  one is
looking  at, this  final  WI maybe  activated  immediately (as  always
happens at the one loop order), or as the final outcome of a sequential
triggering of intermediate WIs (as happens at two and more loops).  Of
the  two  terms appearing  in  the above  STI,  the  first one  remove
(``pinches''  out)  the internal  bare  fermion propagator  $S(p_1+k)$
(thus generating  a propagator-like piece), while the  second one will
vanish on-shell.  The propagator-like  pieces obtained in this way are
next reassigned  to the usual gauge bosons  self-energies, giving rise
to the corresponding PT self-energies.

In this paper, however, we stay general on the matter content of the SM,
allowing for massive fermions, and thus non conserved currents. Now the
application of the PT in such a theory turns out to be rather more
involved with respect to what we have just outlined. 
One of the main differences is that 
the charged $W^\pm$ gauge bosons will couple to fermions
with different masses, consequently modifying the WI of
Eq.(\ref{ccWI}) to
\be
k_\mu\gamma^\mu P_L=S^{-1}(p_1+k)P_L+P_RS^{-1}(p_1)+m_1P_L-mP_R.
\label{nccWI}
\ee
As before, the first two terms will pinch and vanish on-shell
respectively, but the extra terms appearing in Eq.(\ref{nccWI}) give
rise to additional propagator- and vertex-like contributions not
present in the massless case, which are ultimately related to the
presence in the theory of the would-be Goldstone's bosons $\phi^\pm$
and $\chi$.

An important step in
the PT procedure is then clearly the identification of 
all the {\it longitudinal} momenta  involved, {\it i.e.} the momenta which
can trigger the elementary WI above. The
possible sources of longitudinal momenta are two: the bare tree-level
gauge-boson propagators, and some of 
the trilinear vertices appearing in the theory.
As far as the first ones are concerned, one has that the
 $\Rxi$ bare tree-level gauge-boson propagator reads
($\xi_W\equiv\xi_V=\xi$ from now on)
\be
\Delta^{\mu\nu}_{\cal V}(k)=-\frac i{k^2-M^2_{\cal
V}}\left[g^{\mu\nu}-(1-\xi)\frac{k^\mu k^\nu}{k^2-M^2_{\cal
V}}\right],
\ee
and  the  longitudinal  momenta   are  simply  those  multiplying  the
$(1-\xi)$ term (and thus notice that they are not 
present in the $\Rxi$FG case $\xi = 1$). \\
For isolating the longitudinal momenta coming from the trilinear
 vertices instead, one start noticing that
the bare tree-level trilinear gauge boson vertex read
(all the momenta are taken to be incoming, {\it i.e.}, $q+k_1+k_2 = 0$)\\
\bpi(0,130)(-190,-70)
\Photon(-17,0)(20,0){2}{5}
\Photon(20,0)(45,30){2}{5}
\Photon(20,0)(45,-30){2}{5}
\Text(-17,10)[l]{${\cal V}_\alpha$}
\Text(50,-35)[l]{$ {\cal V}_{1\mu}$}
\Text(50,35)[l]{${\cal V}_{2\nu}$}
\Text(-17,-10)[l]{$q$}
\Text(40,-35)[r]{$k_1$}
\Text(40,35)[r]{$k_2$}
\Text(65,0)[l]{$=-i\gw \cv\Gamma^{(0)}_{\alpha\mu\nu}(q,k_1,k_2)$}
\epi
where ${\cal V}=\{W^\pm,Z,A\}$, $\cv$ is defined in (\ref{CV}) and
\be
\Gamma_{\alpha \mu \nu}^{(0)}(q,k_1,k_2)= 
(q-k_1)_{\nu}g_{\alpha\mu} + (k_1-k_2)_{\alpha}g_{\mu\nu}
 + (k_2-q)_{\mu}g_{\alpha\nu}.
\ee
The Lorentz structure $\Gamma_{\alpha \mu \nu}^{(0)}(q,k_1,k_2)$
may be split into two parts \cite{Cornwall:1982zr}
\be
\Gamma_{\alpha \mu \nu}^{(0)}(q,k_1,k_2) 
= 
\Gamma_{\alpha \mu \nu}^{{\rm F}}(q,k_1,k_2) + 
\Gamma_{\alpha \mu \nu}^{{\rm P}}(q,k_1,k_2),
\label{decomp}
\ee
with 
\bea
\Gamma_{\alpha \mu \nu}^{{\rm F}}(q,k_1,k_2) &=& 
(k_1-k_2)_{\alpha} g_{\mu\nu} + 2q_{\nu}g_{\alpha\mu} 
- 2q_{\mu}g_{\alpha\nu} \, , \nonumber\\
\Gamma_{\alpha \mu \nu}^{{\rm P}}(q,k_1,k_2) &=&
 k_{2\nu} g_{\alpha\mu} - k_{1\mu}g_{\alpha\nu}.  
\label{GFGP}
\eea
The above decomposition  
allows $\Gamma_{\alpha \mu \nu}^{{\rm F}}$ to satisfy the WI
\be 
q^{\alpha} \Gamma_{\alpha \mu \nu}^{{\rm F}}(q,k_1,k_2) = 
[(k_2^2 - M_{{\cal V}_2}^2) - (k_1^2 -M_{{\cal V}_1}^2) 
+ ( M_{{\cal V}_1}^2 - M_{{\cal V}_2}^2)] g_{\mu\nu},
\label{WI2B}
\ee
where the first two terms on the 
right-hand side are the difference of 
the two-inverse
propagators appearing inside the one-loop vertex graphs
(in the $\Rxi$FG), while
the last term accounts for the 
difference in their masses, and is associated to the 
coupling of the corresponding would-be Goldstone bosons. 
(A completely analogous, $\xi$-dependent 
separation of the three--gauge-boson vertex may be carried out, such that 
the divergence of the non-pinching part 
will equal to the difference of two inverse tree-level propagators 
written for arbitrary value of $\xi$. This decomposition 
appears for the first time in Eq.(4.4) of \cite{Cornwall:1976ii}, 
and has also been employed in \cite{Haeri:1988af}).\\
Equation (\ref{WI2B}) has to be compared with the usual tree level WI
\bea
q^{\alpha} \Gamma_{\alpha \mu \nu}^{(0)}(q,k_1,k_2) &=& 
[(k_2^2 - M_{{\cal V}_2}^2) - (k_1^2 -M_{{\cal V}_1}^2) 
+ ( M_{{\cal V}_1}^2 - M_{{\cal V}_2}^2)] g_{\mu\nu}\nonumber \\
&+& k_{1\mu}k_{1\nu}-k_{2\mu}k_{2\nu}.
\label{WItgbv}
\eea

The term $\Gamma_{\alpha \mu \nu}^{{\rm P}}$, which in configuration space corresponds to a pure divergence, is the interesting one:
in fact, it
contains the longitudinal momenta, which will eventually trigger 
the PT rearrangements.

However, when considering the non-conserved current case, this is not
the end of the story,
since additional graphs involving the would-be Goldstone's bosons
$\phi^\pm$, $\chi$ and the physical Higgs boson $H$ (which do not couple to
massless fermions), {\it must} now be included: these diagrams give
rise, when considering the scalar sector of the theory, to new pinch contributions as a result of
the longitudinal momenta carried by the trilinear 
vertices of the type ${\cal
S}^2{\cal V}$ \cite{Papavassiliou:1994pr}. These bare tree-level vertices read in fact
(all momenta incoming)\\
\bpi(0,130)(-190,-70)
\DashLine(-17,0)(20,0){5}
\DashLine(20,0)(45,30){5}
\Photon(20,0)(45,-30){2}{5}
\Text(-17,10)[l]{${\cal S}_1$}
\Text(50,-35)[l]{$ {\cal V}_\mu$}
\Text(50,35)[l]{${\cal S}_2$}
\Text(-17,-10)[l]{$q$}
\Text(40,-35)[r]{$k_1$}
\Text(40,35)[r]{$k_2$}
\Text(65,0)[l]{$=i\gw C\Gamma^{(0)}_\mu(q,k_1,k_2)$}
\epi
where $C$ is a coefficient depending on the
actual particle content of the vertex, and we have  
\be
\Gamma_\mu^{(0)}(q,k_1,k_2)=(q-k_2)_\mu.
\ee
From the Lorentz structure above, we can then isolate the longitudinal
momentum $k_1$, writing
\be
\Gamma_\mu^{(0)}(q,k_1,k_2)=\Gamma_\mu^{\rm F}(q,k_1,k_2)+
\Gamma_\mu^{\rm P}(q,k_1,k_2)
\label{sdeco}
\ee
with
\bea
\Gamma_\mu^{\rm F}(q,k_1,k_2) &=& 2q_\mu,\nonumber\\
\Gamma_\mu^{\rm P}(q,k_1,k_2) &=& k_{1\mu}.
\eea
and the $\Gamma_\mu^{\rm P}(q,k_1,k_2)$ part will be then the one that
trigger the relevant STIs for the scalar case.

Before concluding this section, 
we would like to comment on an additional subtle point.
One of the main obstacle related to the 
generalization of the PT beyond one-loop 
has been the issue of whether or not a splitting analogous to that
of Eqs.(\ref{decomp}) and (\ref{sdeco}) should take place for the
{\it internal} three--gauge-boson
vertices, {\it i.e.}, vertices where all the three
legs are irrigated by virtual momenta (so that $q$ never enters  
{\it alone} into any of the legs).
This issue has been resolved by resorting to the 
special unitarity properties satisfied by the PT Green's functions.
The final answer (put forward in \cite{Papavassiliou:1999az}) 
is that no splitting should take place 
for {\it any} of these internal trilinear vertices; this will also be the strategy adopted in all what follows.

\section{\label{Canc}The fundamental cancellations}
 
In what follows we will explain in detail
how one has to deal with the longitudinal momenta we
have been isolating in the previous section.
In particular, we will 
show that on the one hand, 
as far as the longitudinal momenta coming
from the propagators are concerned, one can effectively 
work without loss of generality in the $\Rxi$FG, 
so that they can be completely neglected; 
on the other hand for the longitudinal momenta coming
from the trilinear vertices (which are present also in the $\Rxi$FG),
we will explain how the PT rearrangements enforced by the WI 
of Eq.(\ref{nccWI}), can be collectively captured at any order through
the judicious exploitation of the STIs satisfied by some special Green's
functions, which serve as a common kernel to all higher order
self-energies and vertex diagrams.

\subsection{Gauge-boson propagators: NIs\label{NISec}}

Has already noticed the longitudinal momenta coming from
the gauge-boson propagator vanish in the $\Rxi$FG.
In fact, provided that one is studying the entire $S$-matrix (as we
do), one could  in
principle  start  directly in  the  $\Rxi$FG,  since  that the  entire
$S$-matrix written in  the $\Rxi$FG is equal to  the same entire $S$-matrix
written in  any other gauge  have been shown  long ago.  What  is less
obvious is that all the relevant cancellations of the gauge parameter 
dependent pieces proceed without the need of
carrying  out  integrations  over   the  virtual  loop  momenta,  thus
maintaining the  kinematic identity  of the various  Green's functions
intact. This constitutes in fact a point of central 
importance within the PT philosophy, and 
has been shown to be indeed the case by explicit calculations
at   one  \cite{Cornwall:1982zr}  and    two 
\cite{Binosi:2001hy} loops.

Here
we will show explicitly that this assumption is true to all orders
(thus justifying once and for all the PT projection to the $\Rxi$FG), 
by resorting to the NIs, following closely \cite{Gambino:1999ai}. 
The only hypothesis we make is 
that in the renormalization procedure we remove all the tadpoles and 
fix the parameters of ${\cal L}_{\rm SM}^{\rm cl}$ using
physical observables: then all the possible deformations of Eq.(\ref{NI})
are bound to drop out from the amplitude, and the following
proof goes through to all orders
(provided the STIs have been restored order by order as we always assume)
independently both of the specific choice of the renormalization of
unphysical parameters, as well as of the regularization scheme
adopted.
  
Let then $Z_{f'_1\bar f'_2f_1\bar f_2}^{\rm tr}$ be the
truncated Green's 
function associated to our four fermion process, and let us decompose
it as
\be
Z_{f'_1\bar f'_2f_1\bar f_2}^{\rm tr}=i\Gamma_{f'_1\bar f'_2f_1\bar
  f_2}-
\left(\Gamma_{f'_1\bar f'_2\Phi}\Delta_{\Phi\Phi'}
\Gamma_{\Phi'f_1\bar f_2}+\Gamma_{f'_1\bar f_2\Phi}
\Delta_{\Phi\Phi'}
\Gamma_{\Phi'f_1\bar f'_2}\right),
\ee
where a sum over repeated fields (running over all the allowed SM 
combinations) is intended, 
$\Delta_{\Phi\Phi'}(q)$ indicates a (full) propagator
between the SM fields $\Phi$ and $\Phi'$, and
we have omitted the momentum
dependence of the Green's functions as well as Lorentz indices.
Then, 
\bea
\fl
\partial_\xi Z_{f'_1\bar f'_2f_1\bar f_2}^{\rm tr}&=&
i\partial_\xi\Gamma_{f'_1\bar f'_2f_1\bar
  f_2}\nonumber \\
&-& 
\partial_\xi(\Gamma_{f'_1\bar
    f'_2\Phi}) \Delta_{\Phi\Phi'}
\Gamma_{\Phi'f_1\bar f_2}-\Gamma_{f'_1\bar f'_2\Phi}
\partial_\xi(\Delta_{\Phi\Phi'})
\Gamma_{\Phi'f_1\bar f_2}-\Gamma_{f'_1\bar f'_2\Phi}\Delta_{\Phi\Phi'}
\partial_\xi(\Gamma_{\Phi'f_1\bar f_2})
\nonumber \\
&-& 
\partial_\xi(\Gamma_{f'_1\bar f_2\Phi})
\Delta_{\Phi\Phi'}
\Gamma_{\Phi'f_1\bar f'_2}-\Gamma_{f'_1\bar f_2\Phi}
\partial_\xi(\Delta_{\Phi\Phi'})
\Gamma_{\Phi'f_1\bar f'_2}-\Gamma_{f'_1\bar f_2\Phi}
\Delta_{\Phi\Phi'}
\partial_\xi(\Gamma_{\Phi'f_1\bar f'_2})
\nonumber \\
&=&0.
\label{N1}
\eea
The NIs allow then to uncover the patterns of the rearrangements
needed to get the above equality, and these will reveal to actually be
PT patterns. 
Under our assumptions, the NIs for the various terms that appear
in Eq.(\ref{N1}) can be derived from the master equation (\ref{NI})
and read (neglecting terms that either vanish due to the on-shell
conditions of the external fermions or cancel when using
the LSZ reduction formula)
\bea
\fl
\partial_\xi\Delta_{\Phi\Phi'}=
\Delta_{\Phi\Phi''}\Gamma_{\eta\Phi''{\Phi'}^{*}}+
\Gamma_{\eta\Phi^{*}\Phi''}\Delta_{\Phi''\Phi'},
\nonumber \\
\fl
-\partial_\xi\Gamma_{f'_1\bar f'_2\Phi}=
\Gamma_{\eta{\Phi''}^{*}f'_1\bar f'_2}\Gamma_{\Phi''\Phi}
+\Gamma_{\eta\Phi{\Phi''}^{*}}\Gamma_{\Phi''f'_1\bar f'_2},
\nonumber \\
\fl-\partial_\xi\Gamma_{f'_1\bar f'_2f_1\bar
  f_2}=
\Gamma_{f'_1\bar f'_2\Phi''}
\Gamma_{\eta{\Phi''}^{*}f_1\bar f_2}+\Gamma_{f'_1  \bar
  f_2\Phi''}\Gamma_{\eta{\Phi''}^{*}f_1\bar f'_2}+
\Gamma_{\eta{\Phi''}^{*}f'_1\bar f'_2}\Gamma_{\Phi''f_1\bar
  f_2}
+\Gamma_{\eta{\Phi''}^{*}f'_1\bar f_2}\Gamma_{\Phi''f_1\bar f'_2}.\nonumber \\
\label{NIBVP}
\eea

Then, 
we see immediately that the boxes are not needed for removing the
gauge fixing parameter dependence of the internal self-energies, since 
the latter is exactly canceled from the vertices alone,
according to the pattern
\bea
&&
-\Gamma_{f'_1\bar f'_2\Phi}
(
\Delta_{\Phi\Phi''}\Gamma_{\eta\Phi''{\Phi'}^{*}}+
\Gamma_{\eta\Phi^{*}\Phi''}\Delta_{\Phi''\Phi'})
\Gamma_{\Phi'f_1\bar f_2}\nonumber \\
&+&
(\Gamma_{\eta\Phi{\Phi''}^{*}}
\Gamma_{\Phi''f'_1\bar f'_2})
\Delta_{\Phi\Phi'}\Gamma_{\Phi'f_1\bar f_2}+\Gamma_{f'_1\bar f'_2\Phi}
\Delta_{\Phi\Phi'}
(\Gamma_{\eta\Phi'{\Phi''}^{*}}\Gamma_{\Phi''f_1\bar f_2})
=0.
\eea
Finally, using the relation $
\Delta_{\Phi\Phi''}\Gamma_{\Phi''\Phi'}=i\delta_{\Phi\Phi'}$,
one can uncover the cancellation happening between the boxes and vertices,
according to the rule
\be
-i
\Gamma_{\eta{\Phi''}^{*}f'_1\bar f'_2}\Gamma_{f_1\bar f_2\Phi''}
+
\Gamma_{\eta{\Phi''}^{*}f'_1\bar f'_2}\Gamma_{\Phi''\Phi}\Delta_{\Phi\Phi'}
\Gamma_{\Phi'f_1\bar f_2}=0.
\ee

From the above patterns one conclude that not only the gauge
cancellations go through without the need of integration over the
virtual momenta, but also that they follow the $s$-$t$ cancellations 
characteristic of the PT (which has in a sense to be expected since
both the PT cancellations and the NIs are BRST-driven).
Actually, it is possible with a bit of work 
to explicitly identify the  auxiliary Green's
functions appearing in the NIs of Eq.(\ref{NIBVP}), with the
corresponding pieces cancelled in the PT procedure.

We end up observing that the above proof is totally general: it allows
for all possible mixings and and does not depend on the fermions being
massive or massless. The latter means in turn that the PT algorithm
will go through irrespectively of the latter property.

\subsection{Trilinear vertices: STIs}

From the previous section, we know that we can work
without loss of generality in the $\Rxi$FG, 
eliminating in this way 
the bare tree-level gauge-boson propagators as a source
of longitudinal momenta. We can then focus our attention on 
the (all-order) study of the longitudinal momenta coming from
the trilinear vertices.

In particular, 
we now come to the important 
observation that will allow for the (electroweak) PT
generalization to all orders. The key step has been to realize that
in the QCD (or conserved current) case the PT 
rearrangements induced when triggering the elementary WI
of Eq.(\ref{ccWI}) are but lower order manifestations of a fundamental
cancellation taking place between graph of distinct kinematic nature
(the so-called $s$-$t$ cancellation). As we will show, 
this will continue to be true also in the SM case, where the
elementary WI triggered is that of Eq.(\ref{nccWI}).

Let us concentrate on the gauge boson sector, and 
see how the aforementioned $s$-$t$ cancellation is enforced in the
charged case. Thus, consider 
the process $V_\mu W^+_\nu\to\bar f_uf_d$ with
the gauge-bosons off-shell and the fermions on-shell,
whose tree-level amplitude will be denoted by ${\cal T}^{(0)}_{V_\mu W^+_\nu
  \bar f_uf_d}$.
From now on we will restrict our
attention to the case in which the fermions are leptons; 
moreover we will define the quantities 
\bea
\Gamma^{(0)}_{A_\mu \bar ff}=\gamma_\mu\Gamma^{(0)}_{A\bar ff}
&\qquad&\Gamma^{(0)}_{A\bar ff}=-i\gw\sw Q_f,\nonumber\\
\Gamma^{(0)}_{Z_\mu \bar ff}=\gamma_\mu\Gamma^{(0)}_{Z\bar ff}
&\qquad&\Gamma^{(0)}_{Z\bar ff}=
-i\left(\frac\gw\cw\right)
\left[\sw^2 Q_f-I^{3}_{W,f}P_L\right].
\eea
In the above formulas, $Q_f$ the electric charge of the fermion $f$
and $I^3_{W,f}$ is the third component of the weak isospin [which is
  $+(-) 1/2$ for up (down) leptons]. As described in Section~\ref{EwPT},
the amplitude  ${\cal T}^{(0)}_{V_\mu W_\nu
  \bar ff}$ allow for a decomposition into distinct classes:
graphs which do not contain information
about  the  kinematic  details   of  the  incoming  test-quarks  are
self-energy graphs,  whereas those which  display a dependence  on the
test quarks are vertex graphs. Sice the  former depend only on the variable
$s$,  whereas the latter  on both  $s$ and  the mass  $m$ of  the test
quarks, we  will  refer to  them  as $s$-channel  and
$t$-channel  graphs,   respectively. 
For the amplitude at hands, we then have
\be
\fl
{\cal T}^{(0)}_{V_\mu W^+_\nu
  \bar f_uf_d}(k_1,k_2,p_2,p_1)={\cal T}^{(0),s}_{V_\mu W^+_\nu
  \bar f_uf_d}(k_1,k_2,p_2,p_1)+{\cal T}^{(0),t}_{V_\mu W^+_\nu
  \bar f_uf_d}(k_1,k_2,p_2,p_1),
\ee
where
\bea
\fl{\cal T}^{(0),s}_{V_\mu W^+_\nu
  \bar f_uf_d}(k_1,k_2,p_2,p_1)=i\gw\cv\Gamma^{(0)}_{\mu\nu\alpha}
(-k_1,k_2,-q)\frac i{q^2-M^2_W}
J_{W^+_\alpha \bar f_uf_d}(p_2,p_1),\nonumber \\
\fl{\cal T}^{(0),t}_{V_\mu W^+_\nu
  \bar f_uf_d}(k_1,k_2,p_2,p_1)=\frac{i\gw}{\sqrt2}\bar v_{f_u}(p_2)
\gamma_\nu
P_L S^{(0)}_{f_d}(p_1-k_1)\Gamma^{(0)}_{V_\mu\bar
f_df_d}u_{f_d}(p_1)\nonumber \\
\mbox{}\hspace{1.65cm}
+\frac{i\gw}{\sqrt2}\bar v_{f_u}(p_2)\Gamma^{(0)}_{V_\mu\bar
f_uf_u}S^{(0)}_{f_u}(p_2+k_1)
\gamma_\nu
P_L u_{f_d}(p_1),
\eea
and we have introduced the current
\be
J_{W^+_\alpha \bar f_uf_d}(p_2,p_1)=i\frac\gw{\sqrt2}
\bar v_{f_u}(p_2)\gamma_\alpha P_Lu_{f_d}(p_1).
\label{usualWcur}
\ee
\begin{center}
\begin{figure}
\includegraphics[width=15.6cm]{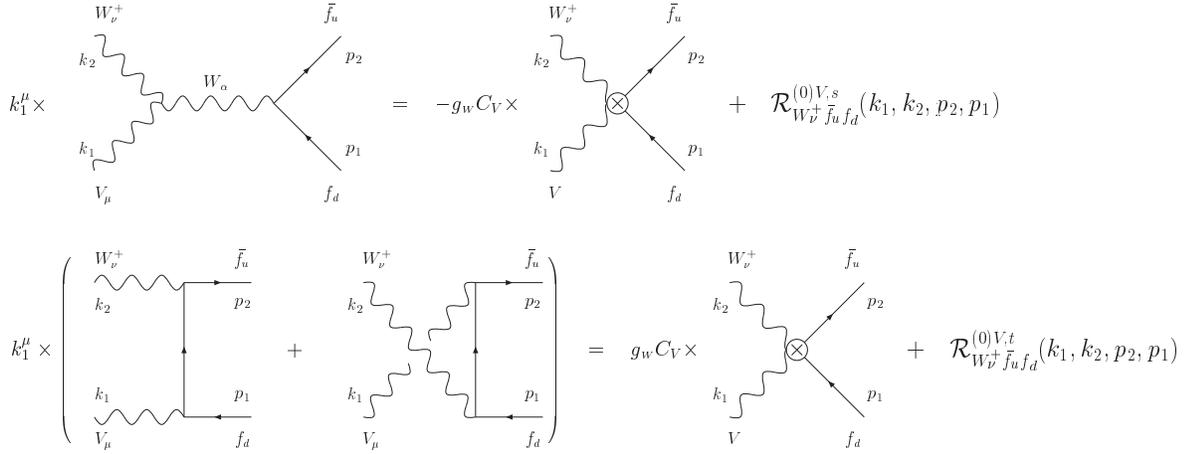}
\caption{\label{Chargedst} The fundamental PT (tree-level) $s$-$t$ cancellation in the charged case. The effective vertex appearing in the above pictures, corresponds to the insertion of the current $J_{W^+_\nu \bar f_uf_d}$ of Eq.(\ref{usualWcur}).
}
\end{figure}
\end{center} 
Notice that the above $s$-$t$ decomposition is ambiguous in the
presence of longitudinal momenta, since their action can change the
topology of a given Feynman diagram.
Let us consider in fact the action of a longitudinal momentum,
say $k_1^\mu$, on the amplitude 
${\cal T}^{(0)}_{V_\mu W^+_\nu
\bar f_uf_d}$. In
this case, the WI of Eq.(\ref{WItgbv}) will be triggered in the
$s$-channel graph, while that of Eq.(\ref{nccWI}) in the $t$-channel
one. In particular, isolating the ``pinched'' terms, one has (Fig.\ref{Chargedst})
\bea
\fl k_1^\mu{\cal T}^{(0),s}_{V_\mu W^+_\nu
  \bar f_uf_d}(k_1,k_2,p_2,p_1)=
-\gw\cv J_{W^+_\nu\bar f_uf_d}(p_2,p_1)+
{\cal R}^{(0)V,s}_{W^+_\nu
  \bar f_uf_d}(k_1,k_2,p_2,p_1),\nonumber \\
\fl
k_1^\mu{\cal T}^{(0),t}_{V_\mu W^+_\nu
  \bar f_uf_d}(k_1,k_2,p_2,p_1)=
\frac\gw{\sqrt 2}\bar v_{f_u}(p_2)\gamma_\nu(P_L\Gamma^{(0)}_{V\bar
f_df_d}-\Gamma^{(0)}_{V\bar
f_uf_u}P_L)u_{f_d}(p_1)\nonumber \\
\mbox{}\hspace{2.1cm}+{\cal R}^{(0)V,t}_{W^+_\nu
  \bar f_uf_d}(k_1,k_2,p_2,p_1)\nonumber \\
\mbox{}\hspace{2.1cm}
=\gw\cv J_{W^+_\nu\bar f_uf_d}(p_2,p_1)+
{\cal R}^{(0)V,t}_{W^+_\nu
  \bar f_uf_d}(k_1,k_2,p_2,p_1).
\eea
Notice that the first term in the first equation above, even if it
comes from a $t$-channel graph, is in fact propagator-like.
Adding by parts the two equations we enforce the PT $s$-$t$
cancellation, thus getting the result
\be
\fl
[k_1^\mu{\cal T}^{(0)}_{V_\mu W^+_\nu
  \bar f_uf_d}(k_1,k_2,p_2,p_1)]^{\rm PT}={\cal
R}^{(0)V,s}_{W^+_\nu
  \bar f_uf_d}(k_1,k_2,p_2,p_1)+
{\cal R}^{(0)V,t}_{W^+_\nu
  \bar f_uf_d}(k_1,k_2,p_2,p_1).
\ee

\begin{center}
\begin{figure}
\includegraphics[width=15.6cm]{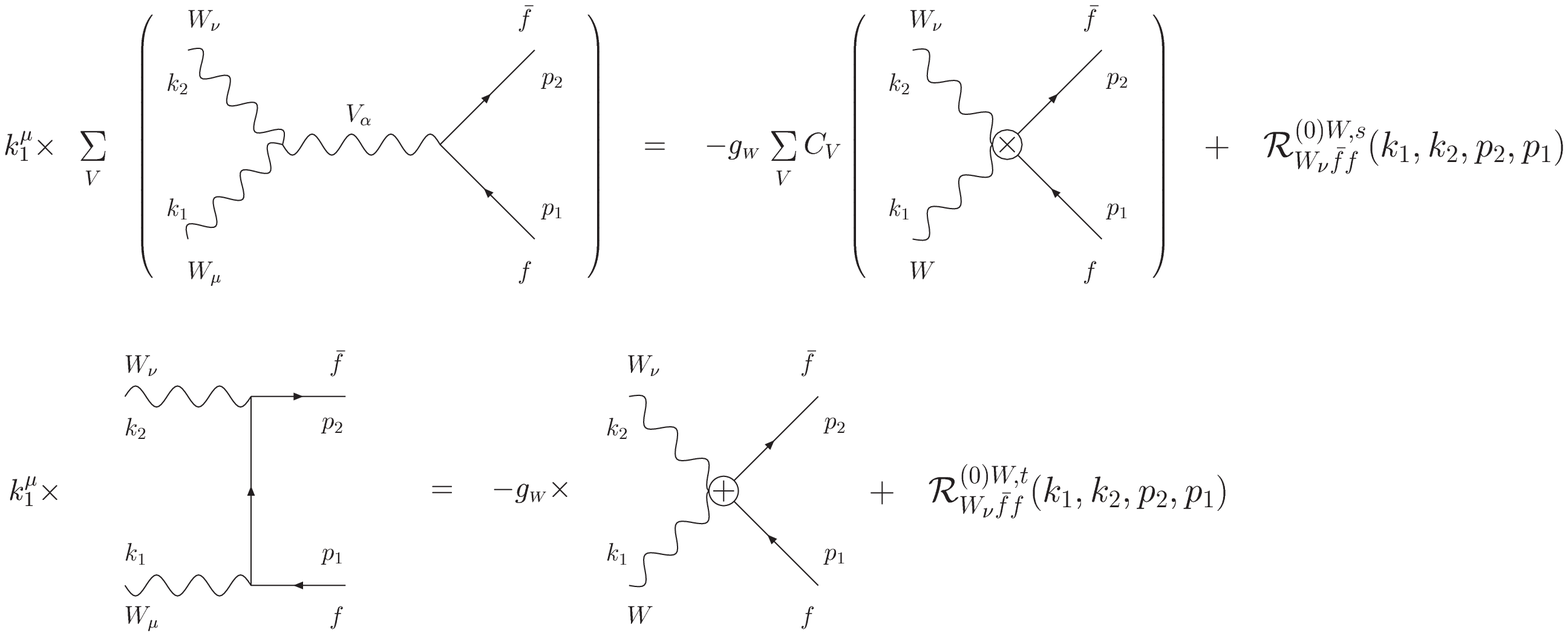}
\caption{\label{Neutralst} The fundamental PT (tree-level) $s$-$t$ cancellation in the neutral case.
Despite its appearence, the effective vertex in the $s$-channel corresponds to the insertion of the current $J_{V_\nu\bar ff}$ of Eq.(\ref{Vcur}), while the $t$-channel one corresponds to the insertion of the current
$J^{\rm eff}_{W_\nu\bar ff}$ of Eq.(\ref{newWcur}).}
\end{figure}
\end{center}

It is interesting to see how the very same cancellation applies also 
in the neutral gauge-boson sector. 
Consider in fact the
tree-level amplitude ${\cal T}^{(0)}_{W_\mu W_\nu\bar ff}$:
once again we can decompose the amplitude
into its $s$- and $t$-channel parts, given by
\bea
\fl
{\cal T}^{(0),s}_{W_\mu W_\nu
  \bar ff}(k_1,k_2,p_2,p_1)=i\gw\sumv\cv\Gamma^{(0)}_{\mu\nu\alpha}
(-k_1,k_2,-q)\frac i{q^2-M^2_V}
J_{V_\alpha \bar ff}(p_2,p_1),\nonumber \\
\fl {\cal T}^{(0),t}_{W_\mu W_\nu
  \bar ff}(k_1,k_2,p_2,p_1)=-\frac{\gw^2}2\bar v_f(p_2)\gamma_\nu
P_L S^{(0)}_{f'}(p_1-k_1)\gamma_\mu P_Lu_f(p_1),
\eea
where
\be
J_{V_\alpha \bar ff}(p_2,p_1)=\bar v_f(p_2)\Gamma^{(0)}_{V_\alpha\bar
ff}u_f(p_1).
\label{Vcur}
\ee
When contracted by a longitudinal momentum $k_1^\mu$ we will then have
the results (Fig.\ref{Neutralst})
\bea
\fl
k_1^\mu{\cal T}^{(0),s}_{W_\mu W_\nu
  \bar ff}(k_1,k_2,p_2,p_1)= -\gw\sumv\cv 
J_{V_\nu\bar ff}(p_2,p_1)+{\cal R}^{(0)W,s}_{W_\nu
  \bar ff}(k_1,k_2,p_2,p_1), \nonumber \\
\fl
k_1^\mu{\cal T}^{(0),t}_{W_\mu W_\nu
  \bar ff}(k_1,k_2,p_2,p_1)= -\gw
J^{\mathrm{eff}}_{W_\nu\bar ff}(p_2,p_1)+{\cal R}^{(0)W,t}_{W_\nu
  \bar ff}(k_1,k_2,p_2,p_1),
\label{PTterms}
\eea
where we have defined the effective $W$ current as
\be
J^{\rm eff}_{W_\nu\bar ff}(p_2,p_1)=-i\frac\gw2\bar v_f(p_2)\gamma_\nu
  P_Lu_f(p_1),
\label{newWcur}
\ee
which should not be confused with the usual $W$ current of
Eq.(\ref{usualWcur}). 
Adding by parts Eq.(\ref{PTterms}), we get finally
\bea
\fl
[k_1^\mu{\cal T}^{(0)}_{W_\mu W_\nu
  \bar ff}(k_1,k_2,p_2,p_1)]^{{\rm PT}}= -\gw\Big[
J^{\rm eff}_{W_\nu\bar ff}(p_2,p_1)+\sumv\cv 
J_{V_\nu\bar ff}(p_1,p_2)\bigg]\nonumber \\
\mbox{}\hspace{2.5cm}+{\cal R}^{(0)W,s}_{W_\nu
  \bar ff}(k_1,k_2,p_2,p_1)+{\cal R}^{(0)W,t}_{W_\nu
  \bar ff}(k_1,k_2,p_2,p_1).
\label{stcanc}
\eea

The first line of the right-hand side (rhs) of the above equation is
always zero, 
independently on the external (on-shell) leptons, reflecting the well
known property of process independence of the PT algorithm. If we
choose down leptons, the zero is due
to the identity
\bea
J^{\rm eff}_{W_\nu\bar f_df_d}(p_2,p_1)&=&\cw J_{Z_\nu\bar f_df_d}(p_2,p_1)-\sw
J_{A_\nu\bar f_df_d}(p_2,p_1)\nonumber \\
&=&-\sumv\cv 
J_{V_\nu\bar f_df_d}(p_2,p_1).
\eea
In the case we choose right-handed polarized down leptons instead, while
there is no $J^{\rm eff}_{W_\nu\bar f_df_d}$ term (there are no boxes
in such case), the cancellation continue to be true due to the
identity
\be
\sumv\cv 
J_{V_\nu\bar f_Rf_R}(p_2,p_1)=0.
\ee
Finally, if we consider up leptons there is no coupling with the
photon in the $s-$ channel, but still the cancellation goes through
since one has that
\be
 J^{\rm eff}_{W_\nu\bar f_uf_u}(p_2,p_1)=-\cw 
J_{Z_\nu\bar f_uf_u}(p_2,p_1),
\ee
(notice that in this case the contribution of the $s$-channel graph
has an extra minus sign due to a permutation of the $W$ gauge bosons in
the trilinear vertex).

The reader can easily convince him-/her-self that the very 
same kind of $s$-$t$ cancellation manifest itself in exactly the same
way when considering the SM scalar
sector (both in the charged as well as in the neutral case), by
calculating the divergence of 
the corresponding amplitudes ${\cal T}_{V_\mu\phi\bar
f_uf_d}$, ${\cal T}_{W_\mu{\rm S}\bar
f_uf_d}$, ${\cal T}_{V_\mu{\rm S}\bar ff}$ and 
${\cal T}_{W_\mu\phi\bar ff}$.
\begin{center}
\begin{figure}
\includegraphics[width=10.0cm]{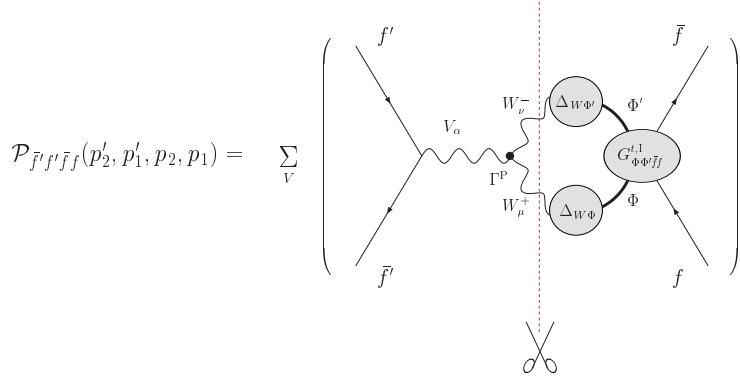}
\caption{\label{PTsubset} The subset of graphs of our $2\to2$ process 
that will receive the action of the
longitudinal momenta stemming from the pinching part of the
gauge-boson trilinear vertices, when considering the neutral gauge-boson sector.
}
\end{figure}
\end{center} 
The (tree-level) $s$-$t$ cancellations outlined above are in fact a consequence
of the underlying BRST
symmetry, and as such they will continue to occur at higher orders; thus
the correct way to look at them is through the use of the STIs.
To fix the ideas let us consider the neutral gauge-boson sector, and
denote by
${\cal P}_{\bar f'f'\bar ff}$ the subset of graphs of our $2\to2$ process 
that will receive the action of the
longitudinal momenta stemming from the pinching part of the
gauge-boson trilinear vertices 
$\Gamma_{\alpha \mu \nu}^{P}(q,k_1,k_2)$. 
Since the $\Gamma^F+\Gamma^P$ decomposition is carried out only on the
external vertices, we have that (see Fig.\ref{PTsubset})
\bea
{\cal P}_{\bar f'f'\bar ff}(p'_2,p'_1,p_2,p_1)&=&-i\gw\!
\sumv\!\cv J_{V^\alpha\bar f'f'}(p'_2,p'_1)
\!\int\!\!\Gamma_{\alpha}^{{\rm P}\,\mu\nu}(q,k-q,-k)\nonumber \\
&\times&{\cal
T}_{ W^+_\mu W^-_\nu\bar ff}(q-k,k,p_2,p_1),
\eea
where we have defined the integral measure
\be
\int\equiv\mu^{2\varepsilon}\int\!\frac{d^dk}{(2\pi)^d},
\label{im}
\ee
with $d=D-2\varepsilon$, $D$ the space-time dimension, 
and $\mu$ the 't~Hooft mass.
Clearly,  there  is  an  equal contribution from the  $\Gamma^{{\rm
P}}$ situated on the right hand-side of ${\cal T}$.

Let us focus on the (on-shell) STI satisfied by the amplitude 
${\cal T}_{W^+_\mu W^-_\nu\bar ff}$; the latter is listed 
in the appendix [Eq.(\ref{stingbs})] and it is of the form
\be
k_1^\mu {\cal T}_{W^+_\mu W^-_\nu\bar ff}(k_1,k_2,p_2,p_1)=
{\cal R}^{W^+}_{W^-_\nu\bar ff}(k_1,k_2,p_2,p_1), 
\label{onshSTI}
\ee

Now, as already discussed in our tree-level examples,
in   perturbation   theory  both   ${\cal T}_{W^+_\mu W^-_\nu\bar ff}$  
and   ${\cal R}^{W^+}_{W^-_\nu\bar ff}$
are given by  Feynman diagrams, which can be separated
into  distinct classes,  depending on  their kinematic  dependence and
their geometrical properties. In  addition   to  the aforementioned $s$-$t$
decomposition,  Feynman diagrams  can be also separated   into one-particle
irreducible (1PI) and one-particle  
reducible (1PR) ones.  In particular, 1PR graphs are  those which, after cutting one line, get
disconnected into two subgraphs none of which is a tree-level graph; if
this does  not happen, then the graph  is 1PI.
The crucial
point is  that the action of  the momenta $k_1^\mu$  or $k_2^\nu$ on
${\cal T}^{ab}_{\mu\nu}$  does {\it not} respect, in  general, the original
$s$-$t$ and 1PI-1PR separation furnished by the Feynman diagrams (see
third paper of~\cite{Papavassiliou:1995fq}).

In other
words, even though Eq.(\ref{onshSTI}) holds for the 
entire amplitude, 
it is not true for the individual sub-amplitudes, 
{\it i.e.},
\bea
\fl
k_1^\mu \left[{\cal T}_{W^+_\mu W^-_\nu\bar ff}(k_1,k_2,p_2,p_1)
\right]_{x,{\rm Y}} \neq  
\left[{\cal R}^{W^+}_{W^-_\nu\bar ff}(k_1,k_2,p_2,p_1)\right]_{x, {\rm
Y}}, \qquad x=s,t; \quad
{\rm Y=I,R},\nonumber \\
\label{INEQ}
\eea
where I (respectively R) indicates the one-particle {\it irreducible}
(respectively {\it reducible}) parts of the amplitude involved.
Evidently,   whereas   the   characterization   of   graphs   as
propagator- and  vertex-like is  unambiguous  in  the absence  of
longitudinal momenta  ({\it e.g.}, in a scalar theory),  their presence tends
to mix propagator- and  vertex-like graphs.  Similarly, 1PR graphs are
effectively converted into 1PI ones (the opposite cannot happen).  The
reason  for  the  inequality   of  Eq.(\ref{INEQ})  are  precisely  the
propagator-like  terms, such  as  those encountered in our tree-level examples
and  in  the one-  and
two-loop PT calculations \cite{Papavassiliou:1999az}; 
they have the characteristic feature that, when
depicted  by means  of Feynman  diagrams contain  unphysical vertices,
{\it   i.e.},  vertices   which   do  not   exist   in  the   original
Lagrangian.  All such diagrams  cancel {\it
diagrammatically} against  
each other.  Thus,  after the PT cancellations have  been enforced, we find
that the $t$-channel irreducible part satisfies the identity
\be 
\left[k_1^\mu 
{\cal T}_{W^+_\mu W^-_\nu\bar ff}(k_1,k_2,p_2,p_1)
\right]_{t,{\rm I}}^{{\rm  PT}}  \equiv
\left[{\cal R}^{W^+}_{W^-_\nu\bar ff}(k_1,k_2,p_2,p_1)
\right]_{t,{\rm I}},
\label{EQPT}
\ee 
since, in this case, all the possible mixing due to the presence of
the longitudinal momenta, has been taken into account.

Of course these observations apply also to the gauge-boson charged
sector case as well as to the scalar sector. 
The non-trivial step for generalizing the PT to all orders is then the
following: Instead  of going through the arduous  task of manipulating
the  left  hand-side of  Eq.(\ref{EQPT})  in  order  to determine  the
pinching parts and explicitly enforce their cancellation, use directly
the right-hand  side, which already contains the  answer!  Indeed, the
right-hand side involves  only conventional (ghost) Green's functions,
expressed  in terms  of normal  Feynman  rules, with  no reference  to
unphysical  vertices. This algorithm has been successfully implemented
in the QCD case; in the next two sections we will see that it gives
rise to the PT vertices with the expected properties to all orders.

\section{\label{Gbff}The PT gauge-boson--fermion--fermion vertex}

This section contains one of the central result 
of the present paper, namely the all-order
PT construction of the gauge-boson--fermion--fermion vertex, 
with the gauge-boson off-shell and the fermions on-shell. 
By virtue of the observations  made in the previous section, 
the derivation presented here turns out to be particularly compact. 

Before entering into the detailed calculations, a note on the notation.
To avoid notational clutter we will refrain, in what follows, to
indicate redundant indices in the Green's functions. Thus, we will
remove from them all the references to the external
fermions and their momenta, since the latter are irrelevant for the PT
construction (which is independent of the external particle
chosen). Thus, for example, the four point amplitudes ${\cal T}_{V_\mu W^+_\nu
\bar f_uf_d}(k_1,k_2,p_2,p_1)$ and ${\cal T}_{W^+_\mu W^-_\nu\bar
  ff}(k_1,k_2,p_2,p_1)$, will read   ${\cal T}_{V_\mu
  W^+_\nu}(k_1,k_2)$ and ${\cal T}_{W^+_\mu W^-_\nu}(k_1,k_2)$ respectively.
  
\subsection{Charged gauge-boson sector}

\begin{center}
\begin{figure}
\includegraphics[width=13.0cm]{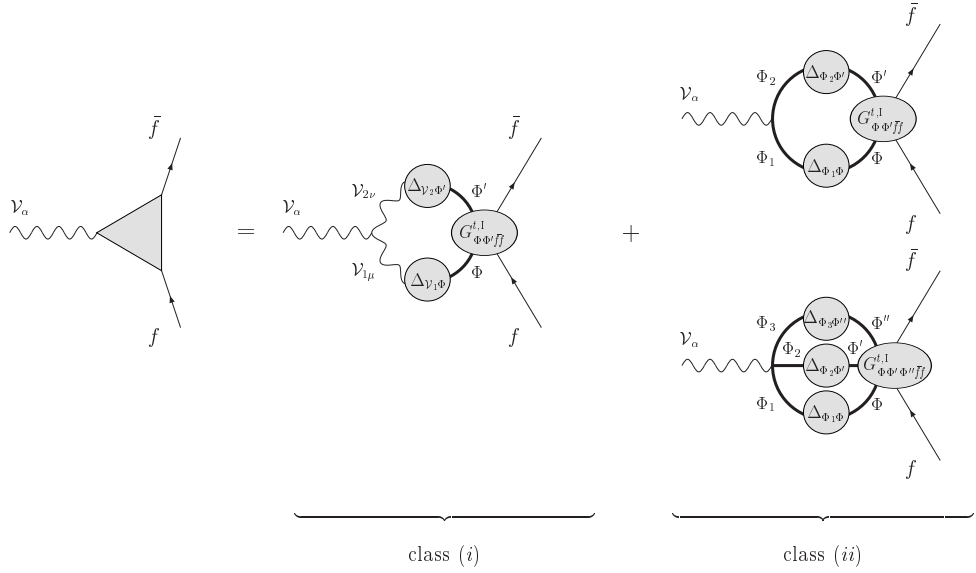}
\caption{\label{GBvertexdec} The gauge-boson sector PT vertex decomposition into class ({\it i}) and class ({\it ii}) diagrams. In the first term of the latter, the fields $\Phi_1$ and $\Phi_2$ can be any allowed combination of SM fields but the $\Phi_1={\cal V}_1$ and $\Phi_2={\cal V}_2$ one.}
\end{figure}
\end{center}

\begin{center}
\begin{figure}
\includegraphics[width=13.0cm]{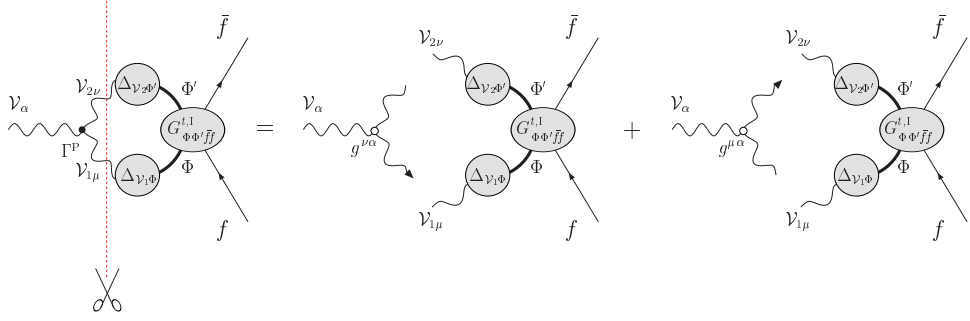}
\caption{\label{GBPTaction} The pinching action of 
the external three-gauge-boson
vertex of the class ({\it i}) diagrams. The arrows indicate the longitudinal momentum 
responsible for the pinching.}
\end{figure}
\end{center}

We begin with the construction of the  PT vertex
$\Gamma_{W^+_\alpha f_1f_2}(q,p_1,p_2)$ ($\Gamma_{W^+_\alpha
}(q,p_i)$ from now on). To achieve this, we   
classify all the diagrams that contribute to this
vertex in the $\Rxi$FG, into the following types (Fig.\ref{GBvertexdec}): ({\it i}) 
those containing an
external (tree-level) three-gauge-boson vertex, {\it i.e.}, those containing a
three-gauge-boson vertex where the momentum $q$ is incoming, 
and ({\it ii}) those which do not have such an external
vertex. This
latter set contains graphs where the incoming gauge-boson couples to the
rest of the diagram with any other type of interaction vertex other
than a three-gauge-boson vertex. Thus we write
\be
\Gamma_{W^+_\alpha }\momo=
\sum_{{\rm class}\ (i)}\Gamma_{W^+_\alpha }^{(i)}\momo+
\sum_{{\rm class}\ (ii)}\Gamma_{W^+_\alpha }^{(ii)}\momo,
\ee
where, according to our definitions,
\bea
\fl
\sum_{{\rm class}\ (i)}\Gamma_{W^+_\alpha }^{(i)}\momo\equiv
\sum_{{\cal V}}\Gamma_{W^+_\alpha }^{{\cal V}^2}\momo \nonumber \\
\fl
\sum_{{\rm class}\ (ii)}\Gamma_{W^+_\alpha }^{(ii)}\momo\equiv
\sum_{{\cal V},\,{\cal S},\,{\cal U},\,f}\left[ 
\Gamma_{W^+_\alpha }^{{\cal S}^2}\momo+
\Gamma_{W^+_\alpha }^{{\cal U}^2}\momo+
\Gamma_{W^+_\alpha }^{f\bar f}\momo+\Gamma_{W^+_\alpha }^{{\cal
V}{\cal S}}\momo+
\Gamma_{W^+_\alpha }^{{\cal V}^3}\momo+
\Gamma_{W^+_\alpha }^{{\cal V}{\cal S}^2}\momo\right].\nonumber \\
\label{f1}
\eea
In Eq.(\ref{f1}) we have ${\cal V}=\{W^\pm,A,Z\}$, 
${\cal S}=\{\phi^\pm,\chi,H\}$, 
${\cal U}=\{u^\pm,\ua,\uz\}$ and  $f=\{\ell,\nu\}$, and the sum is over all the
allowed permutations and combinations of fields.
As a second step, we next carry out  the characteristic PT vertex
decomposition of Eq.(\ref{decomp}) to the external three-gauge-boson
vertex appearing in the class ({\it i}) 
diagrams,
{\it i.e.} we define
\be
\sum_{\cal V}\Gamma_{W^+_\alpha }^{{\cal V}^2}\momo=
\sum_{\cal V}\left[\Gamma_{W^+_\alpha }^{{\rm F},{\cal V}^2}\momo+
\Gamma_{W^+_\alpha }^{{\rm P},{\cal V}^2}\momo\right].
\ee
For the case at hands then 
\bea
\sum_{\cal V}\Gamma_{W^+_\alpha }^{{\rm F},{\cal V}^2}\momo&=&
i\gw\sum_V\cv\int\!\Gamma_{\alpha}^{{\rm F}\,\mu\nu}(q,k-q,-k)\times
\nonumber \\
&\times&
\left\{[{\cal T}_{V_\mu W^-_\nu }\momt]_{t,\rm I}-[{\cal
T}_{W^-_\mu V_\nu } 
\momt]_{t,\rm I}\right\}\nonumber
\\
\sum_{\cal V}\Gamma_{W^+_\alpha }^{{\rm P},{\cal V}^2}\momo
&=& i\gw\sum_V\cv\int\!
[\left(k-q\right)^\mu g^\nu_\alpha+k^\nu g^\mu_\alpha]
\times
\nonumber \\
&\times&
\left\{[{\cal T}_{ V_\mu W^-_\nu }\momt]_{t,\rm I}-
[{\cal T}_{ W^-_\mu V_\nu }\momt]_{t,\rm I}\right\},
\eea
where the integral measure has been defined in (\ref{im}).
Following the discussion presented in the previous section,
the  pinching   action  amounts  then to using the STIs of
Eqs.(\ref{stiex}) and  (\ref{sticgbs1}) for  making the  replacements 
(Fig.\ref{GBPTaction})
\bea
(q-k)^\mu[{\cal T}_{ V_\mu W^-_\nu }
\momt]_{t,{\rm
I}}&\to&[(q-k)^\mu{\cal T}_{ V_\mu W^-_\nu }
\momt]_{t,{\rm I}}\nonumber \\
&\equiv&{\cal
R}^V_{W^-_\nu }\momt]_{t,{\rm  I}},\nonumber \\
k^\nu[{\cal T}_{ V_\mu W^-_\nu }\momt]_{t,{\rm
I}}&\to&[k^\nu{\cal T}_{ V_\mu W^-_\nu }\momt]_{t,{\rm
I}}\nonumber \\
&\equiv&[{\cal
R'}^{W^-}_{V_\mu }\momt]_{t,{\rm
I}} 
\eea
and similarly for the
term involving the  ${\cal T}_{ W^-_\mu V_\nu}$ amplitude, or, equivalently, 
\bea
\sum_{{\cal V}^2}\Gamma_{W^+_\alpha }^{{\rm P},{\cal V}^2}\momo&\to&
i\gw
\sumv\cv\!
\int\!\left\{[{\cal
R'}^{W^-}_{V_\alpha }\momt]_{t,{\rm  I}}-[{\cal
R'}^{V}_{W^-_\alpha }\momt]_{t,{\rm  I}}\right.\nonumber\\
&-&\left.
[{\cal R}^{V}_{W^-_\alpha }\momt]_{t,{\rm  I}}+[{\cal
R}^{W^-}_{V_\alpha }\momt]_{t,{\rm  I}}\right\}.
\label{cgbv}
\eea
At   this  point  the   construction  of   the  effective   PT  vertex
$\widehat\Gamma_{W^+_\alpha }$ has been  completed, and we have 
\bea
\widehat\Gamma_{W^+_\alpha }\momo&=&\sum_{\cal V}
\Gamma_{W^+_\alpha }^{{\rm F},{\cal
V}^2}\momo +\sumii\Gamma_{W^+_\alpha }^{(ii)}\momo
\nonumber \\
&+&i\gw
\sumv\cv\!
\int\!\left\{[{\cal
R'}^{W^-}_{V_\alpha }\momt]_{t,{\rm  I}}-[{\cal
R'}^{V}_{W^-_\alpha }\momt]_{t,{\rm  I}}\right.\nonumber\\
&-&\left.
[{\cal R}^{V}_{W^-_\alpha }\momt]_{t,{\rm  I}}+[{\cal
R}^{W^-}_{V_\alpha }\momt]_{t,{\rm  I}}\right\}.
\label{PTgbcs}
\eea

We pause here to make a comment about the above result.
What we want to stress is that Eq.(\ref{PTgbcs}) is provided {\it as
  it is} by blindly following the PT prescriptions given in the
previous sections. On the other hand, one may now ask if there exists
Feynman rules which can be employed has a shortcut to compute PT
Green's functions, such
as the vertex function of Eq.(\ref{PTgbcs}). It turns out that at the one- 
and two-loop level the answer to this question is positive, and the needed
Feynman rules are in fact provided by the BFM ones at the special
value $\xi_Q=1$ [background Feynman gauge (BFG for short)]. 

In what follows we will show that the effective PT
vertex and the 
gauge-boson--fermion--fermion vertex 
$\Gamma_{\widehat W^+_\alpha }\momo$ written in the BFG are in fact equal.
For doing this we first of all observe that 
part of the type ({\it
ii}) diagrams
contained in the original $\Rxi$FG $\Gamma_{W^+_\alpha }\momo$
vertex, carry
over to the same sub-groups of BFG graphs. In fact, in the BFM all of the
vertices involving fermions have the usual form, so that we have
\be
\sum_{f}\Gamma_{W^+_\alpha }^{f\bar f}\momo\equiv
\sum_{f}\Gamma_{\widehat W^+_\alpha }^{f\bar f}\momo.
\ee
Moreover, since the BFM
gauge-fixing term is quadratic in the quantum fields, apart from
vertices involving ghost fields, only vertices containing exactly two
quantum fields can differ from the conventional $\Rxi$FG ones. Thus 
the vertices $\widehat{\cal V}{\cal V}^3$ (respectively $\widehat{\cal V}{\cal
V}{\cal S}^2$) 
involving one background gauge-boson and three quantum gauge-boson
(respectively one quantum gauge-boson and two quantum scalars) coincide
with the $\Rxi$FG vertices ${\cal V}{\cal V}^3$ (respectively ${\cal VV}{\cal S}^2$),
so that 
\be
\sum_{\cal V}\Gamma_{W^+_\alpha }^{{\cal V}^3}\momo\equiv
\sum_{\cal V}\Gamma_{\widehat W^+_\alpha }^{{\cal V}^3}\momo, \qquad
\sum_{{\cal V},{\cal S}}\Gamma_{W^+_\alpha }^
{{\cal V}{\cal S}^2}\momo\equiv
\sum_{{\cal V},{\cal S}}\Gamma_{\widehat W^+_\alpha }
^{{\cal V}{\cal S}^2}\momo.
\ee
Finally, as far as the vertices involving two quantum fields are concerned, we
have that the vertex
$\widehat{\cal V}{\cal S}^2$ coincides with the
corresponding $\Rxi$FG vertex ${\cal V}{\cal S}^2$, and that the vertex
$\widehat{\cal V}{\cal V}^2$ 
coincides with the $\Gamma^{\rm F}$ part of the PT decomposition
Eq.(\ref{decomp}). Thus we have the identities
\bea
\sum_{{\cal S}}\Gamma_{W^+_\alpha }^{{\cal S}^2}\momo\equiv
\sum_{{\cal S}}\Gamma_{\widehat W^+_\alpha }^{{\cal S}^2}\momo, \qquad
\sum_{\cal V}\Gamma_{W^+_\alpha }^{{\rm F},{\cal V}^2}\momo\equiv
\sum_{\cal V}\Gamma_{\widehat W^+_\alpha }^{{\cal V}^2}\momo.
\eea 

The final step is to recognize that the BFG ghost and
scalar--gauge-boson sectors will be provided precisely by combining the
remaining $\Rxi$FG class ({\it ii}) diagrams with the ${\cal R}$ terms
appearing on the right hand-side of Eq.(\ref{PTgbcs}). Specifically we
will reproduce the {\it symmetric} $\widehat W^+{\cal U}^2$ vertex characteristic
of the BFG, as well as the $\widehat W^+{\cal V}{\cal S}$ and the $\widehat
W^+{\cal V}{\cal U}^2$ vertices 
(the last vertex being totally absent in the $\Rxi$FG). 

Now the form of the relevant ${\cal R}$ terms appearing in
Eq.(\ref{cgbv}), is given in Eqs.(\ref{Rcgbs1}) and (\ref{Rcgbs2}) of the
Appendix. In particular the $t$-channel irreducible part of these
identities has precisely the same form provided that we replace the
${\cal G}$ kernels appearing in the original STIs by the corresponding
$t$-channel irreducible kernels $\gamma=[{\cal G}]_{t,\rm I}$.
Our analysis of the PT terms starts then from the terms which are of
the type $W^+{\cal V}{\cal U}^2$ and that are absent in the $\Rxi$FG
formalism. Then, these terms read
\bea
\fl
\sum_{{\cal V},\,{\cal U}}\widehat\Gamma_{W^+_\alpha }^{{\cal
V}{\cal U}^2}\momo= i\gw^2g^\mu_\alpha\int\!\Delta_{\bar\up\Phi}(k)\left[
\gamma_{\{W^+_\mu\um\}\Phi }\momt-
\gamma_{\{ W^-_\mu\up\}\Phi }\momt\right]
\nonumber \\
\fl
+i\gw^2\sum_{V,\, V'}\cv\cvp g^\mu_\alpha\int\!\Delta_{\bar\uv\Phi}(k)
\left[
\gamma_{\{V'_\mu\um \}\Phi }\momt-
\gamma_{\{ W^-_\mu\uvp\}\Phi }\momt\right]
\nonumber \\
\fl
+i\gw^2\sum_{V,\, V'}\cv\cvp g^\nu_\alpha\!\int\!\Delta_{\bar\uv\Phi}(q-k)
\left[
\gamma_{\Phi\{ W^-_\nu\uvp\} }\momt-\gamma_{\Phi\{V'_\nu\um\}
} \momt\right]\nonumber \\
\fl
+ i\gw^2 g^\nu_\alpha\int\!\Delta_{\bar\up\Phi}(q-k)\left[
\gamma_{\Phi\{ W^-_\nu\up\} }\momt-\gamma_{\Phi\{W^+_\nu\um\}
}\momt\right],
\eea
where we have used the fact that $\sumv\cv^2=1$. It is then
straightforward to check that the above terms correspond precisely 
to the BFG $\widehat W^+{\cal V}{\cal U}^2$ sector of the theory, {\it i.e.},
we have
\be
\sum_{{\cal V},\,{\cal U}}\widehat\Gamma_{W^+_\alpha }^{{\cal
V}{\cal U}^2}\momo\equiv\sum_{{\cal V},\,{\cal U}}\Gamma_{\widehat W^+_\alpha }^{{\cal
V}{\cal U}^2}\momo.
\ee

The remaining PT terms mix instead with the corresponding class ({\it
ii}) contributions. For the ghost sector we have in fact that the
class ({\it ii}) diagrams amount to the contribution
\bea
\sum_{\cal U}\Gamma_{W^+_\alpha  }^{{\cal U}^2}\momo &=& i\gw\sumv\cv\int\!\left[
(k-q)_\alpha\Delta_{\bar\uv\Phi}(q-k)\Delta_{\um\Phi'}(k)
\gamma_{\Phi\Phi' }\momt\right.
\nonumber\\ 
&-&\left.k_\alpha
\Delta_{\uv\Phi}(q-k)\Delta_{\bar\up\Phi'}(k)
\gamma_{\Phi\Phi' }\momt\right]\nonumber \\
&-& i\gw\sumv\cv\int\!\left[(k-q)_\alpha
\Delta_{\bar\up\Phi}(q-k)\Delta_{\uv\Phi'}(k)
\gamma_{\Phi\Phi' }\momt\right.
\nonumber\\ 
&-&\left.
k_\alpha
\Delta_{\um\Phi}(k-q)\Delta_{\bar\uv\Phi'}(k)
\gamma_{\Phi\Phi' }\momt\right].
\eea
which, when added to the corresponding PT terms, gives
\bea
&&\sum_{\cal U}\widehat\Gamma_{W^+_\alpha }^{{\cal U}^2}\momo+
\sum_{\cal U}\Gamma_{W^+_\alpha }^{{\cal U}^2}\momo =\nonumber\\
&=&
i\gw
\sumv\cv\int\!
(2k-q)_\alpha\left[\Delta_{\bar\uv\Phi}(q-k)\Delta_{\um\Phi'}(k)
\gamma_{\Phi\Phi' }\momt\right.
\nonumber \\
&-&\left.
\Delta_{\uv\Phi}(q-k)\Delta_{\bar\up\Phi'}(k)
\gamma_{\Phi\Phi' }\momt\right]\nonumber\\
&+&i\gw
\sumv\cv\int\!(2k-q)_\alpha\left[
\Delta_{\um\Phi}(q-k)\Delta_{\bar\uv\Phi'}(k)
\gamma_{\Phi\Phi' }\momt\right.\nonumber\\
&-&\left.\Delta_{\bar\up\Phi}(q-k)\Delta_{\uv\Phi'}(k)
\gamma_{\Phi\Phi' }\momt\right],
\eea
{\it i.e.}, we have recovered the symmetric BFG ghost sector,
\be
\sum_{\cal U}\widehat\Gamma_{W^+_\alpha }^{{\cal U}^2}\momo
+\sum_{\cal U}\Gamma_{W^+_\alpha }^{{\cal U}^2}\momo
\equiv \sum_{\cal U}\Gamma_{\widehat W^+_\alpha}^{{\cal U}^2}\momo.
\ee

Finally, the class ({\it ii}) contributions for the
gauge-boson--scalar sector, reads
\bea
\sum_{{\cal V},\,{\cal S}}\Gamma_{W^+_\alpha }^{{\cal V}{\cal S}}\momo &=&
i\gw\sumv\cv''
g^\mu_\alpha\int\!\Delta_{V_\mu\Phi}(q-k)\Delta_{\phim\Phi'}(k)
\gamma_{\Phi\Phi' }\momt
\nonumber \\
&+&i\gw\sumv\cv''
g^\nu_\alpha\int\!\Delta_{\phim\Phi}(q-k)\Delta_{V_\nu\Phi'}(k)
\gamma_{\Phi\Phi' }\momt,
\eea
where $C_A''=-\Mw\sw$ and $C_Z''=-\sw^2\Mw/\cw$. One should  
notice that there is no $\Rxi$FG $W^+ W^-\chi$ coupling: the
corresponding BFG vertex $\widehat W^+ W^-\chi$ must be entirely 
generated from the PT terms. Adding in fact the PT terms to the above
contributions we get the result
\bea
&&\sum_{{\cal V},\,{\cal S}}\widehat\Gamma_{W^+_\alpha }
^{{\cal V}{\cal S}}\momo
+\sum_{{\cal V},\,{\cal S}}\Gamma_{W^+_\alpha }
^{{\cal V}{\cal S}}\momo =
\nonumber \\
&=&i\gw(-i\Mw)g^\mu_\alpha\int\!\Delta_{W^-_\mu\Phi}(q-k)
\Delta_{\chi\Phi'}(k)
\gamma_{\Phi\Phi' }\momt\nonumber \\
&+&i\gw(-i\Mw)g^\nu_\alpha\int\!\Delta_{\chi\Phi}(q-k)
\Delta_{W^-_\nu\Phi'}(k)
\gamma_{\Phi\Phi' }\momt\nonumber \\
&+&i\gw\sumv2\Mw\cv'\int\!\Delta_{V_\mu\Phi}(q-k)\Delta_{\phim\Phi'}(k)
\gamma_{\Phi\Phi' }\momt\nonumber \\
&+&i\gw\sumv2\Mw\cv'\int\!\Delta_{\phim\Phi}(q-k)
\Delta_{V_\mu\Phi'}(k)\gamma_{\Phi\Phi' }\momt.
\eea
Now, the first two terms in the above expression 
corresponds precisely to the BFG vertex $\widehat
W^+ W^-\chi$, while $2\Mw\cv'$ represents the BFG $\widehat
W^+V\phi^-$ coupling. Therefore, we find that
\be
\sum_{{\cal V},\,{\cal S}}\widehat\Gamma_{W^+_\alpha }
^{{\cal V}{\cal S}}\momo
+\sum_{{\cal V},\,{\cal S}}\Gamma_{W^+_\alpha }
^{{\cal V}{\cal S}}\momo
\equiv\sum_{{\cal V},\,{\cal S}}
\Gamma_{\widehat W^+_\alpha }
^{{\cal V}{\cal S}}\momo.
\ee
This concludes the proof of the (all-order) identity
(putting back the fermionic indices)
\be
\widehat\Gamma_{W^+_\alpha \bar f_u f_d}(q,p_1,p_2)\equiv
\Gamma_{\widehat W^+_\alpha \bar f_u f_d}(q,p_1,p_2).
\ee

We emphasize that the sole ingredient used in the above construction
has been the STIs of Eqs.(\ref{stiex}) and (\ref{sticgbs1}); 
in particular at no point have we
employed an {\it a priori} knowledge of the background field
formalism. Instead both its special ghost sectors, as well as the
different vertices involving two quantum fields has arisen {\it
  dynamically}, and, at the same time, projected out to the special
value $\xi_Q=1$.  As we will see this will be always the case.    

\subsection{Neutral gauge-boson sector} 

As in the charged case, we start by defining the class ({\it i})
diagrams as
\be
\sumi \Gamma^{(i)}_{V_\alpha }\momo\equiv
\sum_{\cal V}\Gamma^{{\cal V}^2}_{V_\alpha }\momo.
\ee
In the neutral sector there is only one class ({\it i}) term, and,
after carrying out the usual $\Gamma^{\rm F}+\Gamma^{\rm P}$
decomposition, we have 
\bea
\sum_{\cal V}\Gamma^{{\rm F},\,{\cal V}^2}_{V_\alpha }\momo&=&
-i\gw\cv\int\!
\Gamma_{\alpha}^{{\rm F}\,\mu\nu}(q,k-q,-k)[{\cal
T}_{ W^+_\mu W^-_\nu}\momt]_{t,{\rm I}}, \nonumber \\
\sum_{\cal V}\Gamma^{{\rm P},\,{\cal V}^2}_{V_\alpha }\momo&=&
-i\gw\cv\int\!
[\left(k-q\right)^\mu g^\nu_\alpha+k^\nu g^\mu_\alpha]
[{\cal T}_{ W^+_\mu W^-_\nu}\momt]_{t,{\rm I}}.
\eea
Using the STIs of Eq.(\ref{stingbs}), we have that 
the pinching action amounts to the replacement
\be
\fl
\sum_{\cal V}\Gamma^{{\rm P},\,{\cal V}^2}_{V_\alpha }\momo
\to -i\gw\cv\int\!\left\{
[{\cal R}^{W^-}_{W^+_\alpha }\momt]_{t,{\rm I}}-
[{\cal R}^{W^+}_{W^-_\alpha }\momt]_{t,{\rm I}}\right\},
\ee
which gives in turn the PT vertex
\bea
\widehat\Gamma_{V_\alpha }\momo &=&
\sum_{\cal V}\Gamma^{{\rm F},\,{\cal V}^2}_{V_\alpha }\momo
+\sumii \Gamma^{(ii)}_{V_\alpha }\momo \nonumber \\
&-&i\gw\cv\int\!\left\{
[{\cal R}^{W^-}_{W^+_\alpha }\momt]_{t,{\rm I}}-
[{\cal R}^{W^+}_{W^-_\alpha }\momt]_{t,{\rm I}}\right\}. 
\label{ngbptv}
\eea
We can now compare the PT result with the BFG one.
For the same reasons discussed in the charged case, we have the
following identities
\bea
\fl
\sum_{f}\Gamma_{V_\alpha }^{f\bar f}\momo\equiv
\sum_{f}\Gamma_{\widehat V_\alpha }^{f\bar f}\momo
,\qquad
\sum_{\cal V}\Gamma_{V_\alpha }^{{\cal V}^3}\momo\equiv
\sum_{\cal V}\Gamma_{\widehat V_\alpha }^{{\cal V}^3}\momo, \qquad
\sum_{{\cal V},{\cal S}}\Gamma_{V_\alpha }^{{\cal V}{\cal S}^2}\momo\equiv
\sum_{{\cal V},{\cal S}}\Gamma_{\widehat V_\alpha }^{{\cal
    V}{\cal S}^2}\momo,\nonumber \\
\fl
\sum_{S}\Gamma_{V_\alpha }^{{\cal S}^2}\momo\equiv
\sum_{S}\Gamma_{\widehat V_\alpha }^{{\cal S}^2}\momo,\qquad
\sum_{\cal V}\Gamma_{V_\alpha }^{{\rm F},\,{\cal V}^2}\momo\equiv
\sum_{\cal V}\Gamma_{\widehat V_\alpha }^{{\cal V}^2}\momo. 
\eea 

On the other hand once again the BFG ghost sector and
gauge-boson--scalar sector will be dynamically generated through 
PT terms or the combination of PT terms and class ({\it ii}) diagrams.
The relevant ${\cal R}$ terms appearing in Eq.(\ref{ngbptv}) are shown
in Eq.(\ref{Rngbs}); in particular we find that,
as in the previous case, the PT terms of the type $V{\cal V}{\cal U}^2$ are
responsible for the dynamical generation of the corresponding BFG 
$\widehat V{\cal V}{\cal U}^2$ sector, {\it i.e.}, we have
\be
\sum_{{\cal V},\,{\cal U}}\widehat \Gamma_{V_\alpha }^{{\cal
V}{\cal U}^2}\momo
\equiv\sum_{{\cal V},\,{\cal U}}\Gamma_{\widehat V_\alpha }^{{\cal
V}{\cal U}^2}\momo,
\ee
where
\bea
\fl\sum_{{\cal V},\,{\cal U}}\widehat \Gamma_{V_\alpha }^{{\cal
V}{\cal U}^2}\momo=
i\gw^2\cv\sumvp\cvp g^\mu_\alpha\int\!\Delta_{\bar\up\Phi}(k)[
\gamma_{\{W^+_\mu\uvp\}\Phi }\momt-
\gamma_{\{V'_\mu\up\}\Phi }\momt]\nonumber \\
\fl+i\gw^2\cv\sumvp\cvp g^\nu_\alpha\int\!\Delta_{\bar\um\Phi}(q-k)[
\gamma_{\Phi\{V'_\nu\um\}}\momt-\gamma_{\Phi \{W^-_\nu\uvp\}}\momt].
\eea
The remaining PT contributions mixes with the corresponding class
({\it ii}) diagrams, therefore generating the BFG modified 
sector of the theory. In fact,
the class ({\it ii}) diagrams contribution to the ghost sector reads
\bea
\sum_{\cal U}\Gamma_{V_\alpha }^{{\cal U}^2}\momo&=&
i\gw\cv\int\!\left[(q-k)_\alpha\Delta_{\bar\um\Phi}(q-k)
\Delta_{\um\Phi'}(k)
\gamma_{\Phi\Phi' }\momt\right.\nonumber \\
&+&\left.k_\alpha
\Delta_{\up\Phi}(q-k)
\Delta_{\bar\up\Phi'}(k)
\gamma_{\Phi\Phi' }\momt\right],
\eea
so that by adding them to the PT terms, we get
\bea
\fl\sum_{\cal U}\widehat \Gamma_{V_\alpha }^{{\cal U}^2}\momo+
\sum_{\cal U}\Gamma_{V_\alpha }^{{\cal U}^2}\momo
=i\gw\cv\int\!(2k-q)_\alpha\left[
\Delta_{\up\Phi}(q-k)\Delta_{\bar\up\Phi'}(k)
\gamma_{\Phi\Phi' }\momt\right.\nonumber \\
\fl\mbox{}\hspace{4.25cm}-\left.
\Delta_{\bar\um\Phi}(q-k)\Delta_{\um\Phi'}(k)
\gamma_{\Phi\Phi' }\momt\right],
\eea
which represents the BFG symmetric ghost sector, {\it i.e.}
\be
\sum_{\cal U}\widehat \Gamma_{V_\alpha }^{{\cal U}^2}\momo+
\sum_{\cal U}\Gamma_{V_\alpha }^{{\cal U}^2}\momo\equiv
\sum_{\cal U}\Gamma_{\widehat V_\alpha }^{{\cal U}^2}\momo.
\ee

Finally, the class ({\it ii}) diagrams contributing to the
gauge-boson--scalar sector, can be written as
\bea 
\sum_{{\cal V},\,{\cal S}}\Gamma_{V_\alpha }^{{\cal V}{\cal S}}\momo &=& 
i\gw\cv''g^\mu_\alpha\int\!
\Delta_{W^+_\mu\Phi}(q-k)\Delta_{\phim\Phi'}(k)
\gamma_{\Phi\Phi' }\momt \nonumber
\\
&+&i\gw\cv''g^\nu_\alpha\int\!\Delta_{\phip\Phi}(q-k)
\Delta_{W^-_\nu\Phi'}(k)\gamma_{\Phi\Phi' }\momt.
\eea
Notice that in the BFG there is no coupling between the background
photon $\widehat A$ and the $W^\pm\phi^\mp$ fields, so that the above
terms should precisely  cancel (for $V\equiv A$) the PT contributions.
In fact, adding the two terms we find
\bea
\fl
\sum_{{\cal V},\,{\cal S}}\widehat\Gamma_{V_\alpha }^{{\cal V}{\cal S}}\momo
+\sum_{{\cal V},\,{\cal S}}\Gamma_{V_\alpha }^{{\cal V}{\cal S}}\momo
=i\gw\widehat C_Vg^\mu_\alpha\int\! 
\Delta_{W^+_\mu\Phi}(q-k)\Delta_{\phim\Phi'}(k)
\gamma_{\Phi\Phi' }\momt \nonumber
\\\mbox{}\hspace{1.8cm}
+i\gw\widehat C_Vg^\nu_\alpha\int\!\Delta_{\phip\Phi}(q-k)
\Delta_{W^-_\nu\Phi'}(k)\gamma_{\Phi\Phi' }\momt,
\eea
where $\widehat C_V$ turns out to be the 
BFG $\widehat V W^\pm\phi^\pm$ coupling,
{\it i.e.} $\widehat C_A=0$ and $\widehat C_Z=-\Mw/\cw$. Thus we have
the identity
\be
\sum_{{\cal V},\,{\cal S}}\widehat\Gamma_{V_\alpha }^{{\cal V}{\cal S}}\momo
+\sum_{{\cal V},\,{\cal S}}\Gamma_{V_\alpha }^{{\cal V}{\cal S}}\momo\equiv
\sum_{{\cal V},\,{\cal S}}\Gamma_{\widehat V_\alpha }^{{\cal
V}{\cal S}}\momo, 
\ee
which finally show that, also for the neutral gauge boson sector, the
PT result coincides with the BFG ones, {\it i.e.}, putting back the fermion indices,
\be
\widehat\Gamma_{V_\alpha \bar f f}(q,p_1,p_2)\equiv
\Gamma_{\widehat V_\alpha \bar f f}(q,p_1,p_2).
\ee

\section{\label{Sff}The PT scalar--fermion--fermion vertex}

As explained in the earlier, when dealing with spontaneously
broken theories, longitudinal momenta appear also in the vertices
involving two scalar fields and one gauge-boson field, and thus must
be included in the PT procedure. In the next two subsections we discuss
the PT reorganization of the charged and neutral scalar sectors
in details, proving once again the correspondence between the PT effective
vertices and the BFG ones.

\subsection{Charged scalar sector}

\begin{center}
\begin{figure}
\includegraphics[width=13.0cm]{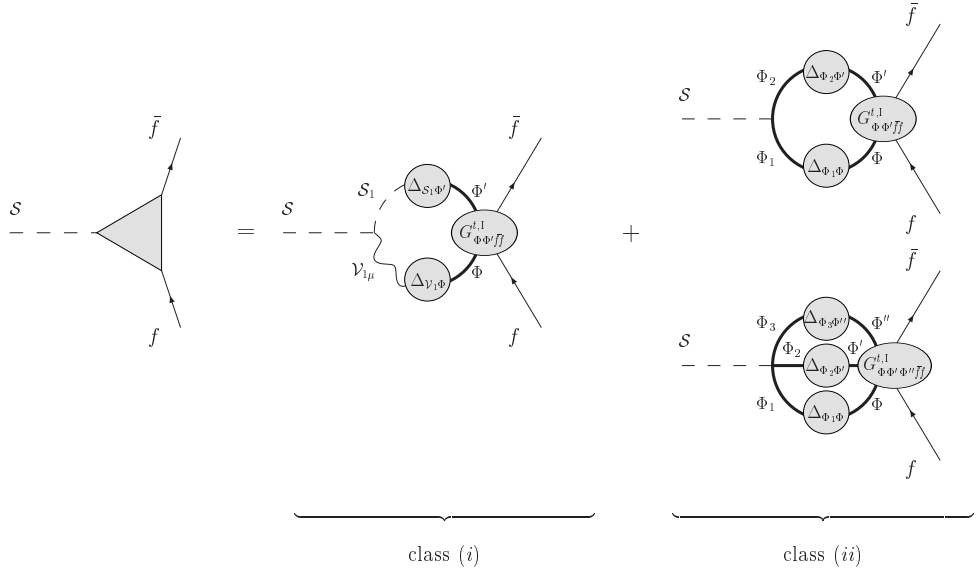}
\caption{\label{Svertexdec} The scalar sector PT vertex decomposition into class ({\it i}) and class ({\it ii}) diagrams. In the first term of the latter, the fields $\Phi_1$ and $\Phi_2$ can be any allowed combination of
SM fields but the $\Phi_1={\cal V}$ and $\Phi_2={\cal S}$ (and viceversa) ones. The symmetric contribution to the class ({\it i}) diagrams (where the internal gauge-boson and scalar legs are inverted) is not shown.}
\end{figure}
\end{center}

The procedure to be applied in the scalar sector 
is very similar to the one used in the gauge-boson sector.
One starts by  
classifying all the diagrams that contribute to this
vertex in the $\Rxi$FG, into the following types (Fig.\ref{Svertexdec}): ({\it i}) 
those containing an
external (tree-level) scalar--scalar--gauge-boson vertex, 
{\it i.e.}, those containing a
scalar--scalar--gauge-boson vertex where the momentum $q$ is incoming, 
and ({\it ii}) those which do not have such an external vertex. This
latter set contains graphs where the incoming gauge-boson couples to the
rest of the diagram with any other type of interaction vertex other
than a scalar--scalar--gauge-boson vertex. Thus, in the charged scalar case, we write
\be
\Gamma_{\phip }\momo=\sum_{{\rm class}\
  (i)}\Gamma_{\phip }^{(i)}\momo+
\sum_{{\rm class}\ (ii)}\Gamma_{\phip }^{(ii)}\momo,
\ee
where, according to our definitions,
\bea
\fl
\sum_{{\rm class}\ (i)}\Gamma_{\phip }^{(i)}\momo\equiv
\sum_{{\cal V},\,{\cal S}}\Gamma_{\phip  }^{{\cal S}{\cal V}}\momo, 
\nonumber \\
\fl
\sum_{{\rm class}\ (ii)}\Gamma_{\phip }^{(ii)}\momo\equiv
\sum_{{\cal V},\,{\cal S},\,{\cal U},\,f}\left[
\Gamma_{\phip }^{{\cal S}^2}\momo+\Gamma_{\phip }^{ {\cal U}^2}\momo+
+\Gamma_{\phip  }^{{\cal V}^2}\momo
+\Gamma_{\phip  }^{f\bar f}\momo+
\Gamma_{\phip }^{{\cal S}^3}\momo+
\Gamma_{\phip }^{{\cal S}{\cal V}^2}\momo\right].\nonumber \\
\eea

\begin{center}
\begin{figure}
\includegraphics[width=9.0cm]{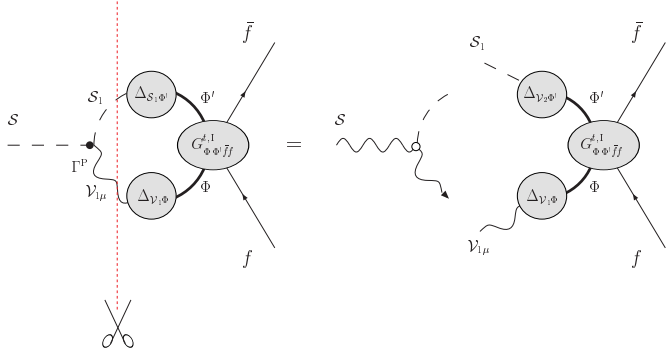}
\caption{\label{SPTaction} The pinching action of 
the external scalar--scalar--gauge-boson 
vertex of the class ({\it i}) diagrams. The arrow indicates the longitudinal momentum 
responsible for the pinching. The symmetric contribution to the class ({\it i}) diagrams (where the internal gauge-boson and scalar legs are inverted) is not shown.}
\end{figure}
\end{center}

As a second step, we next carry out the characteristic PT vertex
decomposition of Eq.(\ref{sdeco}) to the external
scalar--scalar--gauge-boson vertex appearing in the class ({\it i})
diagrams, {\it i.e.} we define
\be
\sum_{{\cal V},\,{\cal S}}\Gamma_{\phip }^{ {\cal S}{\cal V}}\momo=
\sum_{{\cal V},\,{\cal S}}[\Gamma_{\phip }^{{\rm F},\, 
{\cal S}{\cal V}}\momo
+\Gamma_{\phip }^{{\rm P},\, {\cal S}{\cal V}}\momo].
\ee
In the case at hands then
\bea
\fl
\sum_{{\cal V},\,{\cal S}}
\Gamma_{\phip }^{{\rm F},\,S{\cal V}}\momo =
\nonumber \\
\fl i\gw\int\!\Gamma^{{\rm F}\,\mu}  
(q,k-q,-k)\Bigg\{\sumv\cv'[{\cal T}_{V_\mu\phim}\momt]_{t,\rm I}
-\frac i2\sums\cs[{\cal T}_{W^-_\mu \rS}\momt]_{t,\rm I}\Bigg\}\nonumber \\
\fl +i\gw
\int\!\Gamma^{{\rm F}\,\nu}(q,k-q,-k)\Bigg\{\sumv\cv'[{\cal
    T}_{\phim V_\nu }\momt]_{t,\rm I} 
-\frac i2\sums\cs[{\cal T}_{\rS W^-_\nu }\momt]_{t,\rm
  I}\Bigg\},\nonumber \\ 
\fl \sum_{{\cal V},\,S}
\Gamma_{\phip }^{{\rm P},\, S{\cal V}}\momo = 
\nonumber \\ 
\fl i\gw
\int\!(k-q)^\mu\Bigg\{\sumv\cv'[{\cal T}_{V_\mu\phim }\momt]_{t,\rm I}
-\frac i2\sums\cs[{\cal T}^\mu_{W^-_\mu \rS }\momt]_{t,\rm I}\Bigg\}
\nonumber \\
 \fl -i\gw
\int\!k^\nu\Bigg\{\sumv\cv'[{\cal T}_{\phim V_\nu }\momt]_{t,\rm I}
-\frac i2\sums\cs[{\cal T}_{\rS W^-_\nu }\momt]_{t,\rm
  I}\Bigg\},
\eea
where the index $\rS$ runs over the neutral fields only, {\it i.e.}
$\rS=\{\chi,H\}$, with $C_\chi=1$ and $C_H=i$. 
The pinching action amounts then to using the STIs of
Eq.(\ref{sticss}) for making the replacement (Fig.\ref{SPTaction})
\bea
\fl
\sum_{{\cal V},S}\Gamma_{\phip  }^{{\rm P},\,S{\cal V}}\momo\to-i\gw\int\!
\Bigg\{\sumv\cv'[{\cal R}^{V}_{\phim }\momt]_{t,\rm I}
- \frac i2 \sums\cs[{\cal R}^{\Wm}_{\rS }\momt]_{t,\rm I}\nonumber \\
\mbox{}\hspace{0.3cm}+\sumv\cv'[{\cal R}_{\phim }^{'V}\momt]_{t,\rm I}
- \frac i2 \sums\cs[{\cal R}_{\rS }^{'\Wm}\momt]_{t,\rm I}\Bigg\},
\eea
so that the effective PT vertex will be given by
\bea
\fl
\widehat\Gamma_{\phip }\momo=\sum_{{\cal V},S}\Gamma_{\phip }^{
  S{\cal V}}\momo+ 
\sumii\Gamma^{(ii)}_{\phip }\momo
-i\gw\int\!
\Bigg\{\sumv\cv'[{\cal R}^{V}_{\phim }\momt]_{t,\rm I}\nonumber \\
\fl- \frac i2 \sums\cs[{\cal R}^{\Wm}_{\rS }\momt]_{t,\rm I}
+\sumv\cv'[{\cal R}_{\phim }^{'V}\momt]_{t,\rm I}
- \frac i2 \sums\cs[{\cal R}_{\rS }^{'\Wm}\momt]_{t,\rm
  I}\Bigg\}.\nonumber \\
\eea
We then can proceed to the comparison with the corresponding BFG
vertex $\Gamma_{\widehat \phi^+}\momo$. We start by observing that the
vertices $\widehat{\cal  S}{\cal  S}^3$, $\widehat{\cal  S} {\cal S}
{\cal V}^2$ and 
$\widehat {\cal S}{\cal V}^2$ coincide with the corresponding $\Rxi$FG
vertices ${\cal S} {\cal S}^3$, ${\cal S}{\cal S}{\cal V}^2$ and 
${\cal S}{\cal V}^2$; moreover, 
the vertices $\widehat{\cal S}{\cal  S}{\cal V}$
coincide with the $\Gamma^{\rm F}$ part of the corresponding PT
decomposition of Eq.(\ref{sdeco}). Thus we have the following
identities
\bea
\fl
\sum_f\Gamma_{\phip
  }^{f\bar f}\momo\equiv\sum_f\Gamma_{\widehat\phi^+ 
  }^{f\bar f}\momo, \qquad
\sum_{\cal S}\Gamma_{\phip }^{ {\cal S}^3}\momo\equiv\sum_{\cal S}
\Gamma_{\widehat\phi^+  }^{{\cal S}^3}\momo,
\qquad
\sum_{{\cal V},\,{\cal S}}\Gamma_{\phip }^{{\cal  S}{\cal V}^2}\momo
\equiv\sum_{{\cal V},\,{\cal S}}\Gamma_{\widehat\phi^+ }^{ {\cal S}
{\cal V}^2}\momo,
\nonumber \\
\fl
\sum_{\cal V}\Gamma_{\phip }^{
  {\cal V}^2}\momo\equiv\sum_{\cal V}\Gamma_{\widehat\phi^+
  }^{{\cal V}^2}\momo, 
\qquad
\sum_{{\cal V},\,{\cal S}}\Gamma_{\phip }^{{\rm F},\,{\cal S}{\cal V}}\momo
\equiv\sum_{{\cal V},\,{\cal S}}\Gamma_{\widehat\phi^+ }^{{\cal S}
{\cal V}}\momo.
\eea

Let us once again turn our attention on the PT terms given in
Eqs.(\ref{Rcss1}) and (\ref{Rcss2}), and start our analysis from the
four particle sector of the $\phi^+{\cal S}{\cal U}^2$ type. Now,
as far as the corresponding BFG $\widehat\phi^+{\cal S}{\cal U}^2$ sector is 
concerned, one should
notice that there is no BFG coupling such as
$\widehat\phi^+\phi^+\um\bar\up$: in fact, there is a perfect 
cancellation between the PT terms involving the kernel
$\gamma_{\{\phip\um\}\Phi}$ appearing in ${\cal R}_{\chi }^{'\Wm}$ and ${\cal
R}_{\sH }^{'\Wm}$ (${\cal R}_{\chi }^{\Wm}$ and ${\cal
R}_{\sH }^{\Wm}$ respectively), while the terms involving the kernel 
$\gamma_{\{\phim\up\}\Phi}$ carry a half of the corresponding BFG
coupling and add up.  
It is then straightforward to check that the PT terms
\bea
\sum_{{\cal S},\,{\cal U}}\widehat \Gamma_{\phi^+  }^{ {\cal S}
{\cal U}^2}\momo&=&-
i\gw^2\sum_{V,\,V'}\cv'\cvp'\int\!\Delta_{\bar\uv\Phi}(q-k)\gamma_{\Phi
\{\phim\uvp\} }\momt
\nonumber \\
&+&i\gw^2\sum_{V,\,\rS}\frac i2\cv'\cs\int\!\Delta_{\bar\uv\Phi}(q-k)
\gamma_{\Phi \{\rS\um\} }\momt\nonumber \\
&-&i\gw^2\frac12\int\!\Delta_{\bar\up\Phi}(q-k)
\gamma_{\Phi\{\phim\up\} }\momt
\nonumber \\
&+&i\gw^2\sums\frac{\uno}{4\cw}\cs\int\!\Delta_{\bar\up\Phi}(k-q)
\gamma_{\Phi\{\overline \rS\uz\} }\momt \nonumber \\
&+& i\gw^2\sum_{V,\,V'}\cv'\cvp'\int\!\Delta_{\bar\uv\Phi}(k)\gamma_{
\{\phim\uvp\}\Phi }\momt\nonumber \\
&-&i\gw^2\sum_{V,\,\rS}\frac{i}2\cv'\cs\int\!\Delta_{\bar\uv\Phi}(k)
\gamma_{\{\rS\um\}\Phi }\momt\nonumber \\
&+&i\gw^2\frac12\int\!\Delta_{\bar\up\Phi}(k)
\gamma_{\{\phim\up\} \Phi}\momt\nonumber \\
&-&i\gw^2\sums\frac{\uno}{4\cw}\cs\int\!\Delta_{\bar\up\Phi}(k)
\gamma_{\{\overline \rS\uz\}\Phi }\momt,
\eea
give rise to the correct BFG four particle ghost sector, {\it i.e.}, that
\be
\sum_{{\cal S},\,{\cal U}}\widehat \Gamma_{\phi^+ }^{{\cal S}
{\cal U}^2}\momo\equiv\sum_{{\cal S},\,{\cal U}}\Gamma_{\widehat \phi^+ }^{{\cal S}
{\cal U}^2}\momo.
\ee

For getting the remaining BFG sectors of the theory, one has also to
consider the corresponding class ({\it ii}) diagrams.
The $\Rxi$FG contributions to the ghost sector reads in fact
\bea
\sum_{{\cal U}}\Gamma_{\phip }^{
{\cal U}^2}\momo&=&
i\gw\sumv\frac12\int\!\left[\Mv\Delta_{\bar\uv\Phi}(q-k)\Delta_{\um\Phi'}(k)
\gamma_{\Phi\Phi' }\momt\right.\nonumber \\
&+&\left.2\Mw\cv'\int\!
\Delta_{\uv\Phi}(q-k)\Delta_{\bar\up\Phi'}(k)
\gamma_{\Phi\Phi' }\momt\right]\nonumber\\
&-&i\gw\sumv\frac12\int\!\left[\Mv
\Delta_{\um\Phi}(q-k)\Delta_{\bar\uv\Phi'}(k)
\gamma_{\Phi\Phi' }\momt\right.\nonumber \\
&+&\left.
2\Mw\cv'\int\!\Delta_{\bar\up\Phi}(q-k)\Delta_{\uv\Phi'}(k)
\gamma_{\Phi\Phi' }\momt\right].
\eea
Now we notice that on the one hand there is no $\Rxi$FG
$\phip\um\bar\ua$ coupling, so 
that the BFG $\widehat\phi^+\um\bar\ua$ will be generated entirely from
the PT terms; on the other hand the PT terms do not involve the kernel
$\gamma_{\bar\up\ua }$, which tell us that the $\Rxi$FG coupling
$\phip\ua\bar\up$ and the BFG 
one $\widehat\phi^+\ua\bar\up$ should coincide (as indeed happens to
be true). Thus adding the two contributions we get 
\bea
\fl
\sum_{{\cal U}}\widehat \Gamma_{\phip }^{
{\cal U}^2}\momo+\sum_{{\cal U}}\Gamma_{\phi^+ }^{
{\cal U}^2}\momo=i\gw\sumv\widehat C_V'\int\!\left[
\Delta_{\bar\uv\Phi}(q-k)\Delta_{\um\Phi'}(k)
\gamma_{\Phi\Phi' }\momt\right.\nonumber \\
\fl
\mbox{}\hspace{4.3cm}
-\left.\Delta_{\uv\Phi}(q-k)\Delta_{\bar\up\Phi'}(k)
\gamma_{\Phi\Phi' }\momt\right]\nonumber\\
\fl\mbox{}\hspace{4.3cm}
-i\gw\sumv\widehat C_V'\int\!\left[
\Delta_{\um\Phi}(q-k)\Delta_{\bar\uv\Phi'}(k)
\gamma_{\Phi\Phi' }\momt\right.\nonumber \\
\fl\mbox{}\hspace{4.3cm}
-\left.\Delta_{\bar\up\Phi}(q-k)\Delta_{\uv\Phi'}(k)
\gamma_{\Phi\Phi' }\momt\right],
\eea
where $\widehat C'_V$ is such that
$\widehat C'_A=\Mw\sw$ and $\widehat C'_Z=\Mw\sw^2/\cw$. But the
latter couplings are
precisely the corresponding BFG ones so that we have the identity
\be
\sum_{{\cal U}}\widehat \Gamma_{\phip }^{
{\cal U}^2}\momo+\sum_{{\cal U}}\Gamma_{\phi^+ }^{
{\cal U}^2}\momo\equiv\sum_{{\cal U}}\Gamma_{\widehat\phi^+ }^{
{\cal U}^2}\momo.
\ee

The last sector one needs to check is the scalar--scalar one. To this end
we start by noticing that the $\Rxi$FG contributions to this sector
will contain only (external) Higgs fields, and read 
\bea
\sum_{\cal S}\Gamma_{\phip }^{ {\cal S}^2}\momo&=&
-i\gw\frac{M_H^2}{2\Mw}\int\!\Delta_{H\Phi}(q-k)
\Delta_{\phim\Phi'}(k)\gamma_{\Phi\Phi' }\momt \nonumber \\
&-&i\gw\frac{M_H^2}{2\Mw}\int\!\Delta_{\phim\Phi}(q-k)
\Delta_{H\Phi'}(k)\gamma_{\Phi\Phi' }\momt.
\eea
The BFG coupling
$\widehat\phi^+\phim\chi$ will be thus completely generated from the PT
terms, combining the terms proportional to the (external) $\chi$ field 
kernel appearing in ${\cal R}^{V}_{\phim }$
${\cal R}_{\rS}^{'\Wm}$, ${\cal R}_{\phim}^{'V}$
and ${\cal R}^{'\Wm}_{\rS}$.
Adding all the terms we 
find in fact
\bea
\fl
\sum_{{\cal S}}\widehat\Gamma_{\phi^+ }^{ {\cal S}^2}\momo+
\sum_{\cal S}\Gamma_{\phip }^{ {\cal S}^2}\momo=
i\gw\sums\widehat C_\rS\int\!\Delta_{\rS\Phi}(q-k)
\Delta_{\phim\Phi'}(k)\gamma_{\Phi\Phi' }\momt \nonumber \\
\fl\mbox{}\hspace{4.3cm}
+i\gw\sums\widehat C_\rS\int\!\Delta_{\phim\Phi}(q-k)\Delta_{\rS\Phi'}(k)
\gamma_{\Phi\Phi' }\momt, 
\eea
where $\widehat C_\rS$ represents the corresponding BFG coupling, {\it i.e.},
$\widehat C_\chi=-i\Mw\sw^2/2\cw^2$ and $\widehat
C_H=-M^2_H/2\Mw+\Mw/2$. We thus have that
\be
\sum_{{\cal S}}\widehat\Gamma_{\phi^+ }^{ {\cal S}^2}\momo+
\sum_{\cal S}\Gamma_{\phip }^{ {\cal S}^2}\momo\equiv
\sum_{{\cal S}}\Gamma_{\widehat\phi^+ }^{ {\cal S}^2}\momo,
\ee
which represents the last identity we need for proving that
(putting back the fermionic indices)
\be
\widehat\Gamma_{\phip \bar f_u f_d}(q,p_1,p_2)\equiv
\Gamma_{\widehat\phi^+ \bar f_u f_d}(q,p_1,p_2).
\ee

\subsection{Neutral scalar sector}

In the neutral scalar sector case, the class ({\it i}) diagrams allow
for the following PT decomposition
\be
\sum_{{\cal V},\,{\cal S}}\Gamma_{\rS }^{{\cal S}{\cal V}}\momo=
\sum_{{\cal V},\,{\cal S}}\Gamma_{\rS }^{{\rm F},{\cal S}{\cal V}}\momo+
\sum_{{\cal V},\,{\cal S}}\Gamma_{\rS }^{{\rm P},{\cal S}{\cal V}}\momo,
\ee
with
\bea
\fl
\sum_{{\cal V},\,{\cal S}}\Gamma_{\rS }^{{\rm F},\,{\cal S}{\cal V}}\momo=
\nonumber \\
\fl
\frac\gw2\int\!\Gamma^{{\rm F}\,\mu}(q,k-q,-k)\uno
\Bigg\{-\cs[{\cal
T}_{W^+_\mu\phim }\momt]_{t,\rm I}
+
\frac1\cw[{\cal T}_{Z_\mu\overline\rS }\momt]_{t,\rm I}\Bigg\} \nonumber \\
\fl+\frac\gw2\int\!\Gamma^{{\rm F}\,\nu}(q,k-q,-k)\Bigg\{\cs[{\cal
T}_{\phip W^-_\nu }\momt]_{t,\rm I}
-\uno\frac1\cw
[{\cal T}_{\overline\rS Z_\nu }\momt]_{t,\rm I}\Bigg\},\nonumber \\
\fl
\sum_{{\cal V},\,{\cal S}}\Gamma_{\rS }^{{\rm P},\,{\cal S}{\cal V}}\momo=
\frac\gw2\int\!(k-q)^\mu\uno\Bigg\{-\cs[{\cal
T}_{W^+_\mu\phim }\momt]_{t,\rm I}
+
\frac1\cw[{\cal T}_{Z_\mu\overline\rS }\momt]_{t,\rm I}\Bigg\} \nonumber \\
\fl+\frac\gw2\int\!k^\nu\Bigg\{\cs[{\cal
T}_{\phip W^-_\nu }\momt]_{t,\rm I}
-\uno\frac1\cw
[{\cal T}_{\overline\rS Z_\nu }\momt]_{t,\rm I}\Bigg\}.
\eea
In the above expressions 
we have that $\uno$ is $1$ (respectively, $-1$) when $\rS=\chi$ (respectively, $\rS=H$), and that
$\overline\rS=H$ (respectively, $\overline\rS=\chi$) when $\rS=\chi$ 
(respectively, $\rS=H$). 

Through the use of the STIs of Eq.(\ref{stinss}), the pinching action
amount in this case to the following
replacement
\bea
\sum_{{\cal V},\,{\cal S}}\Gamma_{\rS}^{{\rm P,}{\cal S}{\cal V}}\momo&\to&
\uno\frac\gw2\int\!\Bigg\{\cs[{\cal
R}^{W^+}_{\phim}\momt]_{t,\rm I}
-\frac1\cw[{\cal R}^{Z}_{\overline\rS}\momt]_{t,\rm I} \nonumber \\
&+&\uno\cs[{\cal
R}_{\phip}^{W^-}\momt]_{t,\rm I}
-\frac1\cw
[{{\cal R}'}_{\overline\rS}^{Z}\momt]_{t,\rm I}\Bigg\},
\eea
so that the effective PT vertex will be given by
\bea
\widehat\Gamma_{\rS}\momo &=&
\sum_{{\cal V},\,{\cal S}}\Gamma_{\rS{\cal S}{\cal V}}^{\rm F}\momo+
\sumii\Gamma_{\rS}^{(ii)}\momo\nonumber \\
&+&\uno\frac\gw2\int\!\Bigg\{\cs[{\cal
R}^{W^+}_{\phim}\momt]_{t,\rm I}
-\frac1\cw[{\cal R}^{Z}_{\overline\rS}\momt]_{t,\rm I} \nonumber \\
&+&\uno\cs[{\cal
R}_{\phip}^{W^-}\momt]_{t,\rm I}
-\frac1\cw
[{{\cal R}'}_{\overline\rS}^{Z}\momt]_{t,\rm I}\Bigg\}.
\eea

We can now proceed to the comparison with the corresponding BFG
$\Gamma_{\widehat\rS}\momo$ vertex. We first of all notice that
as in the charged scalar case we have the following identities
\bea
\fl\sum_f\Gamma_{\rS }^{f\bar f}\momo\equiv\sum_f\Gamma_{\widehat\rS 
}^{f\bar f}\momo, \qquad
\sum_{\cal S}\Gamma_{\rS }^{ {\cal S}^3}\momo\equiv
\sum_S\Gamma_{\widehat\rS }^{ {\cal S}^3}\momo, \qquad
\sum_{{\cal V},\,{\cal S}}\Gamma_{\rS }^{ {\cal S}{\cal V}^2}\momo
\equiv\sum_{{\cal V},\,{\cal S}}\Gamma_{\widehat\rS }^{
 {\cal S}{\cal V}^2}\momo,
\nonumber \\
\fl\sum_{\cal V}\Gamma_{\rS }^{
  {\cal V}^2}\momo\equiv\sum_{\cal V}\Gamma_{\widehat\rS }
^{{\cal V}^2}\momo,
\qquad
\sum_{{\cal V},\,{\cal S}}\Gamma_{\rS }^{{\rm F},\, {\cal S}{\cal V}}\momo
\equiv\sum_{{\cal V},\,{\cal S}}\Gamma_{\widehat\rS }^{{\cal S}{\cal V}}
\momo.
\eea
Next, we concentrate on the PT terms [shown in Eqs.(\ref{Rnss1}) and
(\ref{Rnss2})], starting from the four particle sector of the
$\rS{\cal S}{\cal U}^2$ type.
It is then a long but straightforward exercise 
to check that the PT terms 
\bea
\fl
\sum_{{\cal S},\,{\cal U}}\widehat \Gamma_{\rS }^{{\cal S}
{\cal U}^2}\momo=
-i\gw^2
\frac{\uno}4\cs\sumsp\csp\int\!\Delta_{\bar\um\Phi}(q-k)
\gamma_{\Phi\{\rS'\um\}}
\momt \nonumber \\
\fl
-i\gw^2\frac{\uno}2i\cs\sumv\cv'\int\!\Delta_{\bar\um\Phi}(q-k)
\gamma_{\Phi\{\phim\uv\}}
\momt\nonumber \\
\fl
-i\gw^2\frac{\uno}{4\cw}C_{\overline S}\!\int\!\!\Delta_{\bar\uz\Phi}
(q-k)\left[
\gamma_{\Phi\{\phip\um\}}\momt+(-1)^{\overline
S}\gamma_{\Phi\{\phim\up\}}\momt\right]\nonumber \\
\fl
-i\gw^2\frac1{4\cw^2}\int\!\Delta_{\bar\uz\Phi}(q-k)
\gamma_{\Phi\{\rS\uz\} }
\momt\nonumber \\
\fl+
i\gw^2\frac\cs4\sumsp(-1)^{\rS'}
\csp\int\!\Delta_{\bar\up\Phi}(k)
\gamma_{\{\rS'\up\}\Phi }\momt\nonumber \\
\fl
-i\gw^2\frac{i}2\cs\sumv\cv'
\int\!\Delta_{\bar\up\Phi}(k)\gamma_{\{\phip\uv\}\Phi }
\momt\nonumber \\
\fl+ i\gw^2\frac{\uno}{4\cw}C_{\overline S}\int\!\Delta_{\bar\uz\Phi}(k)\left[
\gamma_{\{\phip\um\}\Phi }\momt+(-1)^{\overline
S}\gamma_{\{\phim\up\}\Phi }\momt\right]\nonumber \\
\fl
+i\gw^2\frac1{4\cw^2}\int\!\Delta_{\bar\uz\Phi}(k)
\gamma_{\{\rS\uz\}\Phi }
\momt,
\eea
generate the four particles BFG ghost sector, {\it i.e.},
\be
\sum_{{\cal S},\,{\cal U}}\widehat \Gamma_{\rS }^{{\cal S}
{\cal U}^2}\momo\equiv\sum_{{\cal S},\,{\cal U}} \Gamma_{\widehat\rS }^{{\cal S}
{\cal U}^2}\momo.
\ee

The remaining PT terms will mix with the corresponding class ({\it ii})
diagrams. In particular, as far as the remaining part of the ghost
sector is concerned, the $\Rxi$FG contributions reads
\bea
\sum_{\cal U}\Gamma_{\rS }^{ {\cal U}^2}\momo&=&i\gw\frac i2\Mw\cs\int\!\left[
\Delta_{\bar\um\Phi}(q-k)
\Delta_{\um\Phi'}(k)\gamma_{\Phi\Phi' }\momt\right.\nonumber \\
&+&\left.
\uno\Delta_{\up\Phi}(q-k)
\Delta_{\bar\up\Phi'}(k)
\gamma_{\Phi\Phi' }\momt\right]\nonumber \\
&-&i\gw\frac{\Mw}{2\cw^2}\delta_{\rS H}
\int\!\left[
\Delta_{\bar\uz\Phi}(q-k)\Delta_{\uz\Phi'}(k)
\gamma_{\Phi\Phi' }\momt\right.\nonumber
\\
&-&\left. 
\Delta_{\uz\Phi}(q-k)\Delta_{\bar\uz\Phi'}(k)
\gamma_{\Phi\Phi' }\momt\right].
\eea
At this point we observe that the background field $\widehat\chi$ does
not couple with (two) ghosts, so that the PT terms must precisely
cancel the $\Rxi$FG ones. In fact, after adding the two contributions,
one can easily check that
\bea
\fl
\sum_{\cal U}\widehat\Gamma_{\rS }^{ {\cal U}^2}\momo+
\sum_{\cal U}\Gamma_{\rS }^{ {\cal U}^2}\momo
=i\gw\frac i2\Mw\cs[1-\uno]\!\!\int\!\!\left[
\Delta_{\bar\um\Phi}(q-k)
\Delta_{\um\Phi'}(k)
\gamma_{\Phi\Phi' }\momt\right.\nonumber \\
\fl-\left.\Delta_{\up\Phi}(q-k)
\Delta_{\bar\up\Phi'}(k)
\gamma_{\Phi\Phi' }\momt\right]\nonumber \\
\fl-i\gw\frac{\Mw}{\cw^2}\delta_{\rS H}
\int\!\left[\Delta_{\bar\uz\Phi}(q-k)
\Delta_{\uz\Phi'}(k)
\gamma_{\Phi\Phi' }\momt\right.\nonumber \\
\fl-\left. \Delta_{\uz\Phi}(q-k)
\Delta_{\bar\uz\Phi'}(k)
\gamma_{\Phi\Phi' }\momt\right],
\eea
correctly reproduces the BFG $\widehat\rS {\cal U}^2$ sector, {\it i.e.},
\be
\sum_{\cal U}\widehat\Gamma_{\rS }^{ {\cal U}^2}\momo+
\sum_{\cal U}\Gamma_{\rS }^{ {\cal U}^2}\momo\equiv
\sum_{\cal U}\Gamma_{\widehat\rS }^{ {\cal U}^2}\momo.
\ee

We finally need to consider the scalar--scalar sector. The $\Rxi$FG 
contributions to the latter read
\bea
\sum_{{\cal S}}\Gamma_{\rS }^{{\cal S}^2}\momo&=&
-i\gw\delta_{\rS H}\frac{3M_H^2}{2\Mw}
\int\!\Delta_{H\Phi}(q-k)\Delta_{H\Phi'}(k)\gamma_{\Phi\Phi' }
\momt\nonumber \\
&-&i\gw\delta_{\rS H}\frac{M_H^2}{2\Mw}
\int\!\Delta_{\phip\Phi}(q-k)
\Delta_{\phim\Phi'}(k)\gamma_{\Phi\Phi' }\momt\nonumber \\
&-&i\gw\frac{M_H^2}{2\Mw}\int\!\Delta_{\chi\Phi}(q-k)
\Delta_{\overline\rS\Phi'}(k)
\gamma_{\Phi\Phi' }\momt\nonumber \\
&-&i\gw\frac{M_H^2}{2\Mw}\int\!\Delta_{\overline\rS\Phi}(q-k)
\Delta_{\chi\Phi'}(k)
\gamma_{\Phi\Phi' }\momt.
\eea
Now, we first of all notice that the BFG vertex $\widehat HHH$ and the BFG one
$HHH$ coincide; moreover since there is no BFG
$\widehat\chi\phip\phim$ coupling, the PT terms should exactly vanish
in this case, since there is no $\Rxi$FG $\chi\phip\phim$ coupling
either. After adding the PT contributions to the above terms, one
has the result
\bea
\fl\sum_{{\cal S}}\widehat\Gamma_{\rS }^{{\cal S}^2}\momo
+\sum_{{\cal S}}\Gamma_{\rS }^{{\cal S}^2}\momo=\nonumber \\
\fl-i\gw\delta_{\rS H}\frac{3M_H^2}{2\Mw}
\int\!\Delta_{H\Phi}(q-k)\Delta_{H\Phi'}(k)\gamma_{\Phi\Phi' }
\momt\nonumber \\
\fl
i\gw\left\{\frac i2\cs\Mw[1-\uno]-\delta_{\rS H}\frac{M_H^2}{2\Mw}\right\}
\int\!\Delta_{\phip\Phi}(q-k)
\Delta_{\phim\Phi'}(k)\gamma_{\Phi\Phi' }\momt\nonumber
\\
\fl+ i\gw\left\{(-1)^{\rS}\frac\Mw{2\cw^2}-\frac{M^2_H}{2\Mw}\right\}
\int\!\Delta_{\chi\Phi}(q-k)
\Delta_{\overline\rS\Phi'}(k)
\gamma_{\Phi\Phi' }\momt\nonumber \\
\fl+ i\gw\left\{(-1)^{\rS}\frac\Mw{2\cw^2}
-\frac{M^2_H}{2\Mw}\right\}
\int\!\Delta_{\overline\rS\Phi}(q-k)\Delta_{\chi\phi'}(k)
\gamma_{\Phi\Phi' }\momt,
\eea
which precisely coincide with the BFG one, {\it i.e} we have
\be
\sum_{{\cal S}}\widehat\Gamma_{\rS }^{{\cal S}^2}\momo
+\sum_{{\cal S}}\Gamma_{\rS }^{{\cal S}^2}\momo\equiv
\sum_{{\cal S}}\Gamma_{\widehat\rS }^{{\cal S}^2}\momo.
\ee
This concludes the proof that (putting back fermionic indices)
\be
\widehat\Gamma_{\rS \bar f f}(q,p_1,p_2)\equiv
\Gamma_{\widehat\rS \bar f f}(q,p_1,p_2).
\ee  

\section{\label{Rec}Reconstruction of the PT terms: two-point functions}
 
Of course at this point one would expect that the two-point functions
too coincide with the BFG ones, since both the boxes as
well as the vertices coincide with the corresponding quantities in the
BFG, and the $S$-matrix is unique. A proof based on the strong
induction principle along the same lines of the one carried out in 
\cite{Binosi:2002ft} for the QCD case can be easily
carried out.  

In this section however, 
we are going to briefly address a slightly different question that is:
can one {\it explicitly} reconstruct the pinching parts that {\it
implicitly}  cancel in our all-order generalization of the PT
procedure? or, equivalently, can one explicitly construct the
two-point PT functions?
The use of the BQIs of Eq.(\ref{APBQI}) and (\ref{AVBQI}) 
will allow a positive answer to both questions.

To fix the ideas we will hereafter consider the charged electroweak
sector, {\it i.e.}, we choose ${\cal V}_\alpha\equiv W^+_\alpha$ and
${\cal S}\equiv\phi^+$, so that the BQIs of Eq.(\ref{AVBQI}) read
\bea
\Gamma_{\widehat W^+_\alpha \bar f_u f_d }&=&
\Gamma_{W^+_\alpha \bar f_u f_d }+
\Gamma_{\Omega^+_\alpha W^{*-}_\mu}
\Gamma_{W^{+\mu}\bar f_u f_d }+
\Gamma_{\Omega^+_\alpha\phi^{*-}}
\Gamma_{\phi^+\bar f_u f_d }, \nonumber \\
\Gamma_{\widehat\phi^+\bar f_u f_d }&=&
\Gamma_{\phi^+\bar f_u f_d }+
\Gamma_{\Omega^+{\phi}^{*-}}
\Gamma_{\phi^+\bar f_u f_d }+
\Gamma_{\Omega^+W^{*-}_\mu}
\Gamma_{W^{+\mu} \bar f_u f_d },
\label{WBQI}
\eea
where we have omitted the momentum dependence of the various Green's
function as for the rest of this section [the latter can be easily reconstructed
from the appendix equations (\ref{APBQI}) and (\ref{AVBQI})].

On the other  hand, due to our explicit construction,  we know that in
general        $\Gamma_{\widehat{\cal       V}_\alpha \bar f f     }\equiv
\widehat\Gamma_{{\cal   V}_\alpha  \bar f f }$   and  $\Gamma_{\widehat{\cal
S}\bar f f }\equiv\widehat\Gamma_{{\cal  S}\bar f f }$,  so  that the  above  BQIs
express  (to all  orders) the  relations  between the  PT three  point
Green's functions and the  normal ones, and (upon inversion) viceversa.

To extract the propagator-like pieces  from the above BQIs, one has to
isolate  all  the  terms   proportional  to  the  tree-level  vertices
$\Gamma^{(0)}_{W^+ \bar f_u f_d} $  and $\Gamma^{(0)}_{\phi^+\bar f_u f_d} $. 
The complete
set of these  terms can be found inverting  at each perturbative order
Eqs.(\ref{WBQI})  thus   writing  the  $\Rxi$   three  points  Green's
functions in  terms of the PT  ones, and observing  that at tree-level
they coincide.  It is clear from the structure of Eq.(\ref{WBQI}) that
the propagator-like  pieces extracted from the $\Gamma_{W^+_\alpha
\bar f_u f_d} $ BQI (respectively, $\Gamma_{\phi^+\bar f_u f_d} $) 
will  contain in general both the two
points    functions    $\Gamma_{\Omega^+_\alpha    W^{*-}_\mu}$    and
$\Gamma_{\Omega^+_\alpha\phi^{*-}}$     (respectively, $\Gamma_{\Omega^+{\phi}^{*-}}$
and $\Gamma_{\Omega^+W^{*-}_\mu}$).

After isolating  the propagator-like  terms one has  to decide  how to
allot   them  among   the   available  $\Rxi$   two  point   functions
$\Gamma_{W^+_\alpha      W^-_\beta}$,     $\Gamma_{W^+_\alpha\phi^-}$,
$\Gamma_{\phi^+ W^-_\beta}$  and $\Gamma_{\phi^+\phi^-}$.  
At  a first
sight,   it  is  tempting   to  assign   the  $\Gamma_{\Omega^+_\alpha
W^{*-}_\mu}$ ($\Gamma_{\Omega^+{\phi}^{*-}}$) part entirely to the $W$
($\phi$) self-energy,  while   giving   the  term
proportional           to          $\Gamma_{\Omega^+_\alpha\phi^{*-}}$
($\Gamma_{\Omega^+W^{*-}_\mu}$)     to      the     mixed     function
$\Gamma_{W^+_\alpha\phi^-}$ ($\Gamma_{\phi^+ W^-_\beta}$). To see that
this is however not the case, we notice that at
one-loop     $\Gamma^{(1)}_{\Omega^{\cal    V}_\alpha{\cal    S}^*}=0$
(respectively, $\Gamma^{(1)}_{\Omega^{\cal S}{\cal V}^*_\mu}=0$), but still the BQIs
should  provide contribution  for  both the  $\Gamma^{(1)}_{W^+_\alpha
W^-_\beta}$   as  well   as  the   $\Gamma^{(1)}_{W^+_\alpha  \phi^-}$
two-point       functions      (respectively, 
$\Gamma^{(1)}_{\phi^+\phi^-}$      and $\Gamma^{(1)}_{\phi^+ W^-_\beta}$).

The correct procedure is instead the following \cite{Papavassiliou:1994pr}:
\begin{itemize}
\item[{\it i})] To isolate from the terms proportional to
$\Gamma^{(0)}_{W^+_\alpha \bar f_u f_d}$  the corresponding two type of
contributions, we insert the identity
\bea
 g^\mu_\beta&=&[\Delta_{W^{+\mu}W^{-\nu}}^{(0)}(q)]^{-1}\Delta^{(0)}_{W^+_\nu W^-_\beta}(q)
\nonumber \\
&=& -\Gamma^{(0)}_{W^{+\mu}W^{-\nu}}(q)\Delta^{(0)}_{W^+_\nu W^-_\beta}(q)-
\frac{q_\beta}{M_W}\Gamma^{(0)}_{W^{+\mu}\phi^-}(q) \Delta^{(0)}_{\phi^+\bar f_u f_d}(q).
\label{tr1}
\eea
When  looking at the BQI for $\widehat\Gamma_{W^+ \bar f_u f_d}$
(respectively, $\widehat\Gamma_{\phi^+ \bar f_u f_d}$) the first term will contribute to the
$\widehat\Gamma_ {W^+W^-}$ (respectively, $\widehat\Gamma_ {\phi^+W^-}$) PT two point
function, while 
the second to the $\widehat\Gamma_ {W^+\phi^-}$ (respectively, $\widehat\Gamma_
{\phi^+\phi^-}$) 
one.

\item[{\it ii})] To isolate from the terms proportional to
  $\Gamma^{(0)}_{\phi^+\bar f_u f_d}$  the corresponding two type of 
contributions, we instead make use of the following relation, holding
  when contracted with on-shell spinors
\bea
\fl
\Gamma^{(0)}_{\phi^+\bar f_u f_d}(q,p_1,p_2)&=&\frac{q^\rho}{M_W}\Gamma^{(0)}_{W^+_\rho\bar f_u f_d
  } (q,p_1,p_2)\nonumber \\
&=&-\Gamma^{(0)}_{\phi^+\phi^-}(q)\Delta^{(0)}_{\phi^+\phi^-}(q)
\Gamma^{(0)}_{\phi^+\bar f_u f_d}(q,p_1,p_2) \nonumber \\
&-&\Gamma^{(0)}_{\phi^+W^-_\nu}(q) 
\Delta_{W^{+\rho} W^{-\nu}}^{(0)}(q)
\Gamma^{(0)}_{W^+_\rho \bar f_u f_d}(q,p_1,p_2).
\label{tr2}
\eea 
When  looking at the BQI for $\widehat\Gamma_{\phi^+\bar f_u f_d}$
(respectively, $\widehat\Gamma_{W^+ \bar f_u f_d}$) the first term will contribute to the
$\widehat\Gamma_ {\phi^+\phi^-}$ (respectively, $\widehat\Gamma_ {\phi^+W^-}$) PT two point
  function, while 
the second to the $\widehat\Gamma_ {\phi^+W^-}$ (respectively, $\widehat\Gamma_
  {W^+W^-}$) one.
\end{itemize}

At the one-loop level one has for example that
\bea
\Gamma^{(1)}_{W^+_\alpha \bar f_u f_d}&=&
\widehat\Gamma^{(1)}_{W^+_\alpha \bar f_u f_d}-
\Gamma^{(1)}_{\Omega^+_\alpha
  W^{*-}_\mu}\Gamma^{(0)}_{W^{+\mu}\bar f_u f_d},
\nonumber \\
\Gamma^{(1)}_{\phi^+ \bar f_u f_d}&=&
\widehat\Gamma^{(1)}_{\phi^+ \bar f_u f_d}-
\Gamma^{(1)}_{\Omega^+
 \phi^{*-}}\Gamma^{(0)}_{\phi^{+}\bar f_u f_d},
\eea
Making use of Eqs.(\ref{tr1}) and (\ref{tr2}) we then find
\bea
-\Gamma^{(1)}_{\Omega^+_\alpha
  W^{*-}_\mu}\Gamma^{(0)}_{W^{+\mu}\bar f_u f_d}
&=&[\Gamma^{(1)}_{\Omega^+_\alpha
W^{*-}_\mu}\Gamma^{(0)}_{W^+_\mu W^-_\nu}]
\Delta_{W^{+\nu}W^{-\beta}}^{(0)\,\nu\beta}\Gamma^{(0)}_{W^+_\beta
 \bar f_u f_d }\nonumber \\
&+&[\Gamma^{(1)}_{\Omega^+_\alpha W^{*-}_\mu}
\Gamma^{(0)}_{W^{+\mu}\phi^-}]\Delta^{(0)}_{\phi^+\phi^-}
\Gamma^{(0)}_{\phi^+\bar f_u f_d},\nonumber\\
-\Gamma^{(1)}_{\Omega^+
 \phi^{*-}}\Gamma^{(0)}_{\phi^{+}\bar f_u f_d}&=&[
\Gamma^{(1)}_{\Omega^+\phi^{*-}}
\Gamma^{(0)}_{\phi^+\phi^-}]\Delta^{(0)}_{\phi^+\phi^-}
\Gamma^{(0)}_{\phi^+\bar f_u f_d}\nonumber \\
&+&[\Gamma^{(1)}_{\Omega^+\phi^{*-}}
\Gamma^{(0)}_{\phi^+W^-_\nu}]
\Delta_{W^{+\nu}W^{-\beta}}^{(0)}
\Gamma^{(0)}_{W^+_\beta\bar f_u f_d}.
\eea
Therefore, taking into account the mirror vertices, we end up with the results
\bea
\widehat\Gamma^{(1)}_{W^+_\alpha W^-_\beta}&=&
\Gamma^{(1)}_{W^+_\alpha W^-_\beta}+
\Gamma^{(1)}_{\Omega^+_\alpha W^{*-}_\mu}
\Gamma^{(0)}_{W^{+\mu}W^-_\beta}+
\Gamma^{(1)}_{\Omega^-_\beta W^{*-}_\mu}
\Gamma^{(0)}_{W^+_\alpha W^{-\mu}},\nonumber \\
\widehat\Gamma^{(1)}_{W^+_\alpha\phi^-}&=&
\Gamma^{(1)}_{W^+_\alpha\phi^-}+\Gamma^{(1)}_{\Omega^+_\alpha W^{*-}_\mu}
\Gamma^{(0)}_{W^{+\mu}\phi^-}+\Gamma^{(1)}_{\Omega^-\phi^{*+}}
\Gamma^{(0)}_{W^+_\alpha\phi^-}, \nonumber \\
\widehat\Gamma^{(1)}_{\phi^+W^-_\beta}&=&
\Gamma^{(1)}_{\phi^+W^-_\beta}+
\Gamma^{(1)}_{\Omega^+ \phi^{*-}}
\Gamma^{(0)}_{\phi^+W^-_\beta}+\Gamma^{(1)}_{\Omega^-_\beta W^{*+}_\mu}
\Gamma^{(0)}_{\phi^+W^{-\mu}}, \nonumber \\
\widehat\Gamma^{(1)}_{\phi^+ \phi^-}&=&
\Gamma^{(1)}_{\phi^+ \phi^-}+
\Gamma^{(1)}_{\Omega^+ \phi^{*-}}
\Gamma^{(0)}_{\phi^+\phi^-}+\Gamma^{(1)}_{\Omega^- \phi^{*+}}
\Gamma^{(0)}_{\phi^+\phi^-}.
\eea
Inspection of the above expressions shows that they coincide precisely
with the one-loop expansion of the BQIs of Eq.(\ref{APBQI}), thus we
recover the well known results
\bea
\widehat\Gamma^{(1)}_{W^+_\alpha W^-_\beta}\equiv
\Gamma^{(1)}_{\widehat W^+_\alpha \widehat W^-_\beta}, &\qquad&
\widehat\Gamma^{(1)}_{W^+_\alpha\phi^-}\equiv
\Gamma^{(1)}_{\widehat W^+_\alpha\widehat \phi^-}, \nonumber \\
\widehat\Gamma^{(1)}_{\phi^+W^-_\beta}\equiv
\Gamma^{(1)}_{\widehat\phi^+\widehat W^-_\beta}, &\qquad&
\widehat\Gamma^{(1)}_{\phi^+ \phi^-}\equiv
\Gamma^{(1)}_{\widehat \phi^+ \widehat\phi^-}.
\eea

More involved is the analysis in the two-loop case, where
the BQIs for
the vertex functions read
\bea 
\Gamma^{(2)}_{W^+_\alpha \bar f_u f_d} &=& 
\widehat\Gamma^{(2)}_{W^+_\alpha \bar f_u f_d}-
\Gamma^{(2)}_{\Omega^+_\alpha W^{*-}_\mu}
\widehat\Gamma^{(1)}_{W^{+\mu} \bar f_u f_d} \nonumber \\
&-&\left[\Gamma^{(2)}_{\Omega^+_\alpha W^{*-}_\rho}-
\Gamma^{(1)}_{\Omega^+_\alpha W^{*-}_\mu}
\Gamma^{(1)}_{\Omega^{+\mu} W^{*-}_\rho}\right]
\Gamma^{(0)}_{W^{+\rho}\bar f_u f_d}-\Gamma^{(2)}_{\Omega^+_\alpha \phi^{*-}}(q)
\Gamma^{(0)}_{\phi^+\bar f_u f_d}, \nonumber \\
\Gamma^{(2)}_{\phi^+ \bar f_u f_d} &=& 
\widehat\Gamma^{(2)}_{\phi^+ \bar f_u f_d}-
\Gamma^{(2)}_{\Omega^+ \phi^{*-}}
\widehat\Gamma^{(1)}_{\phi^+ \bar f_u f_d}\momo \nonumber \\
&-&\left[\Gamma^{(2)}_{\Omega^+ \phi^{*-}}-
\Gamma^{(1)}_{\Omega^+ \phi^{*-}}
\Gamma^{(1)}_{\Omega^+ \phi^{*-}}\right]
\Gamma^{(0)}_{\phi^{+}\bar f_u f_d}-\Gamma^{(2)}_{\Omega^+ W^{*-}_\mu}
\Gamma^{(0)}_{W^{+\mu}\bar f_u f_d}. 
\eea
To extract the corresponding propagator-like pieces we can make use, as in
the previous  case, of the identities of Eq.(\ref{tr1}) and
(\ref{tr2}).
However beyond the one-loop level this is not the end of the story: the
conversion of the 1PR string $\mathbb{S}^{(2)}$ 
of (normal) one-loop self-energies into the
corresponding string $\widehat{\mathbb{S}}^{(2)}$ 
of one-loop PT self-energies has to be taken into
account. This will generated the 1PI
contributions $\mathbb{S}^{(2),\rm{1PI}}$ 
that have to be allotted to the corresponding two-loop PT
two-point functions. 

Therefore, after adding the mirror vertex contributions, we have the results
\bea
\fl
\widehat \Gamma^{(2)}_{W^+_\alpha W^-_\beta} =
\Gamma^{(2)}_{W^+_\alpha W^-_\beta}+\mathbb{S}^{(2),{\rm
    1PI}}_{W^+_\alpha W^-_\beta} +\Gamma^{(2)}_{\Omega^+_\alpha
  W^{*-}_\mu}\Gamma^{(0)}_{W^{+\mu}W^-_\beta}
+\Gamma^{(2)}_{\Omega^-_\beta W^{*+}_\mu}\Gamma^{(0)}_{W^+_\alpha
  W^{-\mu}} 
+ \Gamma^{(2)}_{\Omega^+_\alpha
  \phi^{*-}}\Gamma^{(0)}_{\phi^+W^-_\beta}\nonumber \\
\fl \mbox{}\hspace{1.4cm} 
+\Gamma^{(2)}_{\Omega^-_\beta \phi^{*+}}\Gamma^{(0)}_{W^+_\alpha
  \phi^-} 
-\Gamma^{(1)}_{\Omega^+_\alpha
  W^{*-}_\mu}\Gamma^{(1)}_{\Omega^{+_\mu}
  W^{*-}_\rho}\Gamma^{(0)}_{W^{+\rho}W^-_\beta}-
\Gamma^{(1)}_{\Omega^-_\beta
  W^{*+}_\mu}\Gamma^{(1)}_{\Omega^{-_\mu}
  W^{*+}_\rho}\Gamma^{(0)}_{W^+_\alpha W^{-_\rho}}, \nonumber\\
\fl
\widehat \Gamma^{(2)}_{W^+_\alpha \phi^-} \hspace{0.15cm}=
\Gamma^{(2)}_{W^+_\alpha \phi^-}+\mathbb{S}^{(2),{\rm
    1PI}}_{W^+_\alpha \phi^-} +\Gamma^{(2)}_{\Omega^+_\alpha
  W^{*-}_\mu}\Gamma^{(0)}_{W^{+\mu}\phi^-}
+\Gamma^{(2)}_{\Omega^- \phi^{*+}}(q)\Gamma^{(0)}_{W^+_\alpha
  \phi^-}
+ \Gamma^{(2)}_{\Omega^+_\alpha
  \phi^{*-}}\Gamma^{(0)}_{\phi^+\phi^-}\nonumber \\
\fl \mbox{}\hspace{1.4cm}
+\Gamma^{(2)}_{\Omega^- W^{*+}_\mu}\Gamma^{(0)}_{W^+_\alpha
  W^{-\mu}}  
-\Gamma^{(1)}_{\Omega^+_\alpha
  W^{*-}_\mu}\Gamma^{(1)}_{\Omega^{+_\mu}
  W^{*-}_\rho}\Gamma^{(0)}_{W^{+\rho}\phi^-}-
\Gamma^{(1)}_{\Omega^-
  \phi^{*+}}\Gamma^{(1)}_{\Omega^-
  \phi^{*+}}\Gamma^{(0)}_{W^+_\alpha \phi^-}, \nonumber\\
\fl
\widehat \Gamma^{(2)}_{\phi^+ W^-_\beta} \hspace{0.15cm}=
\Gamma^{(2)}_{\phi^+ W^-_\beta}+\mathbb{S}^{(2),{\rm
    1PI}}_{\phi^+ W^-_\beta} +\Gamma^{(2)}_{\Omega^+
  \phi^{*-}}\Gamma^{(0)}_{\phi^+W^-_\beta}(q)
+\Gamma^{(2)}_{\Omega^-_\beta W^{*+}_\mu}\Gamma^{(0)}_{\phi^+
  W^{-\mu}} 
+ \Gamma^{(2)}_{\Omega^+
  W^{*,-}_\mu}\Gamma^{(0)}_{W^{+\mu}W^-_\beta}\nonumber \\
\fl \mbox{}\hspace{1.4cm} 
+\Gamma^{(2)}_{\Omega^-_\beta \phi^{*+}}\Gamma^{(0)}_{\phi^+
  \phi^-} 
-\Gamma^{(1)}_{\Omega^+
  \phi^{*-}}\Gamma^{(1)}_{\Omega^+
  \phi^{*-}}\Gamma^{(0)}_{\phi^+W^-_\beta}-
\Gamma^{(1)}_{\Omega^-_\beta
  W^{*+}_\mu}\Gamma^{(1)}_{\Omega^{-_\mu}
  W^{*+}_\rho}\Gamma^{(0)}_{\phi^+ W^-_\rho}, \nonumber\\
\fl
\widehat \Gamma^{(2)}_{\phi^+ \phi^-} \hspace{0.25cm}=
\Gamma^{(2)}_{\phi^+ \phi^-}+\mathbb{S}^{(2),{\rm
    1PI}}_{\phi^+ \phi^-} +\Gamma^{(2)}_{\Omega^+
  \phi^{*-}}\Gamma^{(0)}_{\phi^+\phi^-}
+\Gamma^{(2)}_{\Omega^- \phi^{*+}}\Gamma^{(0)}_{\phi^+
  \phi^-}+
  \Gamma^{(2)}_{\Omega^+
  W^{*-}_\mu}\Gamma^{(0)}_{W^{+\mu}\phi^-}\nonumber\\
\fl \mbox{}\hspace{1.4cm} 
+\Gamma^{(2)}_{\Omega^- W^{*+}_\mu}(q)\Gamma^{(0)}_{\phi^+
  W^{-\mu}}  
-\Gamma^{(1)}_{\Omega^+
  \phi^{*-}}\Gamma^{(1)}_{\Omega^+
  \phi^{*-}}\Gamma^{(0)}_{\phi^+\phi^-}-
\Gamma^{(1)}_{\Omega^-
  \phi^{*+}}\Gamma^{(1)}_{\Omega^-
  \phi^{*+}}\Gamma^{(0)}_{\phi^+ \phi^-}. 
\label{TL1}
\eea 
Moreover, following the second paper of \cite{Papavassiliou:1995fq} we find,
\bea
\mathbb{S}^{(2),{\rm
    1PI}}_{W^+_\alpha W^-_\beta} =
\Gamma^{(1)}_{\Omega^+_\alpha W^{*-}_\mu} \Gamma^{(1)}_{W^{+\mu}W^-_\beta} 
+\Gamma^{(1)}_{\Omega^-_\beta W^{*+}_\mu} 
\Gamma^{(1)}_{W^+_\alpha W^{-\mu}} 
+\Gamma^{(1)}_{\Omega^+_\alpha
  W^{*-}_\mu} \Gamma^{(1)}_{\Omega^{+_\mu}
  W^{*-}_\rho} \Gamma^{(0)}_{W^{+\rho}W^-_\beta} \nonumber \\
\mbox{}\hspace{1.4cm}  +
\Gamma^{(1)}_{\Omega^-_\beta
  W^{*+}_\mu} \Gamma^{(1)}_{\Omega^{-_\mu}
  W^{*+}_\rho} \Gamma^{(0)}_{W^+_\alpha W^-_\rho} 
+ \Gamma^{(1)}_{\Omega^-_\beta
  W^{*+}_\mu} \Gamma^{(1)}_{\Omega^+_\alpha
  W^{*-}_\rho} \Gamma^{(0)}_{W^{+\rho}W^{-\mu}} , \nonumber \\
\mathbb{S}^{(2),{\rm
    1PI}}_{W^+_\alpha \phi^-} \hspace{.15cm}=
\Gamma^{(1)}_{\Omega^+_\alpha W^{*-}_\mu} 
\Gamma^{(1)}_{W^{+\mu}\phi^-} +\Gamma^{(1)}_{\Omega^-\phi^{*+}}
\Gamma^{(1)}_{W^+_\alpha\phi^-}
+\Gamma^{(1)}_{\Omega^+_\alpha
  W^{*-}_\mu} \Gamma^{(1)}_{\Omega^{+_\mu}
  W^{*-}_\rho} \Gamma^{(0)}_{W^{+\rho}\phi^-} \nonumber \\
 \mbox{}\hspace{1.4cm}  +
\Gamma^{(1)}_{\Omega^-
  \phi^{*+}} \Gamma^{(1)}_{\Omega^-
  \phi^{*+}} \Gamma^{(0)}_{W^+_\alpha \phi^-}+ 
\Gamma^{(1)}_{\Omega^-\phi^{*+}} 
\Gamma^{(1)}_{\Omega^+_\alpha W^{*-}_\mu} \Gamma^{(0)}_{W^{+\mu}\phi^-} ,
\nonumber\\
\mathbb{S}^{(2),{\rm
    1PI}}_{\phi^+ W^-_\beta} \hspace{.15cm}=
\Gamma^{(1)}_{\Omega^+
  \phi^{*-}} \Gamma^{(1)}_{\phi^+W^-_\beta} 
+\Gamma^{(1)}_{\Omega^-_\beta
  W^{*+}_\mu} \Gamma^{(0)}_{\phi^+ W^{-\mu}} 
+\Gamma^{(1)}_{\Omega^+
  \phi^{*-}} \Gamma^{(1)}_{\Omega^+
  \phi^{*-}} \Gamma^{(0)}_{\phi^+W^-_\beta} \nonumber \\
\mbox{}\hspace{1.4cm}  +
\Gamma^{(1)}_{\Omega^-_\beta
  W^{*+}_\mu} \Gamma^{(1)}_{\Omega^{-_\mu}
  W^{*+}_\rho} \Gamma^{(0)}_{\phi^+ W^{-\rho}} 
+\Gamma^{(1)}_{\Omega^-_\beta W^{*+}_\mu} \Gamma^{(1)}_{\Omega^+
  \phi^{*-}} \Gamma^{(0)}_{\phi^+W^{-\mu}} , \nonumber \\
\mathbb{S}^{(2),{\rm
    1PI}}_{\phi^+ \phi^-} \hspace{.15cm}=
\Gamma^{(1)}_{\Omega^+\phi^{*-}} \Gamma^{(1)}_{\phi^+ \phi^-} +
\Gamma^{(1)}_{\Omega^-\phi^{*+}} \Gamma^{(1)}_{\phi^+ \phi^-} 
+\Gamma^{(1)}_{\Omega^+
  \phi^{*-}} \Gamma^{(1)}_{\Omega^+
  \phi^{*-}} \Gamma^{(0)}_{\phi^+\phi^-} \nonumber\\
\mbox{}\hspace{1.4cm}   +
\Gamma^{(1)}_{\Omega^-
  \phi^{*+}} \Gamma^{(1)}_{\Omega^-
  \phi^{*+}} \Gamma^{(0)}_{\phi^+ \phi^-}  +
  \Gamma^{(1)}_{\Omega^-
  \phi^{*+}} \Gamma^{(1)}_{\Omega^+
  \phi^{*-}} \Gamma^{(0)}_{\phi^+ \phi^-} .
\eea

Adding the above contributions to Eq.(\ref{TL1}), we see that the
resulting expressions coincide precisely with the two-loop expansion
of the BQIs of Eq.(\ref{APBQI}), thus providing us with the (expected) 
result
\bea
\widehat\Gamma^{(2)}_{W^+_\alpha W^-_\beta} \equiv
\Gamma^{(2)}_{\widehat W^+_\alpha \widehat W^-_\beta}, &\qquad&
\widehat\Gamma^{(2)}_{W^+_\alpha\phi^-} \equiv
\Gamma^{(2)}_{\widehat W^+_\alpha\widehat \phi^-}, \nonumber \\
\widehat\Gamma^{(2)}_{\phi^+W^-_\beta} \equiv
\Gamma^{(2)}_{\widehat\phi^+\widehat W^-_\beta}, &\qquad&
\widehat\Gamma^{(2)}_{\phi^+ \phi^-} \equiv
\Gamma^{(2)}_{\widehat \phi^+ \widehat\phi^-},
\eea
which extends to the full SM the one proved in \cite{Binosi:2002bs}
for the case of conserved currents.

\section{\label{Conc}Conclusions}

In the present paper we have extended the algorithm presented in
\cite{Binosi:2002ft} for generalizing the PT to all orders
in QCD, to the case of the electroweak sector of the SM.
This  generalization has
been a  pending problem,  mainly due both to  the proliferation  of Feynman
diagrams as compared  to the QCD case, as well  as to the
complication arising  from the presence of Goldtone's bosons in the theory, 
which implies that the BRST symmetry (and therefore the STIs) are now realized 
through them. These problems have been
solved  by   resorting  to   the  recently  introduced  PT
construction by  means of the STI satisfied by 
special four-point functions which serve as  a  common kernel  
to  all  higher  order
self-energy and vertex diagrams. Thus,   instead  of   manipulating  
algebraically
individual  Feynman  diagrams,  all the pinching action could be 
simultaneously  addressed.
In particular, we have shown that without any modification (apart from
the obvious one of the inclusion of the vertices of the type ${\cal
  S}^2{\cal V}$ as sources of pinching momenta in the scalar sector), the QCD algorithm
goes through in the SM case, allowing for the all-order generalization
of the PT. 

It should be clear by now that, being valid to all orders, the PT is a
procedure intrinsic to any gauge theory (Abelian and non-Abelian), 
that can be applied to obtain
Green's functions possessing many of the properties of $S$-matrix
elements. This is particularly important in the SM case, in which the consistent
description of unstable particles necessitates the definition and resummation
of off-shell (two-point) Green's functions, which must respect the crucial
physical requirements of resummability, analiticity, unitarity, gauge
invariance, multiplicative renormalization and no shifting of the
position of the pole \cite{Papavassiliou:1995fq}. 
This is naturally provided by the PT two-point functions to all orders.

The correspondence between the PT Green's functions and the BFG ones, that we have shown to persist to all order, has been a source of considerable confusion in the literature; in particular it has been argued that the PT is but a special case of the BFM, representing one out of an infinite number of equivalent choices parametrized by the  $\xi_Q$ gauge fixing parameter. One should however recall that 
for a general value of the latter parameter, the BFM Green's functions shows a residual dependence on it.
Thus, from the PT point of view there is no difference between a theory quantized in the BFM or $R_\xi$ gauges: in fact, to eliminate this residual $\xi_Q$ dependence of the BFM Green's functions, one would apply the very same PT algorithm discussed in this paper (but with the STIs written down in this gauge), and arrive at precisely the same vertices and propagators 
we have described. The BFG has only the special property that pinching contribution vanishes, and thus is the most economical way of obtainig the PT results. In addition, the PT construction goes through unaltered, under 
circumstances where the BFM Feynman rules cannot even be applied.
Specifically, if instead of an $S$-matrix element one were to 
consider a different observable, such as a current-current correlation
function or a Wilson loop, as was in fact done by Cornwall in the original 
formulation  \cite{Cornwall:1982zr} (and, more recently, in 
\cite{Binosi:2001hy}), one could not start out using the 
BFM Feynman rules, because {\it all} fields appearing inside the 
first non-trivial loop are quantum ones. Instead, by following the 
PT rearrangement inside these physical amplitudes 
one would {\it dynamically} arrive at the BFM answer. 

Notice that the renormalization program will not spoil the PT construction presented here. 
Of course there is no doubt that if one supplies the correct set of counterterms within the conventional formulation the entire $S$-matrix will continue to be renormalized, even after the PT rearrangements of the (unrenormalized) Feynman graphs. The question is eventually if the new Green's functions constructed through the PT rearrangements are {\it individually} renormalizable (a classic counter example being the unitary gauge of the SM, where the entire $S$-matrix is renormalizable, whereas the individual Green's functions computed in this gauge are not). The general methodology for dealing with this issue has been established in the second paper of \cite{Papavassiliou:1999az}, where the two-loop QCD
case was studied in detail, and consist of two steps: one 
should first of all start out with the counterterms
which  are  necessary  to  renormalize individually  the  conventional
Green's functions contributing to  the $n$-loop $S$-matrix in the RFG;
then, one should show  that, by simply rearranging these counterterms,
following the PT rules, we arrive at the renormalized $n$-loop PT Green's
functions. This analysis was extended to the (QCD) all order case in the second paper of \cite{Binosi:2002ft},
where it was shown that renormalization poses no problems whatsoever to the PT construction. The generalization of the above analysis to the SM case does not present any additional conceptual complication.

The extension of the PT to theories beyond the SM should pose (at
least at the one-loop level) no problems at all, and has been recently
pursued in the MSSM \cite{Espinosa:2002cd}, in the context of finding a 
scale and gauge independent definition of mixing angles for scalar
particles. It remains to be seen what Feynman rules give rise to these 
super-PT 
Green's functions in theories possessing many scalars in the spectrum
(such as the MSSM), since in the latter case (contrary to the SM one) 
one has, in principle, 
the freedom to include (or not) in the BFM gauge fixing function
also scalars that do not posses a vev (such as squarks), therefore
symmetrizing (or not) the corresponding ${\cal S}^2{\cal V}$ coupling.
To the best of our knowledge, it is not yet clear
which one (if any) of the above possible choices will
constitute the super-BFM gauge of the MSSM. 
However notice that the PT procedure has no
arbitrariness whatsoever (contrary to the claims in \cite{Espinosa:2002cd}), 
and moreover naturally select a super-BFM gauge fixing 
function where all the scalars 
have been put, since we have seen that one should pinch whenever a longitudinal
momenta appears in the external tree-level vertex of type ({\it i})
diagrams (and this will happen for all the vertices of the type 
${\cal S}^2{\cal V}$, irrespectively from whether or not 
the scalar ${\cal S}$ has a vev). 
We conjecture that the MSSM super-BFM gauge will coincide with
the one dynamically projected out by the super-PT procedure.

Clearly, it would also be very interesting  to reach a
deeper understanding of why the PT dynamically singles out the BFG value  
$\xi_Q = 1$. There are several pieces of evidence that the BFG is somehow special; 
for example, the anomalous one-loop SUSY breaking in Yang-Mills theories 
with local coupling
is uniquely determined by the BFG value of a certain vertex function
\cite{Kraus:2001id}.
One possible direction to explore 
would be to look  for special properties of the BFM action
at  $\xi_Q  =   1$;  an  interesting  3D   example  of  a
field-theory, which,  when formulated  in the background  Landau gauge
($\xi_Q  =  0$),  displays  an  additional  (non-BRST  related)  rigid
super-symmetry, is given in \cite{Birmingham:1990ap}. 

Concluding, we think that
many of the open questions related to the PT procedure, such as its 
uniqueness, independence from
the test particles and especially its generalization beyond leading order,
have been fully addressed and solved, both
in theories with, as well as without, spontaneous symmetry breaking.
Moreover, a deep connection between the PT and powerful BRST related 
algebraic formalisms (such as the ones of Batalin-Vilkovisky and Nielsen) has
been unveiled, allowing the PT to acquire the status of a well-defined formal
tool.

\ack
The author is deeply indebted to J. Papavassiliou, for suggesting him
this project, countless discussions about it, constant support, and a critical reading of 
the manuscript.
Useful communications with   
I. Bandos, M. Passera, P. Gambino and G. Barnich, are  also 
greatfully acknowledge.
Finally, he would like to thank the Physics Department of the University 
of Valencia, where the most part of this work has been carried out,
for providing such a stimulating atmosphere, 
and especially Prof. J. Bernab\'{e}u for financial support. \\

All the figures of this paper have been drawn with {\tt JaxoDraw} \cite{Binosi:2003yf}.
Visit the {\tt JaxoDraw} home page 
\verb|http://altair.ific.uv.es/~JaxoDraw/home.html| to know more about it.

\appendix

\section{Slavnov-Taylor identities}

In this appendix we briefly discuss the derivation of the (on-shell)
STIs needed, by taking as a case study the STI satisfied by the ${\cal
T}_{V_\mu W^-_\nu }$ amplitude when hitting from the neutral gauge
boson side.
\begin{center}
\begin{figure}
\includegraphics[width=11.3cm]{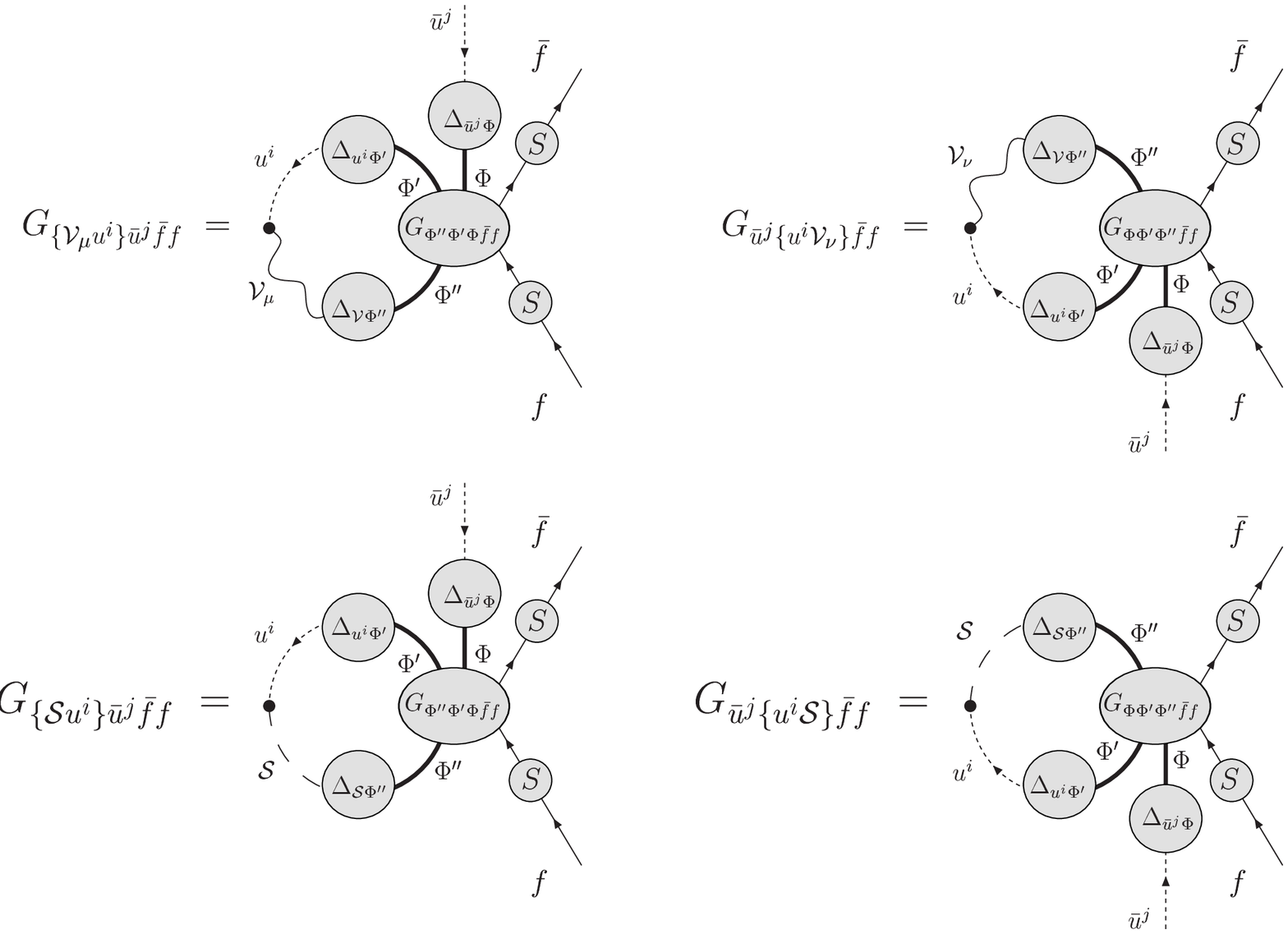}
\includegraphics[width=11.3cm]{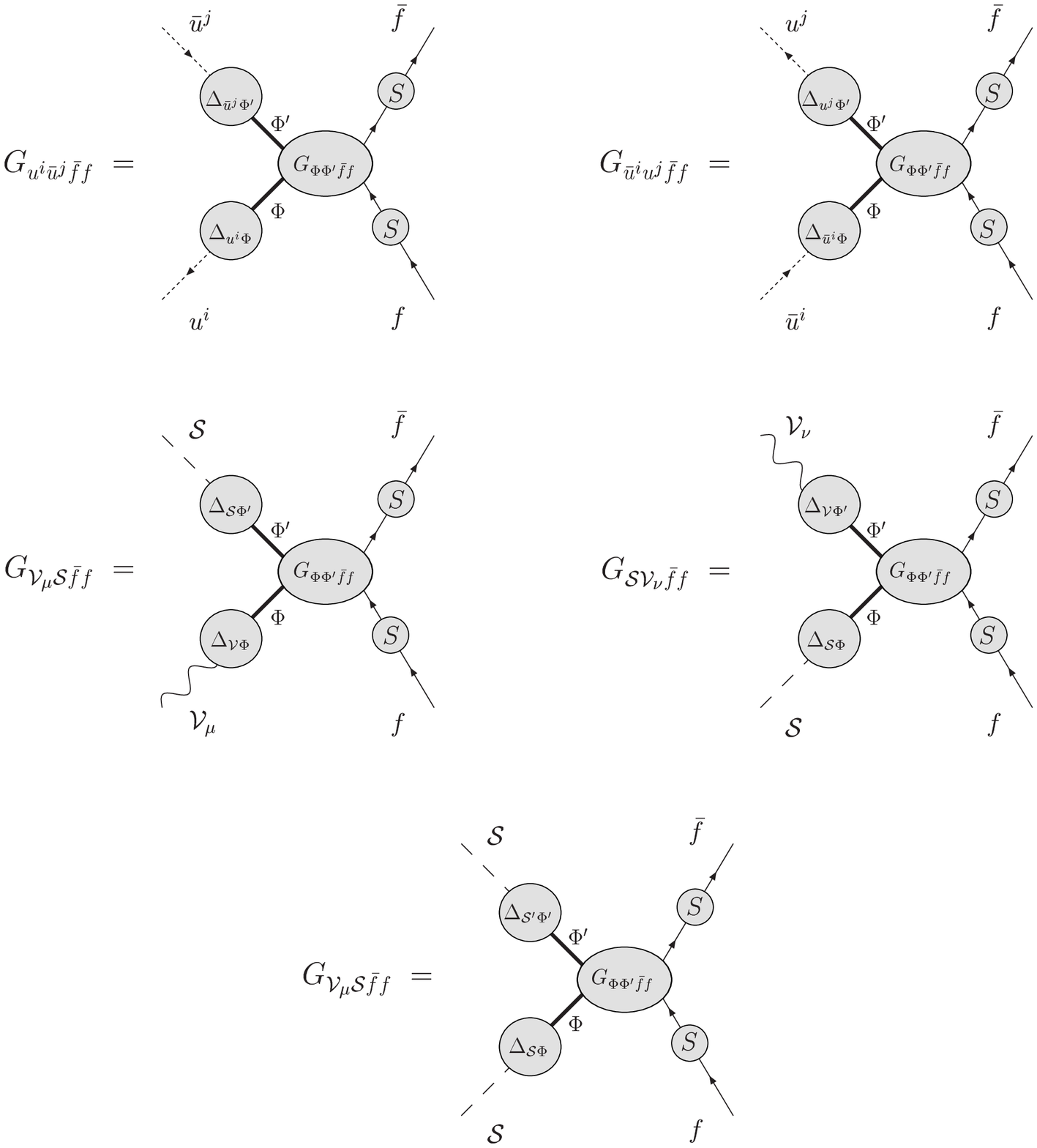}
\caption{\label{figGreen} Diagrammatic representation of the various
Green's function appearing in the STIs. Here the indices $i$ and $j$ run over the four possible values $+,-,A,Z$. The corresponding (on-shell) Green's functions ${\cal G}$ appearing in the formulas, are obtained from the above one by truncating the external legs (with the exclusion the ones that end at a common point).}
\end{figure}
\end{center}
For getting this STI one starts by observing that the BRST
transformation of an antighost-field $\bar u^{\cal V}$ starts with the
term $\partial_\mu{\cal V}^\mu$ (which is $\sim k_\mu{\cal V}^\mu$
in momentum space). In this particular case we then
trade the gauge boson field
$V_\mu$ for the anti-ghost field $\bar\uv$, and consider the trivial 
(due to ghost charge conservation) identity
\be
\left\langle T[\bar\uv(x)W^-_\nu(y)f_1(z)f_2(w)]\right\rangle =0,
\ee
leaving the fermionic fields unspecified. By applying the BRST
operator $s$ to the above identity, and passing to momentum space, we get the 
identity
\bea
k_{1}^{\mu} G_{V_\mu W^-_\nu }\mom&+&k_{2\nu}G_{\bar\uv\um }\mom+i\Mv
G_{\schi W^-_\mu  }\mom \nonumber \\
&+&\gw\sumvp\cvp [G_{\bar\uv \{W^-_\nu \uvp\} }\mom-
G_{\bar\uv\{\um\Vp_\nu\} }\mom]\nonumber\\
&+&X_\nu\mom=0,
\label{stiex}
\eea
where (FT stands for Fourier transform) 
\be
X_\nu\mom\stackrel{\scriptscriptstyle{\mathrm{FT}}}{=}-
\left\langle T\left\{\bar\uv(x)W^-_\nu(y)s[f_1(z)f_2(w)]\right\}
\right\rangle,
\label{x}
\ee
and two fields between brackets are taken to be at the same space-time
position.  
A diagrammatic representation of the Green's functions appearing
above is shown in Fig.(\ref{figGreen}). 

Now let us concentrate on the $X_\nu$ terms. From the BRST variations
of Eq.(\ref{BRST}), it is clear that independently of the type of fermions
appearing in Eq.(\ref{x}),  these terms will miss an
external fermion propagator, and will thus vanish due to the on-shell
condition. At one-loop order, for example, these terms represent simply 
the ones proportional to
the inverse tree-level propagator appearing in the PT
calculations. Indeed, we can multiply both sides of the Eq.(\ref{stiex})
by the product $S_1^{-1}(p_1)S_2^{-1}(p_2)$ of the two inverse propagator
of the external fermions, and then sandwich the amplitude between
their spinors: due to the on-shell conditions the vanishing
of the aforementioned terms follows by virtue of the Dirac equation.
One should observe that the vanishing of these on-shell pieces is in
fact the reason for the (all order) process independence of the PT algorithm 
\cite{Watson:1994tn,Binosi:2002ft}.

Thus one arrives at the following (on-shell) STI
\be
k_{1}^{\mu}{\cal T}_{V_\mu W^-_\nu }\mom={\cal R}^V_{W^-_\nu  }\mom,
\label{sticgbs0}
\ee
where, omitting the spinors,
\bea
\fl
{\cal T}_{V_\mu W^-_\nu  }\mom&=&\Delta_{V_\mu\Phi}(k_1)
\Delta_{W^-_\nu{\Phi'}}(k_2){\cal G}_{\Phi {\Phi'}  }\mom\nonumber \\
\fl
{\cal R}^{V}_{W^-_\nu }\mom&=&
-k_{2\nu}
\Delta_{\bar\uv\Phi}(k_1)\Delta_{\um\Phi'}(k_2)
{\cal G}_{\Phi\Phi' }\mom\nonumber \\
&-& i\Mv \Delta_{\chi\Phi}(k_1)\Delta_{W^-_\nu\Phi'}(k_2)
{\cal G}_{\Phi\Phi' }\mom\nonumber \\
&-&\gw\sumvp\cvp\Delta_{\bar\uv\Phi}(k_1)\left[
{\cal G}_{\Phi \{W^-_\nu \uvp\}}\mom
-{\cal G}_{\Phi \{V'_\nu\um\}}\mom\right].
\label{Rcgbs1}
\eea
In Eq.(\ref{Rcgbs1}) (as in the rest of these Appendices), a sum over
twice repeated fields (running over all the allowed SM combinations)
is intended.

In what follows we list all the remaining on-shell STIs used
throughout the paper.

\subsection{Gauge-boson sector}

\subsubsection{Charged sector}

In addition to the STI of Eq.(\ref{stiex}) we need the following
identities
\bea
k_{2}^{\nu}{\cal T}_{\sV_\mu W^-_\nu  }\mom&=&
{\cal R}_{\sV_\mu }^{'W^- }\mom, \nonumber \\
k_{1}^{\mu}{\cal T}_{ W^-_\mu \sV_\nu }\mom&=&
{\cal R}^{W^-}_{\sV_\nu }\mom,\nonumber \\
k_{2}^{\nu}{\cal T}_{ W^-_\mu \sV_\nu }\mom&=&
{\cal R}^{ ' \sV}_{W^-_\mu }\mom,
\label{sticgbs1}
\eea
where
\bea
\fl
{\cal R}_{\sV_\mu }^{'W^- }\mom
&=&
k_{1\mu}
\Delta_{\uv\Phi}(k_1)\Delta_{\bar\up{\Phi'}}(k_2){\cal G}_{\Phi{\Phi'}
  }\mom \nonumber \\
&-&\Mw \Delta_{V_\mu\Phi}(k_1)\Delta_{\phim{\Phi'}}(k_2)
{\cal G}_{\Phi{\Phi'} }\mom\nonumber \\
&+&\gw\cv\Delta_{\bar\up\Phi}(k_2)\left[
{\cal G}_{\{W^+_\mu\um\}\Phi }\mom-{\cal
G}_{\{ W^-_\mu \up\}\Phi }\mom\right],\nonumber \\
\fl
{\cal R}^{W^-}_{\sV_\nu }\mom
&=&
-k_{2\nu}
\Delta_{\bar\up\Phi}(k_1)\Delta_{\uv{\Phi'}}(k_2)
{\cal G}_{\Phi{\Phi'} }\mom\nonumber \\
&-&\Mw\Delta_{\phim\Phi}(k_1)\Delta_{V_\nu{\Phi'}}(k_2)
{\cal G}_{\Phi{\Phi'} }\mom\nonumber \\
&-&\gw\cv\Delta_{\bar\up\Phi}(k_1)\left[{\cal G}_{\Phi\{ W^+_\nu \um\} }\mom-
{\cal G}_{\Phi \{W^-_\nu \up\}}\mom\right],\nonumber \\
\fl
{\cal R}^{ ' \sV}_{W^-_\mu }
\mom&=&
k_{1\mu}
\Delta_{\um\Phi}(k_1)\Delta_{\bar\uv{\Phi'}}(k_2)
{\cal G}_{\Phi{\Phi'} }\momt\nonumber \\
&-&i\Mv
\Delta_{W^-_\mu\Phi}(k_1)\Delta_{\chi{\Phi'}}(k_2)
{\cal G}_{\Phi{\Phi'} }\mom\nonumber\\
&+&\gw\sumvp\cvp\Delta_{\bar\uv\Phi}(k_2)\left[
{\cal G}_{ \{W^-_\mu \uvp\}\Phi }\mom
-{\cal G}_{\{V'_\mu\um\}\Phi }\mom\right].
\label{Rcgbs2}
\eea

\subsubsection{Neutral sector}

In this case the STIs needed are just
\bea
k_{1}^{\mu}{\cal T}_{W^+_\mu W^-_\nu }\mom&=&
{\cal R}^{W^+}_{W^-_\nu }\mom,\nonumber \\
k_{2}^{\nu}{\cal T}_{  W^+_\mu W^-_\nu }\mom&=&
{\cal R}^{W^-}_{W^+_\mu }\mom,
\label{stingbs}
\eea
where we have
\bea
\fl
{\cal R}^{W^+}_{W^-_\nu }\mom
&=&
-k_{2\nu}\Delta_{\bar\um\Phi}(k_1)\Delta_{\um{\Phi'}}(k_2) 
{\cal G}_{\Phi{\Phi'} }\mom
\nonumber \\
&+&\Mw\Delta_{\phip\Phi}(k_1)\Delta_{W^-_\nu{\Phi'}}(k_2)
{\cal G}_{\Phi\Phi'  }\mom \nonumber \\
&-&\gw\sumv\cv\Delta_{\bar\um\Phi}(k_1)[
{\cal G}_{\Phi\{W^-_\nu \uv\}}\mom-
{\cal G}_{\Phi\{V_\nu\um\}}\mom],
\nonumber \\
\fl
{\cal R}^{W^-}_{W^+_\mu }\mom
&=&
k_{1\mu}\Delta_{\up\Phi}(k_1)\Delta_{\bar\up{\Phi'}}(k_2) 
{\cal G}_{\Phi{\Phi'} }\mom
\nonumber \\
&-&\Mw\Delta_{W^+_\mu\Phi}(k_1)\Delta_{\phim{\Phi'}}(k_2)
{\cal G}_{\Phi{\Phi'} }\mom \nonumber \\
&-&\gw\sumv\cv\Delta_{\bar\up\Phi}(k_2)[
{\cal G}_{\{ W^+_\mu \uv\}\Phi }\mom-{\cal
G}_{\{V_\mu\up\}\Phi }\mom].
\label{Rngbs}
\eea

\subsection{Scalar bosons sector}

\subsubsection{Charged sector}

The STIs employed in this case reads
\bea
k_{1}^{\mu}{\cal T}_{V_\mu\phim }\mom&=&
{\cal R}^{V}_{\phim }\mom,\nonumber \\
k_{2}^{\nu}{\cal T}_{\phim\sV_\nu }\mom&=&
{\cal R}_{\phim }^{'\sV}\mom,\nonumber \\
k_{1}^{\mu}{\cal T}_{ W^-_\mu \rS }\mom&=&
{\cal R}^{W^-}_{\rS }\mom,\nonumber \\
k_{2}^{\nu}{\cal T}_{\rS W^-_\nu }\mom&=&
{\cal R}_{\rS }^{'W^- }\mom,
\label{sticss}
\eea
where $\rS$ runs over the neutral would-be Goldstone bosons only, {\it
i.e.}, $\rS=\{\chi,H\}$. In particular we have
\bea
{\cal R}^{V}_{\phim }\mom&=&
\Mw\Delta_{\bar\uv\Phi}(k_1)\Delta_{\um{\Phi'}}(k_2)
{\cal G}_{\Phi{\Phi'} }\mom \nonumber \\
&-&
i\Mv\Delta_{\chi\Phi}(k_1)\Delta_{\phim{\Phi'}}(k_2)
{\cal G}_{\Phi{\Phi'} }\mom
\nonumber \\
&-&i\frac\gw2\sums\cs\Delta_{\bar\uv\Phi}(k_1)
  {\cal G}_{\Phi \{\rS\um\}}\mom
\nonumber \\
&+&\gw\sumvp\cvp'\Delta_{\bar\uv\Phi}(k_1)
{\cal G}_{\Phi\{\phim\uvp\}}\mom
\nonumber \\
{\cal R}_{\phim }^{'\sV}\mom&=&
-\Mw
  \Delta_{\um\Phi}(k_1)\Delta_{\bar\uv{\Phi'}}(k_2)
{\cal G}_{\Phi{\Phi'} }\mom 
\nonumber \\
&-&i\Mv\Delta_{\phim\Phi}(k_1)\Delta_{\chi{\Phi'}}(k_2)
{\cal G}_{\Phi{\Phi'} }\mom
\nonumber \\
&+&i\frac\gw2\sums\cs\Delta_{\bar\uv\Phi}(k_2)
  {\cal G}_{\{\rS\um\}\Phi }\mom
\nonumber \\
&-&\gw\sumvp\cvp'\Delta_{\bar\uv\Phi}(k_2)
{\cal G}_{\{\phim\uvp\}\Phi }\mom, 
\label{Rcss1}
\eea
with $C_\chi=1$ and $C_H=i$, and
\bea
\fl
{\cal R}^{W^-}_{ \rS }\mom&=&
-i\Mz\delta_{\rS\schi}\Delta_{\bar\up\Phi}(k_1)\Delta_{\uz{\Phi'}}(k_2)
  {\cal G}_{\Phi{\Phi'} }\mom \nonumber\\
&-&\Mw\Delta_{\phim\Phi}(k_1)\Delta_{\rS{\Phi'}}(k_2)
{\cal G}_{\Phi{\Phi'}  }\mom
  \nonumber\\ 
&+&i\frac\gw2\cs
\Delta_{\bar\up\Phi}(k_1)\left[{\cal G}_{\Phi\{\phip\um\} }\mom+\uno
{\cal G}_{\Phi\{\phim\up\} }\mom\right]\nonumber\\
&-&i\uno \frac\gw{2\cw}
\Delta_{\bar\up\Phi}(k_1)
{\cal G}_{\Phi \{\overline \rS\uz\} }\mom,\nonumber\\ 
\fl
{\cal R}_{\rS }^{'W^- }\mom&=&i\Mz\delta_{\rS\schi}
\Delta_{\uz\Phi}(k_1)\Delta_{\bar\up{\Phi'}}(k_2)
  {\cal G}_{\Phi{\Phi'} }\mom \nonumber\\
&-&\Mw\Delta_{\rS\Phi}(k_1)\Delta_{\phim{\Phi'}}(k_2)
{\cal G}_{\Phi{\Phi'} }\mom
  \nonumber\\ 
&-&i\frac\gw2\cs
\Delta_{\bar\up\Phi}(k_2)\left[{\cal G}_{\{\phip\um\}\Phi }\mom+\uno
{\cal G}_{\{\phim\up\}\Phi }\mom\right]\nonumber\\
&+&i\uno \frac\gw{2\cw}\Delta_{\bar\up\Phi}(k_2)
{\cal G}_{\{\overline \rS\uz\}\Phi }\mom,
\label{Rcss2}
\eea
where $\uno$ is $1$ (respectively, $-1$) when $\rS=\chi$ (respectively, $\rS=H$), and
$\overline\rS=H$ (respectively, $\overline\rS=\chi$) when $\rS=\chi$ 
(respectively, $\rS=H$). 

\subsubsection{Neutral sector}

Finally, in the neutral scalar sector the following STIs appear
\bea
k_{1}^{\mu}{\cal T}_{ W^+_\mu \phim }\mom&=&
{\cal R}^{W^+}_{\phim }\mom,\nonumber \\
k_{2}^{\nu}{\cal T}_{\phip W^-_\nu  }\mom&=&
{\cal R}_{\phip }^{ W^- }\mom,\nonumber \\
k_{1}^{\mu}{\cal T}_{Z_\mu\rS }\mom&=&
{\cal R}^{Z}_{\rS }\mom,\nonumber \\
k_{2}^{\nu}{\cal T}_{\rS Z_\nu }\mom&=&
{\cal R}_{\rS }^{'Z}\mom,
\label{stinss}
\eea
where we have
\bea
{\cal R}^{W^+}_{ \phim }\mom&=&\Mw\Delta_{\bar\um\Phi}(k_1)
\Delta_{\um{\Phi'}}(k_2){\cal G}_{\Phi{\Phi'} }\mom\nonumber \\
&+&\Mw\Delta_{\phip\Phi}(k_1)
\Delta_{\phim{\Phi'}}(k_2){\cal G}_{\Phi{\Phi'} }\mom\nonumber \\
&-&i\frac\gw2\sums\cs\Delta_{\bar\um\Phi}(k_1)
{\cal G}_{\Phi\{\rS\um\} }\mom\nonumber \\
&+&\gw\sumv\cv'\Delta_{\bar\um\Phi}(k_1){\cal G}_{\Phi\{\phim\uv\} }\mom,
\nonumber \\
{\cal R}_{\phip }^{ W^- }\mom&=&\Mw\Delta_{\up\Phi}(k_1)
\Delta_{\bar\up{\Phi'}}(k_2)
{\cal G}_{\Phi{\Phi'} }\mom\nonumber \\
&-&\Mw\Delta_{\phip\Phi}(k_1)
\Delta_{\phim{\Phi'}}(k_2){\cal G}_{\Phi{\Phi'} }\mom\nonumber \\
&+& i\frac\gw2\sums\uno\cs\Delta_{\bar\up\Phi}(k_2)
{\cal G}_{\{\rS\up\}\Phi }\mom\nonumber \\
&+&\gw\sumv\cv'\Delta_{\bar\up\Phi}(k_2){\cal G}_{\{\phip\uv\}\Phi }\mom,
\label{Rnss1}
\eea
and
\bea
\fl
{\cal R}^{Z}_{\rS }\mom&=&-i\delta_{\rS\chi}
\Mz\Delta_{\bar\uz\Phi}(k_1)\Delta_{\uz{\Phi'}}(k_2)
{\cal G}_{\Phi{\Phi'} }\mom
\nonumber \\
&-&
i\Mz\Delta_{\chi\Phi}(k_1)\Delta_{\rS{\Phi'}}(k_2)
{\cal G}_{\Phi{\Phi'} }\mom
\nonumber \\
&+&i\frac\gw2\cs\Delta_{\bar\uz\Phi}(k_1)\left[
{\cal G}_{\Phi\{\phip\um\} }\mom+
\uno{\cal G}_{\Phi\{\phim\up\} }\mom\right]
\nonumber \\
&-&i\uno
\frac\gw{2\cw}\Delta_{\bar\uz\Phi}(k_1)
{\cal G}_{\Phi\{\overline \rS\uz\} }\mom,
\nonumber \\
\fl
{\cal R}_{\rS }^{'Z}\mom&=&
i\delta_{\rS\chi}\Mz\Delta_{\uz\Phi}(k_1)\Delta_{\bar\uz{\Phi'}}(k_2)
{\cal G}_{\Phi{\Phi'} }\mom
\nonumber \\
&-&
i\Mz\Delta_{\rS\Phi}(k_1)\Delta_{\chi{\Phi'}}(k_2)
{\cal G}_{\Phi{\Phi'} }\mom
\nonumber \\
&-&i\frac\gw2\cs\Delta_{\bar\uz\Phi}(k_2)\left[
{\cal G}_{\{\phip\um\}\Phi }\mom+
\uno{\cal G}_{\{\phim\up\}\Phi }\mom\right]
\nonumber \\
&+&i\uno
\frac\gw{2\cw}\Delta_{\bar\uz\Phi}(k_2)
{\cal G}_{\{\overline \rS\uz\}\Phi }\mom.
\label{Rnss2}
\eea

\section{Background quantum identities}

In this appendix we report the BQIs used in the paper. Details on how
to derive them can be found in \cite{Grassi:1999tp,Binosi:2002ez}.

\subsection{Two-point functions}
The BQIs for the two-point functions read
\bea
\fl
\Gamma_{\widehat{\cal V}_\alpha\widehat{{\cal V}'}_\beta}(q)&=&
\Gamma_{{\cal V}_\alpha{{\cal V}'}_\beta}(q)+
\Gamma_{\Omega^{\cal V}_\alpha {{\cal
    V}''}^*_\mu}(q)\Gamma_{{{\cal V}''}^\mu{{\cal V}'}_\beta}(q)+
 \Gamma_{\Omega^{\cal V}_\alpha {{\cal
    S}}^*}(q)\Gamma_{{\cal S}{\cal V}'_\beta}(q)\nonumber \\
\fl    
&+&
\Gamma_{\Omega^{{\cal V}'}_\beta{{\cal V}''}^*_\mu}(q)\left[
\Gamma_{{\cal V}_\alpha{{\cal V}''}^\mu}(q)+\Gamma_{\Omega^{\cal
V}_\alpha{{\cal V}'''}^*_\rho}(q)\Gamma_{{{\cal V}'''}^\rho{{\cal
      V}''}^\mu}(q)+\Gamma_{\Omega^{\cal V}_\alpha{\cal S}^*}(q)
\Gamma_{{\cal S}{{\cal V}''}^\mu}(q)\right]\nonumber \\
\fl
&+&
\Gamma_{\Omega^{{\cal V}'}_\beta{{\cal S}}^*}(q)\left[
\Gamma_{{\cal V}_\alpha{\cal S}}(q)+\Gamma_{{\Omega^{\cal
	V}_\alpha}{{\cal V}''}^*_\mu}(q)\Gamma_{{{\cal V}''}^\mu{\cal
      S}}(q)+\Gamma_{\Omega^{\cal V}_\alpha{{\cal S}'}^*}(q)
\Gamma_{{\cal S}'{\cal S}}(q)\right],\nonumber \\
\fl
\Gamma_{\widehat{\cal V}_\alpha\widehat{\cal S}}(q)&=&
\Gamma_{{\cal V}_\alpha{\cal S}}(q)+
\Gamma_{\Omega^{\cal V}_\alpha {{\cal
    V}'}^*_\mu}(q)\Gamma_{{{\cal V}'}^\mu{\cal S}}(q)+
 \Gamma_{\Omega^{\cal V}_\alpha {{\cal
    S}'}^*}(q)\Gamma_{{\cal S}'{\cal S}}(q)\nonumber \\
\fl
&+&
\Gamma_{\Omega^{\cal S}{{\cal V}'}^*_\mu}(q)\left[
\Gamma_{{\cal V}_\alpha{{\cal V}'}^\mu}(q)+\Gamma_{\Omega^{{\cal
	V}'\mu}{{\cal V}''}^{*\rho}}(q)\Gamma_{{\cal V}''_\rho{\cal
      V}_\alpha}(q)+\Gamma_{\Omega^{{\cal V}'\mu}{{\cal S}'}^*}(q)
\Gamma_{{\cal S}'{\cal V}_\alpha}(q)\right]\nonumber \\
\fl
&+&
\Gamma_{\Omega^{\cal S}{{\cal S}'}^*}(q)\left[
\Gamma_{{\cal V}_\alpha{\cal S}'}(q)+\Gamma_{{\Omega^{{\cal
	V}\alpha}}{{\cal V}'}^*_\mu}(q)\Gamma_{{{\cal V}'}^\mu{\cal
      S}'}(q)+\Gamma_{\Omega^{{\cal V}_\alpha}{{\cal S}''}^*}(q)
\Gamma_{{\cal S}''{\cal S}'}(q)\right],\nonumber \\
\fl
\Gamma_{\widehat{\cal S}\widehat{\cal V}_\beta}(q)&=&
\Gamma_{{\cal S}{\cal V}_\beta}(q)+
\Gamma_{\Omega^{\cal S}{{\cal
    V}'}^*_\mu}(q)\Gamma_{{{\cal V}'}^\mu{\cal V}_\beta}(q)+
 \Gamma_{\Omega^{\cal S}{{\cal
    S}'}^*}(q)\Gamma_{{\cal S}'{\cal V}_\beta}(q)\nonumber \\
\fl
&+&
\Gamma_{\Omega^{\cal V}_\beta{{\cal V}'}^*_\mu}(q)\left[
\Gamma_{{\cal S}{{\cal V}'}^\mu}(q)+\Gamma_{\Omega^{{\cal
	S}}{{\cal V}''}^*_\rho}(q)\Gamma_{{{\cal V}''}^\rho{{\cal
      V}'}^\mu}(q)+\Gamma_{\Omega^{{\cal S}}{{\cal S}'}^*}(q)
\Gamma_{{\cal S}'{{\cal V}'}^\mu}(q)\right]\nonumber \\
\fl
&+&
\Gamma_{\Omega^{\cal V}_\beta{{\cal S}'}^*}(q)\left[
\Gamma_{{\cal S}{\cal S}'}(q)+\Gamma_{{\Omega^{{\cal
	S}}}{{\cal S}''}^*}(q)\Gamma_{{{\cal S}'}{\cal
      S}''}(q)+\Gamma_{\Omega^{{\cal S}}{{\cal V}'}^*_\mu}(q)
\Gamma_{{{\cal V}'}^\mu{\cal S}'}(q)\right], \nonumber \\
\fl
\Gamma_{\widehat{\cal S}\widehat{\cal S}'}(q)&=&\Gamma_{{\cal S}{\cal
    S}'}(q)+
 \Gamma_{\Omega^{\cal S}{\cal
    V}^*_\mu}(q)\Gamma_{{\cal V}^\mu{\cal S}'}(q)+
\Gamma_{\Omega^{\cal S}{{\cal
    S}''}^*}(q)\Gamma_{{{\cal S}''}{\cal S}'}(q)\nonumber \\
\fl    
&+&
\Gamma_{\Omega^{{\cal S}'}{{\cal V}}^*_\mu}(q)\left[
\Gamma_{{\cal S}{{\cal V}}^\mu}(q)+\Gamma_{\Omega^{{\cal
	S}}{{\cal V}'}^*_\rho}(q)\Gamma_{{{\cal V}'}^\rho{{\cal
      V}}^\mu}(q)+\Gamma_{\Omega^{{\cal S}}{{\cal S}''}^*}(q)
\Gamma_{{\cal S}''{{\cal V}}^\mu}(q)\right]\nonumber \\
\fl
&+&
\Gamma_{\Omega^{\cal S}{{\cal S}''}^*}(q)\left[
\Gamma_{{\cal S}{\cal S}''}(q)+\Gamma_{{\Omega^{{\cal
	S}}}{{\cal S}'''}^*}(q)\Gamma_{{{\cal S}'''}{\cal
      S}''}(q)+\Gamma_{\Omega^{{\cal S}}{{\cal V}}^*_\mu}(q)
\Gamma_{{{\cal V}}^\mu{\cal S}''}(q)\right].
\label{APBQI}
\eea

\subsection{Three-point functions}

Neglecting pieces that vanish when considering 
the external fermions to be on-shell,
the BQIs for the three-point functions read
\bea
\fl
\Gamma_{\widehat{\cal V}_\alpha \bar f f}(q,p_1,p_2)&=&
\Gamma_{{\cal V}_\alpha \bar f f}(q,p_1,p_2)+
\Gamma_{\Omega^{\cal V}_\alpha{{\cal V}'}^{*}_\mu}(q)
\Gamma_{{{\cal V}'}^{\mu}\bar f f}(q,p_1,p_2)+
\Gamma_{\Omega^{\cal V}_\alpha{\cal S}^*}(q)
\Gamma_{{\cal S}\bar f f}(q,p_1,p_2), \nonumber \\
\fl
\Gamma_{\widehat{\cal S}\bar f f}(q,p_1,p_2)&=&
\Gamma_{{\cal S}\bar f f}(q,p_1,p_2)+
\Gamma_{\Omega^{\cal S}{{\cal S}'}^{*}}(q)
\Gamma_{{\cal S}'\bar f f}(q,p_1,p_2)+
\Gamma_{\Omega^{\cal S}{\cal V}^*_\mu}(q)
\Gamma_{{\cal V}^\mu \bar f f}(q,p_1,p_2). \nonumber  \\
\label{AVBQI}
\eea
Notice that the Green's functions that provide the reorganization of the
Feynman diagrams for converting the three-point background functions
into the corresponding quantum ones are precisely the same that appear
in the two-point functions of Eq.(\ref{APBQI}).


\section*{References}

\begin{thebibliography}{10}
\expandafter\ifx\csname url\endcsname\relax
  \def\url#1{\texttt{#1}}\fi
\expandafter\ifx\csname urlprefix\endcsname\relax\def\urlprefix{URL }\fi

\bibitem{Cornwall:1982zr}
J.~M. Cornwall, Phys. Rev. D26
  (1982) 1453;
%
J.~M. Cornwall, J.~Papavassiliou,
  Phys. Rev. D40 (1989) 3474;
%
J.~Papavassiliou,  Phys. Rev. D41 (1990) 3179;
%
G.~Degrassi and A.~Sirlin,
Phys.\ Rev.\ D46 (1992) 3104.

\bibitem{Papavassiliou:1999az}
J.~Papavassiliou,  Phys. Rev. Lett. 84 (2000)
  2782;
%
Phys. Rev. D62 (2000) 045006.

\bibitem{Veltman:th}
M.~J.~Veltman,
Physica 29 (1963) 186;
%
W.~W.~Repko and C.~J.~Suchyta,
Phys.\ Rev.\ Lett.\ 62 (1989) 859;
%
S.~Dawson and S.~Willenbrock,
Phys.\ Rev.\ D40 (1989) 2880;
%
A.~Pilaftsis,
Z.\ Phys.\ C47 (1990) 95; 
%
A.~Dobado,
Phys.\ Lett.\ B237 (1990) 457;
%
A.~Sirlin,
Phys.\ Rev.\ Lett.\  67 (1991) 2127;
%
A.~Sirlin,
Phys.\ Lett.\ B267 (1991) 240;
%
R.~G.~Stuart,
Phys.\ Lett.\ B262 (1991) 113;
%
R.~G.~Stuart,
Phys.\ Lett.\ B272 (1991) 353;
%
R.~G.~Stuart,
Phys.\ Rev.\ Lett.\  70 (1993) 3193;
%
M.~Nowakowski and A.~Pilaftsis,
Z.\ Phys.\ C60 (1993) 121;
%
A.~Aeppli, G.~J.~van Oldenborgh and D.~Wyler,
Nucl.\ Phys.\ B428  (1994) 126;
%
U.~Baur and D.~Zeppenfeld,
Phys.\ Rev.\ Lett.\ 75 (1995) 1002;
%
M.~Beuthe, R.~Gonzalez Felipe, G.~Lopez Castro and J.~Pestieau,
Nucl.\ Phys.\ B498 (1997) 55;
%
D.~Wackeroth and W.~Hollik,
Phys.\ Rev.\ D55 (1997) 6788;
%
M.~Passera and A.~Sirlin,
Phys.\ Rev.\ D58 (1998) 113010;
%
B.~A.~Kniehl and A.~Sirlin,
Phys.\ Rev.\ Lett.\  81 (1998) 1373.
  
\bibitem{Papavassiliou:1995fq}
J.~Papavassiliou, A.~Pilaftsis, Phys.
  Rev. Lett. 75 (1995) 3060;
%
 Phys. Rev. D53 (1996) 2128;
%
Phys. Rev. D54 (1996) 5315;
%
Phys. Rev.
  Lett. 80 (1998) 2785;
%
Phys. Rev. D58 (1998) 053002.

\bibitem{Cornwall:1974km}
J.~M.~Cornwall, D.~N.~Levin and G.~Tiktopoulos,
Phys.\ Rev.\ D10 (1974) 1145;
%
C.~E.~Vayonakis,
Lett.\ Nuovo Cim.\ 17 (1976) 383;
%
M.~S.~Chanowitz and M.~K.~Gaillard,
Nucl.\ Phys.\ B261 (1985) 379;
%
G.~J.~Gounaris, R.~Kogerler and H.~Neufeld,
Phys.\ Rev.\ D34 (1986) 3257.

\bibitem{Fujikawa:1972fe}
K.~Fujikawa, B.~W. Lee, A.~I. Sanda, Phys. Rev. D6 (1972)
  2923.

\bibitem{Papavassiliou:1993ex}
J.~Papavassiliou, K.~Philippides,  Phys. Rev. D48 (1993)
  4255.

\bibitem{Papavassiliou:1994qe}
J.~Papavassiliou, C.~Parrinello, Phys. Rev. D50 (1994) 3059.

\bibitem{Bernabeu:2000hf}
J.~Bernabeu, L.~G. Cabral-Rosetti, J.~Papavassiliou, J.~Vidal,  Phys. Rev. D62 (2000) 113012.

\bibitem{Bernabeu:2002nw}
J.~Bernabeu, J.~Papavassiliou, J.~Vidal,  Phys. Rev. Lett. 89 (2002) 101802;
%
arXiv:hep-ph/0210055, Nucl. Phys. B to appear;
%
J.~Papavassiliou, J.~Bernabeu, D.~Binosi and J.~Vidal,
arXiv:hep-ph/0310028.

\bibitem{Degrassi:1993kn}
G.~Degrassi, B.~A. Kniehl, A.~Sirlin, Phys. Rev. D48 (1993) 3963.

\bibitem{Papavassiliou:1996hj}
J.~Papavassiliou, K.~Philippides, K.~Sasaki, Phys. Rev. D53 (1996)
  3942.

\bibitem{Nadkarni:1988ti}
S.~Nadkarni, Phys.
  Rev. Lett. 61 (1988) 396;
%
G.~Alexanian, V.~P. Nair, Phys. Lett. B352 (1995) 435;
%
M.~Passera, K.~Sasaki, Phys. Rev. D54 (1996) 5763;
%
K.~Sasaki, Phys. Lett. B369 (1996) 117.

\bibitem{Papavassiliou:1997fn}
J.~Papavassiliou, E.~de Rafael, N.~J.~Watson,  Nucl. Phys. B503 (1997) 79.
%
D.~Binosi, J.~Papavassiliou, Nucl.
  Phys. Proc. Suppl. 121 (2003) 281.

\bibitem{Hagiwara:1994pw}
K.~Hagiwara, S.~Matsumoto, D.~Haidt, C.~S. Kim,  Z. Phys. C64 (1994) 559.

\bibitem{Pilaftsis:1997dr}
A.~Pilaftsis,  Nucl. Phys. B504 (1997) 61.

\bibitem{Binosi:2001hy}
D.~Binosi, J.~Papavassiliou,  Phys. Rev. D65 (2002) 085003.

\bibitem{Yamada:2001px}
Y.~Yamada,  Phys.
  Rev. D64 (2001) 036008;
%
A.~Pilaftsis, Phys. Rev. D65 (2002) 115013.

\bibitem{Espinosa:2002cd}
J.~R. Espinosa, Y.~Yamada, Phys. Rev. D67 (2003) 036003.

\bibitem{Ferroglia:2003wa}
A.~Ferroglia, G.~Ossola, A.~Silin,
arXiv:hep-ph/0307200.

\bibitem{Binger:2003by}
M.~Binger, S.~J. Brodsky, arXiv:hep-ph/0310322.

\bibitem{Binosi:2002ft}
D.~Binosi, J.~Papavassiliou, Phys. Rev. D66
  (2002) 111901;
%
J. Phys. G30 (2004) 203;
%
arXiv:hep-ph/0310149.

\bibitem{Dewitt:1967ub}
B.~S. Dewitt, 
  Phys. Rev. 162 (1967) 1195;
%
J.~Honerkamp,  Nucl. Phys. B48 (1972)
  269;
%
R.~E. Kallosh,  Nucl. Phys.
  B78 (1974) 293;
%
H.~Kluberg-Stern, J.~B. Zuber,  Phys. Rev. D12 (1975) 482;
%
I.~Y. Arefeva, L.~D. Faddeev, A.~A. Slavnov,  Theor. Math. Phys. 21 (1975) 1165;
%
G.~'t~Hooft,  in: *Karpacz
  1975, Proceedings, Acta Universitatis Wratislaviensis*, Vol. 1, No. 368,
  Wroclaw, 1976, pp. 354;
%
S.~Weinberg, Phys. Lett. B91 (1980) 51;
%
L.~F. Abbott, Nucl. Phys. B185
  (1981) 189;
%
G.~M. Shore,  Ann. Phys. 137 (1981) 262;
%
L.~F. Abbott, M.~T. Grisaru, R.~K. Schaefer, Nucl. Phys. B229 (1983) 372;
%
C.~F. Hart, 
  Phys. Rev. D28 (1983) 1993;
%
A.~Rebhan, G.~Wirthumer,  Z.
  Phys. C28 (1985) 269.

\bibitem{Weinberg:kr}
S.~Weinberg, The Quantum Theory of Fields, Vol.~II, Cambridge University Press,
  New York, 1996.

\bibitem{Denner:1994nn}
A.~Denner, G.~Weiglein, S.~Dittmaier, Phys. Lett. B333 (1994)
  420;
%
S.~Hashimoto, J.~Kodaira, Y.~Yasui, K.~Sasaki, Phys. Rev. D50
  (1994) 7066.

\bibitem{Binosi:2002bs}
D.~Binosi, J.~Papavassiliou, Phys. Rev. D66 (2002) 076010.
 
\bibitem{Denner:1995xt}
A.~Denner, G.~Weiglein, S.~Dittmaier,  Nucl. Phys. B440 (1995) 95.

\bibitem{Batalin:1977pb}
I.~A. Batalin, G.~A. Vilkovisky, Phys. Lett. B69 (1977) 309;
%
Phys. Rev. D28 (1983) 2567.

\bibitem{Grassi:1999tp}
P.~A. Grassi, T.~Hurth, M.~Steinhauser, 
  Annals Phys. 288 (2001) 197;
%
Nucl. Phys. B610
  (2001) 215.

\bibitem{Binosi:2002ez}
D.~Binosi, J.~Papavassiliou, Phys. Rev. D66 (2002) 025024.

\bibitem{Barnich:2000zw}
G.~Barnich, F.~Brandt, M.~Henneaux, 
  Phys. Rept. 338 (2000) 439.

\bibitem{Gambino:1999ai}
P.~Gambino, P.~A. Grassi, Phys. Rev. D62 (2000) 076002.

\bibitem{Nielsen:1975fs}
N.~K. Nielsen, Nucl. Phys. B101 (1975) 173.

\bibitem{Cornwall:1976ii}
J.~M.~Cornwall and G.~Tiktopoulos,
Phys.\ Rev.\ D15 (1997) 2937.
 
\bibitem{Haeri:1988af}
B.~J.~Haeri,
Phys.\ Rev.\ D38 (1988) 3799;
%
A.~Pilaftsis,
Nucl.\ Phys.\ B487 (1997) 467.

\bibitem{Papavassiliou:1994pr}
J.~Papavassiliou,
Phys.\ Rev.\ D50 (1994) 5958.

\bibitem{Kraus:2001id}
E.~Kraus, Phys. Rev. D65 (2002) 105003.

\bibitem{Birmingham:1990ap}
D.~Birmingham, M.~Rakowski, G.~Thompson, Nucl. Phys. B329 (1990) 83.

\bibitem{Binosi:2003yf}
D.~Binosi and L.~Theussl,
arXiv:hep-ph/0309015.

\bibitem{Watson:1994tn}
N.~J.~Watson,
Phys.\ Lett.\ B349 (1995) 155.

\end{thebibliography}
\end{document}